 \documentclass[final,3p,times,twocolumn]{elsarticle}
\usepackage{amssymb}
\usepackage{amsmath}
\usepackage{amssymb}
\usepackage{graphicx}
\usepackage{tikz}
\usepackage[american]{babel}
\usepackage[utf8]{inputenc}
\usepackage[T1]{fontenc}
\usetikzlibrary{shapes}
\usepackage{subcaption}
\usepackage{tabularx}
\usepackage{booktabs}
\usepackage[colorlinks,linkcolor=blue,citecolor=blue,urlcolor=blue]{hyperref}% add hypertext capabilities
\usepackage{dcolumn}% Align table columns on decimal point
\usepackage{bm}% bold math
\usepackage{stmaryrd} % rr/llbracket
\usepackage{multirow}

\newcommand{\markershear}{\raisebox{0.5pt}{\protect\tikz{\protect\node[draw,scale=0.4,circle,fill=none,color=red, line width=0.25mm](){};}}}
\newcommand{\markeracp}{\raisebox{0.5pt}{\protect\tikz{\protect\node[draw,scale=0.25,regular polygon, regular polygon sides = 3,fill=none,color=blue, line width=0.25mm, shape border rotate=-90](){};}}}
\newcommand{\markeracm}{\raisebox{0.5pt}{\protect\tikz{\protect\node[draw,scale=0.25,regular polygon, regular polygon sides = 3,fill=none,color=black!60!green, line width=0.25mm, shape border rotate=90](){};}}}
\newcommand{\markernum}{\raisebox{0.5pt}{\protect\tikz[baseline=-0.4ex]{\protect\draw[color=black, line width=0.5mm] (-0.2,0)--(0.2,0);}}}

\newcommand{\reddashedline}{\raisebox{0.5pt}{\protect\tikz[baseline=-0.4ex]{\protect\draw[color=red, line width=0.3mm, dashed] (-0.4,0)--(0.4,0);}}}
\newcommand{\bluedashedline}{\raisebox{0.5pt}{\protect\tikz[baseline=-0.4ex]{\protect\draw[color=blue, line width=0.3mm, dashed] (-0.4,0)--(0.4,0);}}}
\newcommand{\greendashedline}{\raisebox{0.5pt}{\protect\tikz[baseline=-0.4ex]{\protect\draw[color=black!60!green, line width=0.3mm, dashed] (-0.4,0)--(0.4,0);}}}

\newcommand{\blackdashedlinesquare}{\raisebox{0.5pt}{\protect\tikz{\protect\node[draw,scale=0.3,regular polygon, regular polygon sides = 4, fill=none, color=black, line width=0.3mm](){};\protect\draw[color=black, line width=0.3mm, dashed] (-0.2,0)--(0.3,0);}}}

\newcommand{\greendashedlinestar}{\raisebox{0.5pt}{\protect\tikz{\protect\node[draw,scale=0.2, star, star points=5, star point ratio=2.25, fill=none, color=black!60!green, line width=0.3mm](){};\protect\draw[color=black!60!green, line width=0.3mm, dashed] (-0.2,0)--(0.3,0);}}}

\newcommand{\reddashedlinecircle}{\raisebox{0.5pt}{\protect\tikz{\protect\node[draw,scale=0.3,circle, fill=none, color=red, line width=0.3mm](){};\protect\draw[color=red, line width=0.3mm, dashed] (-0.2,0)--(0.3,0);}}}

\newcommand{\bluedashedlinetriangle}{\raisebox{0.5pt}{\protect\tikz{\protect\node[draw,scale=0.25, regular polygon, regular polygon sides = 3, fill=none, color=blue, line width=0.3mm](){};\protect\draw[color=blue, line width=0.3mm, dashed] (-0.2,0)--(0.3,0);}}}

\definecolor{slateblue}{HTML}{6A5ACD}

\definecolor{purple2}{HTML}{800080}

\newcommand{\blacklinefullcircle}{\raisebox{0.5pt}{\protect\tikz{\protect\node[draw,scale=0.2, circle, fill=black, color=black, line width=0.3mm](){};\protect\draw[color=black, line width=0.2mm] (-0.2,0)--(0.2,0);}}}

\begin{document}

\begin{frontmatter}



\journal{Phys. Rev. E}

\author[label1]{Gauthier Wissocq\corref{cor1}}
\ead{wissocq@cerfacs.fr}
 
\author[label2]{Christophe Coreixas}
\author[label1]{Jean-Fran\c{c}ois Boussuge}

\address[label1]{CERFACS, 42 Avenue G. Coriolis, 31057 Toulouse Cedex, France}
\address[label2]{Department of Computer Science, University of Geneva, 1204 Geneva, Switzerland}

\title{Linear stability of athermal regularized lattice Boltzmann methods}


\author{}

\address{}

\begin{abstract}
The present work is dedicated to a better understanding of the stability properties of regularized lattice Boltzmann (LB) schemes. To this extent, linear stability analyses of two-dimensional models are proposed: the standard Bhatnagar-Gross-Krook (BGK) collision model, the original pre-collision regularization and the recursive regularized model, where off-equilibrium distributions are partially computed thanks to a recursive formula. A systematic identification of the physical content carried by each LB mode is done by analyzing the eigenvectors of the linear systems. Stability results are then numerically confirmed by performing simulations of shear and acoustic waves. This work allows drawing fair conclusions on the stability properties of each model. In particular, recursive regularization turns out to be the most stable model for the D2Q9 lattice, especially in the zero-viscosity limit. Two major properties shared by every regularized model are highlighted: (1) a mode filtering property, and (2) an incorrect, and broadly anisotropic, dissipation rate of the modes carrying physical waves in under-resolved conditions. The first property is the main source of increased stability, especially for the recursive regularization. It is a direct consequence of the reconstruction of off-equilibrium populations before each collision process, decreasing the rank of the system of discrete equations. The second property seems to be related to numerical errors directly induced by the equilibration of high-order moments. In such a case, this property is likely to occur with any collision model %aiming at equilibrating high-order moments.
that follows such a stabilization methodology.
\end{abstract}


\end{frontmatter}

%\linenumbers

\section{\label{sec:intro}Introduction}

The lattice Boltzmann (LB) method is a powerful numerical approach for computational fluid dynamics (CFD)%~\cite{McNamara1988, Qian1992, Chen1992}
~\cite{GUO_Book_2013,HUANG_Book_2015,Kruger2017,SUCCI_Book_2018}. Based on a velocity-space discretization of the Boltzmann equation (BE)~\cite{Boltzmann1872} followed by an appropriate time and space discretization, its ``collide \& stream'' algorithm offers several advantages compared to its Navier-Stokes (NS) counterparts. Its natural Cartesian mesh generation allowing to easily deal with complex geometries~\cite{Touil2014}, together with a low-dissipative scheme~\cite{Marie2009} and an efficient and easily parallelizable algorithm~\cite{Schornbaum2016} have made it appealing for aeronautical applications~\cite{MANOHA_AIAA_2846_2015}.

 Yet, simulations of compressible flows, even for subsonic cases, and flows with large temperature variations are still challenging with a LB approach. This is due to two distinct phenomena. First, the reduction of the velocity space of the BE, together with an adapted choice of equilibrium distribution function towards which the collision process is done, leads to a Galilean invariance error in the equations. On standard lattices (\textit{e.g.} the well-known D2Q9 lattice~\cite{Qian1992}), this error has a macroscopic effect on the Navier-Stokes equations, involving an incorrect energy equation and a cubic error in Mach number in the viscous terms of the momentum equation%~\cite{SHAN2006, Philippi2006}
 ~\cite{QIAN_EPL_21_1993,CHEN_PRE_50_1994}. The latter is at the origin of an anti-dissipative behavior which can lead to negative viscosity when the Mach number becomes too large~\cite{Marie2009, Wissocq2019} and makes the system of equations unstable. Secondly, the most simple LB scheme, based on a Bhatnagar-Gross-Krook (BGK) collision model~\cite{Bhatnagar1954}, is subject to severe numerical instabilities when the Mach number and the temperature fluctuations increase, as well as in the zero-viscosity limit~\cite{Lallemand2000, Siebert2008a}. There is some numerical evidence that the presence of non-hydrodynamic modes, \textit{i.e.} unphysical waves inherited by the BE and unexpected by the NS equations, are responsible for unfortunate instabilities~\cite{Dellar2002, Lallemand2003}. Modal interactions occurring between these waves and hydrodynamic ones are consequent of numerical errors in the time and space discretization, and a so-called \textit{eigenvalue collision} phenomenon can lead to severe linear instabilities even for well-resolved cases~\cite{Wissocq2019}.

In order to handle the first problem due to a reduction of the velocity space, several approaches have been proposed in the literature. Since the error done at the macroscospic level can be explicitly known, \textit{e.g.} thanks to a Chapman-Enskog expansion~\cite{CHAPMAN_Book_3rd_1970}, it can be corrected by appropriate changes in the LB scheme. For instance, one can add a correction as a body-force term~\cite{Prasianakis2007}, or by modi\-fying the relaxation time of the collision model~\cite{Dellar2014}. Another solution is to increase the number of discrete velocities of the lattice, so that the Galilean invariance error does not impact the fluid modelling at the NS level~\cite{SHAN2006}. However, the numerical stability issue is even more prominent in that case because of the large number of non-hydrodynamic modes~\cite{Coreixas2017, Wissocq2019_these}. 

In order to try to get rid of the numerical stability issues, many authors have turned to the development of more sophisticated collision models than the BGK one. They can be classified into three main families. 
First, noticing that numerical issues are due to the presence of non-hydrodynamic modes, several models were built in order to damp them in a simulation. It is the purpose of multiple-relaxation-time (MRT) models, which rely on a specific relaxation time applied to each non-hydrodynamic moment. Various models therefore exist, depending on the adopted definition of moments and the choice of relaxation parameters. The original MRT models are based on a raw-moments formulation~\cite{DHumieres1994, Lallemand2000, DHumieres2002}. However, finding optimal values for each relaxation time rapidly became a tedious task~\cite{XU_JCP_230_2011,Xu2012}. That is why two-relaxation-time (TRT) models~\cite{Ginzburg2008, Ginzburg2010} have later been proposed, where odd and even parts of the distribution functions are relaxed at a specific characteristic time. Other models are based on a definition of central moments, which consists in a shift of the discrete velocities by the local flow velocity, and are also referred to as \textit{cascaded} models~\cite{Geier2006, LYCETTBROWN_CMWA_67_2014, Dubois2015, DeRosis2017,FEI_PRE_97_2018, DeRosis2019}. Finally, the collision process can also be performed on other macroscopic quantities called \textit{cumulants}~\cite{Geier2015}, which can be expressed as non-linear combination of the standard moments~\cite{Coreixas2019}. The second family of collision models is based on a completely different observation: there is no equivalent of the famous Boltzmann's $H$-theorem for the LB method. Indeed, if a $H$ function had a monotonous increase during a simulation, it could be considered as an \textit{entropy}, and numerical stability in the sense of Lyapunov would be ensured~\cite{LaSalle1961}. The purpose of the so-called LB entropic models is therefore to restore an equivalent of the $H$-theorem~\cite{Karlin1998, Boghosian2001, Ansumali2003, Karlin2014, Frapolli2015,ATIF_PRL_119_2017}. They are based on (1) an adapted choice of equilibrium distribution function as the maximal state of a pseudo-entropy $H$, and (2) a modified collision step ensuring the monotony of $H$. The third collision family is referred to as \textit{regularized} models, which is of interest in the present article. %The original model was formerly used by Ladd \& Verberg~\cite{Ladd2001} and later proposed by Latt \& Chopard~\cite{Latt2006} for its stability properties. 
The original model was formerly used by Skordos~\cite{SKORDOS_PRE_49_1993} as a way to (re)construct populations in the context of initial and boundary conditions. Ladd \& Verberg~\cite{Ladd2001} further relied on this approach to reduce memory consumption by only storing macroscopic quantities and their gradients. It is only with the work of Latt \& Chopard~\cite{Latt2006} that its stability properties were highlighted. It aims at filtering out the non-hydrodynamic content in off-equilibrium distribution functions before each collision process. In a sense, it can be viewed as a particular MRT model where relaxation parameters of high-order Hermite moments are set so that they are imposed at their equilibrium value at each collision step~\cite{Latt2007}. This approach has later been extended by Malaspinas~\cite{Malaspinas2015}, who proposed to reconstruct high-order off-equilibrium moments at each iteration thanks to a recursive relation obtained by a Chapman-Enskog expansion. %Such an approach was further assessed on high-order lattices, leading to a considerable increase in numerical stability~\cite{Mattila2017, Coreixas2017}. 
Such an approach was further assessed on high-order lattices, leading to an increase in numerical stability~\cite{Coreixas2017}. Interestingly, the latter recursive approach is equivalent to equilibrating high-order contributions of populations in the central Hermite~\cite{Mattila2017} and temperature-scaled central Hermite~\cite{LI_PRE_100_2019} moment spaces for isothermal and thermal models respectively~\cite{Coreixas2019,COREIXAS_RSTA_378_2020,HOSSEINI_RSTA_378_2020}.
Eventually, Jacob \textit{et al.}~\cite{Jacob2018} proposed a hybrid formulation of this recursive regularization, where the second-order off-equilibrium moment was partially reconstructed thanks to a finite-difference estimation of the shear stress tensor.

Some of these sophisticated collision models have since proven their worth to increase the numerical stability of %the LB method
standard LB models and seem able to overcome the compressible and transonic limit%~\cite{Frapolli2015, Feng2019, Renard2019, HOSSEINI_RSTA_378_2020}
~\cite{Feng2019, Renard2019}. However, and despite recent approaches aiming to compare and drawing links between different collision models%~\cite{Kramer2019, Coreixas2019}
~\cite{Coreixas2019}, the reasons that provide greater numerical stability properties to a given model are still unclear. %This question can especially 
As an example, this question can be raised in the case of the recursive regularized models. Indeed, one can wonder how enriching off-equilibrium distribution functions with high-order moments might lead to an enhanced stability, while the added terms are not supposed to contribute to the NS physics. To the author's knowledge, no convincing explanation has been provided so far regarding this question, and for good reasons: there is some evidence that the effect of a given model is mainly numerical, which makes the understanding of these phenomena very difficult.

A fairly simple method can be systematically employed to study the numerical properties of a given scheme, referred to as linear stability analyses. They rely on a linea\-rization of the algorithm about a mean flow and an investigation of the behavior of linear waves in the spectral space, in terms of propagation and dissipation. Initially proposed by von Neumann~\cite{VonNeumann1950}, this method was first adapted to the LB formalism by Sterling \& Chen~\cite{Sterling1996}. %It was later widely used to exhibit the linear properties of the BGK collision model~\cite{Worthing1997}, its order of precision compared to NS-based solvers~\cite{Marie2009}, and to optimize the choice of parameters of MRT models~\cite{Lallemand2000, Xu2012, ChvezModena2018}.
It was later widely used to exhibit the linear properties of the BGK collision model~\cite{Worthing1997}, its order of precision compared to NS-based solvers~\cite{Marie2009}, to optimize the choice of parameters of MRT models~\cite{Lallemand2000, Xu2012,HOSSEINI_PRE_99_2019b, ChvezModena2018}, and to evaluate the impact of collision models~\cite{Dubois2015,COREIXAS_RSTA_378_2020}, numerical discretizations
~\cite{Wilde2019}, or lattice shifting~\cite{HOSSEINI_PRE_100_2019} on linear stability domains.
An extended linear analysis has been recently proposed~\cite{Wissocq2019}, allowing a systematic identification of the linear modes thanks to the information contained by the eigenvectors of the linear problem. Such a technique highlighted two kinds of modal interactions occurring with the BGK collision model, namely a \textit{curve veering} phenomenon and an \textit{eigenvalue collision}. The former is responsible for the fact that each linear mode of the LB scheme can carry a superposition of physical waves (acoustics and shear), while the latter is the source of severe instabilities of the BGK model. Such analyses are crucial for a better understanding of some local numerical phenomena, for instance occurring at mesh refinement interfaces~\cite{ASTOUL_ARXIV_2020_11863}.

The aim of the present article is to perform such linear stability analyses to the aforementioned regularized collision models in two dimensions. The objectives are multiple: (1) draw fair conclusions regarding the numerical stability of each model, (2) clearly identify the behavior of the physical (acoustic and shear) waves expected by the NS equations in every spatial direction, and (3) provide a better understanding of the stability properties of each collision model. Especially, the effect of the pre-collision regularization on the modal interactions highlighted on the BGK model will be of particular interest.  

The present article is divided as follows. In Sec.~\ref{sec:LBM}, the LB scheme is recalled and the regularized schemes of interest in this work are introduced. In Sec.~\ref{sec:LSA}, the principle of linear stability analyses is recalled and applied to regularized collision models, involving the matrices of each linear system derived in~\ref{app:linear_matrices}. In Sec.~\ref{sec:LSA_results}, the main results obtained with the D2Q9 and D2V17 lattices are provided (see~\ref{app:lattices} for their structure). They are further numerically validated thanks to simulations of shear and acoustic waves in a two-dimensional LB solver in Sec.~\ref{sec:numerical_validation}. Finally, Sec.~\ref{sec:discussion} summarizes two major properties exhibited by the linear analyses and aims at providing theoretical explanations for the observed phenomena. These properties are further highlighted by considering a so-called analytically-regularized scheme.

\section{The lattice Boltzmann method\label{sec:LBM}}

In the following, LB schemes will be recalled for the very standard BGK collision model and regularized ones. All studies will be restricted to athermal LB methods on two-dimensional lattices of $V$ velocities $(\boldsymbol{e_i})_{i\in \llbracket 1, V \rrbracket}$. Note that in all the following, every vector and tensor will be written in bold text. Eventually, if not otherwise stated, the LB unit system is adopted for all formulas below~\cite{Kruger2017}.

\subsection{BGK collision model}

The BGK-LB scheme relies on a particular time and space discretization of the BGK discrete-velocity Boltzmann equations. The resulting numerical scheme, in its dimensionless form, can be decomposed into a collision and a streaming steps~\cite{Kruger2017}
\begin{align}
\label{eq:LBM_BGK_collision}
%    f_i (\boldsymbol{x} + \boldsymbol{e_i}, t+1) = f_i(\boldsymbol{x}, t) - \frac{1}{\overline{\tau}} \left( f_i(\boldsymbol{x}, t) - f_i^{eq}(\boldsymbol{x}, t) \right),
    & f_i^*(\boldsymbol{x}, t) = f_i(\boldsymbol{x}, t) -\frac{1}{\overline{\tau}} \left(f_i(\boldsymbol{x}, t) - f_i^{eq}(\boldsymbol{x}, t) \right), \\
\label{eq:LBM_BGK_streaming}
    & f_i(\boldsymbol{x}+\boldsymbol{e_i}, t+1) = f_i^*(\boldsymbol{x}, t),
\end{align}
where $(f_i)_{i \in \llbracket 1, V \rrbracket}$ is the set of distribution functions associated to the lattice velocities   $(\boldsymbol{e_i})_{i\in \llbracket 1, V \rrbracket}$, $(f_i^{eq})_{i \in \llbracket 1, V \rrbracket}$ is its equilibrium counterpart and $(f_i^{*})_{i \in \llbracket 1, V \rrbracket}$ are the post-collision distributions. In Eqs.~(\ref{eq:LBM_BGK_collision})-(\ref{eq:LBM_BGK_streaming}), $t$ and $\boldsymbol{x}$ respectively stand for the time and spatial coordinates and $\overline{\tau}$ is the dimensionless relaxation time of the BGK collision model. Macroscopic quantities, such as the density field $\rho$ and the velocity field $\boldsymbol{u}$, can be defined as discrete moments of the distribution function:
\begin{align}
    \rho = \sum_{i=1}^V f_i, \qquad \rho \boldsymbol{u} = \sum_{i=1}^V \boldsymbol{e_i} f_i.
\end{align}
Regarding the equilibrium distribution functions $(f_i^{eq})_{i \in \llbracket 1, V \rrbracket}$, they are usually built so that their discrete moments match that of the Maxwell-Boltzmann distribution function $f^{eq}$~\cite{MAXWELL_PTRSL_157_1867}:
\begin{align}
    f^{eq}(\boldsymbol{\xi}) = \frac{\rho}{(2 \pi c_s^2)^{D/2}} \exp \left( -\frac{||\boldsymbol{\xi} - \boldsymbol{u}||^2}{2 c_s^2} \right),
\end{align}
$D$ being the number of spatial dimensions, $\boldsymbol{\xi}$ is the continuous velocity variable and $c_s$ is the lattice constant (\textit{cf.}~\ref{app:lattices}). However, one cannot match an infinite number of continuous equilibrium moments with a discrete set of velocities. It is therefore necessary to restrict the number of preserved moments to a finite number $N$, whose impact on the simulated physics will be further discussed in this section. A systematic way to \textit{exactly} impose the first $N$ equilibrium moments relies on a Gauss-Hermite quadrature together with a Hermite polynomial expansion of the equilibrium distribution function~\cite{Grad1949,SHAN_PRL_80_1998,SHAN2006, Philippi2006}:
\begin{align}
    f_i^{eq, N} = w_i \sum_{n=0}^N \frac{1}{n! c_s^{2n}}\, \boldsymbol{a}_{eq}^{(n)}:\boldsymbol{\mathcal{H}}_i^{(n)}.
\end{align}
In the above equation; `:' stands for the full contraction of indices of two $n^\mathrm{th}$-order tensors, $\boldsymbol{\mathcal{H}}_i^{(n)} =\boldsymbol{\mathcal{H}}^{(n)}(\boldsymbol{e_i})$ where $\boldsymbol{\mathcal{H}}^{(n)}$ is the $n^\mathrm{th}$-order Hermite polynomial defined as
\begin{align}
    \boldsymbol{\mathcal{H}}^{(n)}(\boldsymbol{\xi}) = \frac{(-c_s^2)^n}{w(\boldsymbol{\xi})} \frac{\partial^n w}{\partial \boldsymbol{\xi}^n},\ \ w(\boldsymbol{\xi}) = \frac{1}{(2\pi c_s^2)^{D/2}}\exp \left( \frac{- \xi^2}{2c_s^2} \right),
\end{align}
where $\xi^2 = ||\boldsymbol{\xi}||^2$ and $\boldsymbol{\xi}^n$ denotes the $n^\mathrm{th}$-rank tensor built by $n$ tensor products of $\boldsymbol{\xi}$ . Moreover, $\boldsymbol{a}_{eq}^{(n)}$ are the so-called Hermite moments of the Maxwell-Boltzmann equilibrium distribution
\begin{align}
    \boldsymbol{a}_{eq}^{(n)} = \int \boldsymbol{\mathcal{H}}^{(n)} (\boldsymbol{\xi}) f^{eq}(\boldsymbol{\xi})\, \mathrm{d}\boldsymbol{\xi}.
\end{align}

Finally, $N$ stands for the highest-order Hermite equilibrium moment that can be recovered with such a polynomial expansion. It should obey $2N \leq Q$, where $Q$ is the order of quadrature of the lattice (recalled in~\ref{app:lattices}). Note that, regarding the D2Q9 lattice, even if the quadrature order is $Q=5$ (thus $N \leq 2$), a partial polynomial expansion up to the third and fourth orders can be performed by including the following Hermite polynomials:
\begin{align}
    {\mathcal{H}}^{(3)}_{i,xxy},\  {\mathcal{H}}^{(3)}_{i,xyy}, \ {\mathcal{H}}^{(4)}_{i,xxyy},
\end{align}
leading to improved stability properties~\cite{Dellar2002, Siebert2008a,Malaspinas2015, Coreixas2017, Wissocq2019}. The latter expansions will be referred to as $N=3^*$ and $N=4^*$ in the following.

Hydrodynamic limits of the lattice Boltzmann equations solved by the LBM can be glimpsed by performing a so-called Chapman-Enskog (CE) expansion~\cite{CHAPMAN_Book_3rd_1970}. It consists in expanding the distribution functions around their equilibrium value
\begin{align}
    f_i = f_i^{eq, N} + f_i^{(1)} + f_i^{(2)} + f_i^{(3)} + ...,
\end{align}
where each component $f_i^{(k)}$ is sought in the order $O(\epsilon^k)$, where $\epsilon$ is a smallness parameter assumed to be the Knudsen number. Such an expansion allows linking the maximal equilibrium moment order $N$ with the simulated macroscopic physics. For instance, with $N=2$, the athermal Navier-Stokes equations are modelled with a well-known cubic Mach error in the momentum equation~\cite{Qian1992, SHAN2006}, while, for $N \geq 3$ no such error remains. Furthermore, a CE expansion yields a relation between the relaxation time $\overline{\tau}$ and the dimensionless fluid kinematic viscosity $\nu$~\cite{He1998}:
\begin{align}
    \nu = \left(\overline{\tau}-\frac{1}{2}\right) c_s^2.
\end{align}

\subsection{Regularized collision models}

The principle of regularized collision models is based on the observation that a CE expansion up to the first-order in Knudsen number is sufficient to recover the Navier-Stokes fluid behavior. Hence, the distribution functions can be reconstructed before each collision step as
\begin{align}
    f_i^{reg} = f_i^{eq, N} + f_i^{(1)},
\end{align}
where $f_i^{(1)}$ is a first-order term in Knudsen number that needs to be defined. It leads to the following regularized collision step, replacing Eq.~(\ref{eq:LBM_BGK_collision}):
\begin{align}
    f_i^*(\boldsymbol{x}, t) &= f_i^{reg}(\boldsymbol{x}, t) - \frac{1}{\overline{\tau}} \left( f_i^{reg} - f_i^{eq, N} \right) \\
\label{eq:general_form_regul}
    & = f_i^{eq, N} + \left( 1-\frac{1}{\overline{\tau}} \right) f_i^{(1)}.
\end{align}
The remaining question relies on the way $f_i^{(1)}$ is computed. This is where different methodologies arise. Two approaches will be adopted in the following: the regularization by projection and the recursive regularization.

\subsubsection{Regularization by projection}

In the original model proposed by Skordos~\cite{SKORDOS_PRE_49_1993}, and further investigated by Latt \& Chopard~\cite{Latt2006} for its stability properties, it is noticed that, due to mass and momentum conservation, only the second-order moment of $f_i^{(1)}$ is required to recover the athermal Navier-Stokes behavior. For this reason, it is reduced to its second-order Hermite polynomial expansion:
\begin{align}
    f_i^{(1)} = w_i \frac{1}{2 c_s^4} \boldsymbol{a}_1^{(2)}:\boldsymbol{\mathcal{H}}_i^{(2)},
\end{align}
where $\boldsymbol{a}_1^{(2)}$ is the expansion coefficient at first-order in Knudsen number. It can be approximated by its off-equilibrium counterpart
\begin{align}
\label{eq:a_1^2_projected_off_eq}
    \boldsymbol{a}_1^{(2)} \approx \boldsymbol{a}_{neq}^{(2)} \equiv \sum_{i=1}^V \boldsymbol{\mathcal{H}}_i^{(2)}\, \left(f_i - f_i^{eq, N} \right).
\end{align}
Based on the orthogonality properties of the Hermite polynomials~\cite{Grad1949, SHAN_PRL_80_1998,SHAN2006}, this regularization procedure can be viewed as an orthogonal projection, before each collision step, of the off-equilibrium distribution functions onto the second-order Hermite polynomials, so as to cancel their higher-order contribution. For this reason, it will be referred to as the \textit{projected regularization} (PR) in the rest of the paper.

\subsubsection{Recursive regularization}

Instead of a regularization involving the second-order moments of $f_i^{(1)}$ only, Malaspinas~\cite{Malaspinas2015} proposed a procedure based on a reconstruction of as many off-equilibrium moments as possible. It starts by expanding $f_i^{(1)}$ in Hermite polynomials:
\begin{align}
    f_i^{(1)} = w_i \sum_{n=2}^{N_r} \frac{1}{n! c_s^{2n}} \boldsymbol{a}_1^{(n)}:\boldsymbol{\mathcal{H}}_i^{(n)},
\end{align}
where $N_r$ is the order of the regularization. Note that the sum starts at $n=2$ since $\boldsymbol{a}_1^{(0)} = \boldsymbol{a}_1^{(1)} = 0$ (collision invariants). Thanks to a CE expansion, it can be shown that coefficients $\boldsymbol{a}_1^{(n)}$ are linked with each other through the following recursive relation:
\begin{align}
\label{eq:regul_recursive_formula}
    a_{1, \alpha_1..\alpha_n}^{(n)} &= u_{\alpha_n} a_{1, \alpha_1..\alpha_{n-1}}^{(n-1)} \nonumber \\
    &+ \left( u_{\alpha_1}..u_{\alpha_{n-2}} a_{1, \alpha_{n-1} \alpha_n}^{(2)} + \mathrm{perm}(\alpha_n) \right),
\end{align}
where ``$\mathrm{perm}(\alpha_n)$'' stands for all the cyclic permutations of indexes from $\alpha_1$ to $\alpha_{n-1}$. As with the PR collision, $\boldsymbol{a}_1^{(2)}$, required to initialize the recurrence, is approximated by the projection of the off-equilibrium part provided in Eq.~(\ref{eq:a_1^2_projected_off_eq}). This approach was later extended to high-order lattices~\cite{Coreixas2017}. The corresponding collision model will be referred to as \textit{recursive regularization at order $N_r$} (RR$N_r$) in the following. Note that $N_r$ obeys the same condition as $N$, \textit{i.e.} $2N_r \leq Q$, and one can define partial third and fourth orders with the D2Q9 lattice (resp. $N_r=3^*$ and $N_r=4^*$).

\section{Linear stability analyses}
\label{sec:LSA}

This section is dedicated to the linear analyses, in the von Neumann formalism, of the aforementioned BGK and regularized collision models. Moreover, a systematic modal identification will be performed through the information provided by the eigenvector of each linear mode.

\subsection{Von Neumann formalism}

Sterling and Chen~\cite{Sterling1996} were among the firsts to propose a von Neumann linear analysis of the LB scheme through a linear decomposition of any distribution function as
\begin{align}
\label{eq:linear_decomposition}
    f_i = \overline{f_i} + f'_i,
\end{align}
where the global populations $\overline{f_i}$ are constants (no variation in space and time) and $f'_i$ are fluctuating populations, assumed to be very small compared to the global populations. A linearization about the global equilibrium state is then performed. In the LB scheme, nonlinearities come from the collision step while the streaming step is fully linear in $f'_i$. Hence, post-collision distribution functions $f_i^*$ are linearized as
\begin{align}
    f_i^* (f_j) = f_i^*(\overline{f_j}) + \left. \frac{\partial f_i^*}{\partial f_j} \right|_{f_j=\overline{f_j}} \, f'_j + O({f'_j}^2),
\end{align}
where Einstein's summation convention is adopted on index $j$. Injecting Eq.~(\ref{eq:linear_decomposition}) into the lattice Boltzmann scheme, keeping the zeroth-order in fluctations and cancelling any spatial and temporal derivatives leads to
\begin{align}
    \overline{f_i} = f_i^{eq}(\overline{\rho}, \overline{\boldsymbol{u}}),
\end{align}
where $\overline{\rho}$ and $\overline{\boldsymbol{u}}$ are respectively the mean flow density and the mean flow velocity. On the other hand, keeping the first-order equation in populations yields
\begin{align}
\label{eq:general_linearized_system}
    f_i'(\boldsymbol{x}+\boldsymbol{e_i}, t+1) = \left. \frac{\partial f^*_i}{\partial f_j} \right|_{f_j = \overline{f_j}}\, f'_j.
\end{align}
In the von Neumann analysis, fluctuating populations are sought as complex plane monochromatic waves
\begin{align}
\label{eq:fluctuations_monochromatic_waves}
    f'_i (\boldsymbol{x}, t) = \widehat{f_i} \exp (\mathrm{i}(\boldsymbol{k} \cdot \boldsymbol{x} - \omega t)),
\end{align}
where $\mathrm{i}^2 = -1$, $(\widehat{f_i}) \in \mathbb{C}^V$, $\boldsymbol{k}$ is the dimensionless wavenumber vector and $\omega$ is the dimensionless pulsation of the wave. In the case of a temporal analysis, $\boldsymbol{k} \in \mathbb{R}^D$ and $\omega \in \mathbb{C}$. Finally, injecting Eq.~(\ref{eq:fluctuations_monochromatic_waves}) into the general linearized equations of Eq.~(\ref{eq:general_linearized_system}) yields a linear system of size $V$ that can be written under the following matricial form:
\begin{align}
\label{eq:LSA_matrix_system}
    e^{-\mathrm{i} \omega} \mathbf{\widehat{F}} = \mathbf{M} \mathbf{\widehat{F}},
\end{align}
where $\mathbf{\widehat{F}}=(\widehat{f_i})_{i \in \llbracket 1, V \rrbracket}$ and $\mathbf{M}$ is a square $(V\times V)$-size matrix. For the collision models introduced in Sec.~\ref{sec:LBM}, the expressions of this matrix can be found \textit{e.g.} in~\cite{HOSSEINI_PRE_99_2019b,HOSSEINI_PRE_100_2019}, and are derived in a general way in~\ref{app:linear_matrices}, as \newline

\begin{itemize}
    \item BGK collision model:
\begin{align}
    M_{ij}^{\mathrm{BGK}} = e^{-\mathrm{i} \boldsymbol{k} \cdot \boldsymbol{e_i}} \left[ \delta_{ij} - \frac{1}{\overline{\tau}} \left(\delta_{ij} - J^{eq,N}_{ij} \right) \right],
\end{align}
with
\begin{align}
    J_{ij}^{eq,N} = w_i \sum_{n=0}^N \frac{1}{n! c_s^{2n}} \boldsymbol{\Lambda}^{(n)}_{eq,j}:\boldsymbol{\mathcal{H}}_i^{(n)},
\end{align}
and where coefficients $\boldsymbol{\Lambda}_{eq,j}$ are provided in \ref{app:linear_matrices},
\item projected regularization (PR):
\begin{align}
    M_{ij}^{\mathrm{PR}} = e^{-\mathrm{i} \boldsymbol{k} \cdot \boldsymbol{e_i}} \bigg[ J^{eq,N}_{ij} + \left( 1-\frac{1}{\overline{\tau}} \right)  \left( \delta_{kj} - J^{eq,N}_{kj} \right) h_{ik} \bigg],
\end{align} 
with 
\begin{align}
    h_{ik} = \frac{w_i}{2c_s^4} \, \boldsymbol{\mathcal{H}}_i^{(2)}:\boldsymbol{\mathcal{H}}_k^{(2)},
\end{align}
\item recursive regularization (RR$N_r$):
\begin{align}
    M_{ij}^{\mathrm{RR}N_r} &= M_{ij}^{\mathrm{PR}} \nonumber \\
    & + e^{-\mathrm{i} \boldsymbol{k} \cdot \boldsymbol{e_i}} \left( 1-\frac{1}{\overline{\tau}} \right) \sum_{n=3}^{N_r} \frac{w_i}{n! c_s^{2n}}\, \mathbf{\Lambda}_{1, j}^{(n)}:\boldsymbol{\mathcal{H}}_i^{(n)},
\end{align}
where coefficients $\mathbf{\Lambda}_{1, j}^{(n)}$ are provided in \ref{app:linear_matrices}.
%\item full reconstruction (FR$N_r$):
%\begin{align}
%    M_{ij}^{\mathrm{FR}} = & e^{-i \boldsymbol{k} \cdot \boldsymbol{e_i}} \bigg[ J_{ij}^{eq,N} \nonumber \\
%    & + \left( 1-\frac{1}{\overline{\tau}} \right) \, \sum_{n=2}^{N_r} \frac{w_i}{n!c_s^{2n}} \, \mathbf{\Lambda}_{1, j}^{(n), \mathrm{FR}} : \boldsymbol{\mathcal{H}}_{i}^{(n)} \bigg], 
%\end{align}
%with (\textit{cf.} App.~\ref{app:linear_matrices})
%\begin{align}
%    \left(\mathbf{\Lambda}_{1, j}^{(2), \mathrm{FR}}\right)_{\alpha \beta} = - i\overline{\tau} c_s^2 ( &(e_{j, \alpha} - \overline{u_\alpha}) \sin(k_\beta) \nonumber \\
%    & + (e_{j, \beta} - \overline{u_\beta}) \sin(k_\alpha) ),
%\end{align}
%and with the high-order off-equilibrium coefficients $\left( \boldsymbol{\Lambda}_{1,j}^{(n), \mathrm{FR}} \right)_{n \geq 3}$ provided in App.~\ref{app:linear_matrices}.
\end{itemize}
Note that none of the above expressions involves the mean flow density $\overline{\rho}$, so that $\mathbf{M}$ 
depends on the dimensionless relaxation time $\overline{\tau}$, the mean flow velocity $\overline{\boldsymbol{u}}$ and the wavenumber vector $\boldsymbol{k}$ only.

Regarding the possible values of $\boldsymbol{k}$, a short discussion has to be made. First, according to the Nyquist-Shannon sampling theorem~\cite{Nyquist1928, Shannon1949}, it is sufficient to study wavenumbers for which $|k_x|\leq \pi$, $|k_y| \leq \pi$, which corresponds to considering waves discretized with more than two points per wavelength in each direction. Moreover, only the real part of the fluctuating populations given in Eq.~(\ref{eq:fluctuations_monochromatic_waves}) is of interest:
\begin{align}
\label{eq:fluctuating_populations_re}
    \Re(f'_i) = |\widehat{f_i}| \cos (\boldsymbol{k} \cdot \boldsymbol{x} - \omega_r t + \arg(\widehat{f_i}))e^{\omega_i t},
\end{align}
where $\omega_r=\Re(\omega)$ is related to the wave propagation, $\omega_i=\Im(\omega)$ to its amplification rate and $\arg(\widehat{f_i})$ is the argument of the complex amplitude $\widehat{f_i}$. Thanks to parity properties of the $\cos$ function, it is perfectly equivalent to study:
\begin{align}
    (\boldsymbol{k}, \omega, \widehat{f_i})\qquad \mathrm{or} \qquad (-\boldsymbol{k}, -\omega^\dagger, {\widehat{f_i}}^\dagger),
\end{align}
where the `$\dagger$' superscript stands for the conjugate of a complex number. Hence, it is sufficient to restrict the problem to half of the possible wavenumber vectors $\boldsymbol{k}$, \textit{e.g.} $k_x \in [-\pi, \pi]$, $k_y \in [0, \pi]$ in two dimensions.

In practice, physical phenomena of interest are rarely resolved with less than eight points per wavelength, meaning that investing cases for which $||\boldsymbol{k}|| < \pi/4$ should be sufficient for common purposes. However, any wave with more than two points per wavelength is naturally considered in a numerical simulation. If such under-resolved wave is linearly amplified, the numerical scheme is found unstable. This is why the full range of possible wavenumbers, including under-resolved ones, has to be considered in the stability analyses. 

\subsection{Modal identification through eigenvectors}

In the common von Neumann approach, the eigenvalue problem of Eq.~(\ref{eq:LSA_matrix_system}) is solved for each value of $\boldsymbol{k}$, providing $V$ eigenvalues, then $V$ complex pulsations $\omega$ whose imaginary part $\omega_i$ provides information on the amplification rate of the mode, and its real part $\omega_r$ on the propagation of the mode. Especially, phase velocity $v_\phi$ and group velocity $v_g$ can be defined as
\begin{align}
    v_\phi = \frac{\omega_r}{||\boldsymbol{k}||}, \qquad v_g = \frac{\partial \omega_r}{\partial ||\boldsymbol{k}||}.
\end{align}
However, the eigenvectors $\mathbf{\widehat{F}}$ are usually not exploited, whereas they contain interesting information on the quantity carried by a given mode of the LB method. In the present linear stability analyses, the methodology introduced in a previous article~\cite{Wissocq2019} will be adopted to systematically identify each mode by its macroscopic content. The main steps of this procedure are:
\begin{enumerate}
    \item compute the macroscopic moments of a given eigenvector:
\begin{align}
    \widehat{\rho} = \sum_i \widehat{f_i}, \qquad \widehat{\rho \boldsymbol{u}} = \sum_i \boldsymbol{e_i} \widehat{f_i},
\end{align}
to identify the considered mode as either a \textit{non-observable} one [$(\widehat{\rho}, \widehat{\rho\boldsymbol{u}})=(0, \boldsymbol{0})$] or an \textit{observable} one [$(\widehat{\rho}, \widehat{\rho \boldsymbol{u}}) \neq (0, \boldsymbol{0})$],
\item perform a von Neumann analysis of the Navier-Stokes equations, in order to obtain the eigenvectors of the physical (acoustic and shear) waves
\begin{align}
    \mathbf{\widehat{V_{ac+}}}, \mathbf{\widehat{V_{ac-}}}, \mathbf{\widehat{V_{shear}}},
\end{align}
expressed in the basis of the macroscopic moments $(\widehat{\rho}, \widehat{\rho \boldsymbol{u}})$,
\item find the coefficients of the linear decomposition
\begin{align}
    \mathbf{\widehat{V}} &= (\widehat{\rho}, \widehat{\rho \boldsymbol{u}})^T \nonumber \\
    & = \alpha_1 \mathbf{\widehat{V_{ac+}}} + \alpha_3 \mathbf{\widehat{V_{ac-}}} + \alpha_3 \mathbf{\widehat{V_{shear}}},
\end{align}
thanks to the passage matrix composed of the Navier-Stokes eigenvectors,
\item normalize the coefficients $\alpha_i$ so that $\sum_i |\alpha_i| = 1$, and then systematically identify the physical information carried if
\begin{align}
    |\alpha_i| > \eta,
\end{align} 
where $\eta \leq 1$ is an arbitrary threshold.
\end{enumerate}
Note that in the rest of the article, the following distinction between \textit{modes} and \textit{waves} will be adopted:
\begin{itemize}
    \item the denomination \textit{mode} will refer to the continuous curves (as function of the wavenumber $\boldsymbol{k}$) of the LB linear analyses,
    \item the denomination \textit{wave} will refer to the physical waves expected by NS analyses (acoustics and shear).
\end{itemize}
In particular, as highlighted in a previous work~\cite{Wissocq2019} and due to modal interactions occurring in the numerical scheme, a given LB \textit{mode} can carry different physical \textit{waves} according to the considered range of wavenumbers $\boldsymbol{k}$. This particular property will be illustrated in the analyses of the next section.

\section{Main results}
\label{sec:LSA_results}

The eigenvalue problems detailed in the previous section are discretized for any value of $k_x \in [-\pi, \pi]$ and $k_y \in [0, \pi]$ with a step $\Delta k = 0.001$. Each eigenvalue problem is then solved with the NumPy Python library~\cite{vanderWalt2011}. Note that in the case of a mean flow aligned with the horizontal or vertical direction, lattice symmetry properties allow reducing the study parameters to $k_x \in [0, \pi]$. For each problem, a systematic modal identification thanks to the moments of the eigenvectors, as described above, is performed. For this purpose, the parameter $\eta$ is set to $\eta=0.9$ in order to identify modes carrying more than $90\%$ of a physical wave. Moreover, a validation of every linearized system has been done by initializing a two-dimensional LB solver with the superposition of a mean flow $f_i^{eq}(\overline{\rho}, \overline{\boldsymbol{u}})$ and a fluctuating part given by Eq.~(\ref{eq:fluctuating_populations_re}), where $\widehat{f_i}$ are the populations provided by an eigenvector of the linear analysis. Observing a monochromatic plane wave behavior allows qualitatively validating a correct linearization of the LB scheme.

In the following, the focus will be put on two lattices: the standard D2Q9~\cite{Qian1992} and the multi-speed D2V17 lattices~\cite{Philippi2006, Shan2016}, both described in \ref{app:lattices}. Comparisons will be drawn with linear analyses of the Navier-Stokes equations, which lead to (\textit{cf.}~\cite{LEKKERKERKER_PRA_10_1974, Wissocq2019}):
\begin{align}
    & \omega_{\mathrm{shear}} = \boldsymbol{k} \cdot \overline{\boldsymbol{u}} - \mathrm{i} \nu || \boldsymbol{k}||^2 %+ O(\boldsymbol{k}^3)
    , \nonumber \\
    \label{eq:LSA_Navier_Stokes}
    & \omega_{\mathrm{ac+}} = \boldsymbol{k} \cdot \overline{\boldsymbol{u}} + ||\boldsymbol{k}|| c_s - \mathrm{i} \nu || \boldsymbol{k}||^2 + O(\boldsymbol{k}^3), \\
    & \omega_{\mathrm{ac-}} = \boldsymbol{k} \cdot \overline{\boldsymbol{u}} - ||\boldsymbol{k}|| c_s - \mathrm{i} \nu || \boldsymbol{k}||^2 + O(\boldsymbol{k}^3). \nonumber
\end{align}
Note that these solutions correspond to a fluid modeling including a bulk viscosity $\nu_b = \nu$, as usual in two-dimensional athermal LB methods~\cite{Dellar2001}. In all the cases presented below, a dimensionless relaxation time will be defined as
\begin{align}
    \tau = \overline{\tau} - 1/2 = \nu/c_s^2.
\end{align}

\subsection{Standard D2Q9 lattice}

In this section, the linear behavior of the D2Q9 lattice is investigated. First, reminders are given on the BGK collision model, then regularized ones are investigated. For a sake of clarity and compactness, all the studies of this section are performed with a partial fourth-order equilibrium ($N=4^*$), since it is known that including higher-order equilibrium moments can enhance numerical stability~\cite{Dellar2002, Coreixas2018, Wissocq2019_these}. It is all the more noticed that other forms of Gauss-Hermite based polynomial equilibria (with $N=2$ or $N=3^*$) do not affect the main conclusions drawn below~\cite{Wissocq2019_these}.

%\onecolumngrid
\begin{figure*}[h!]
\begin{minipage}{1.\textwidth}
\centering
\includegraphics[scale=1.]{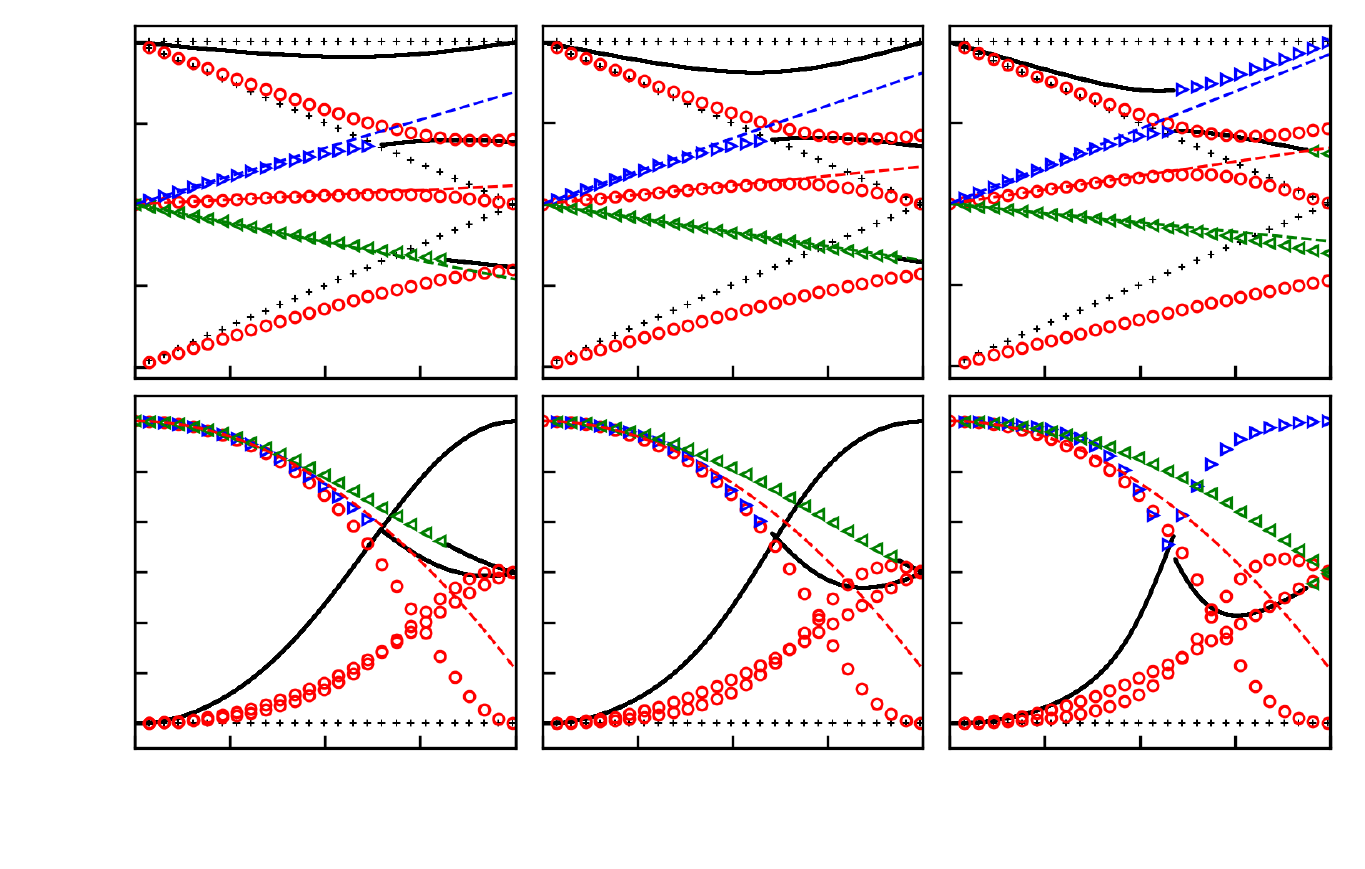}
\put(-389,189){$\omega_r$}
\put(-365,237){$\pi$}
\put(-375,212){$\pi/2$}
\put(-365,189){$0$}
\put(-382,166){$-\pi/2$}
\put(-372,143){$-\pi$}
\put(-400,83){$\omega_i/\nu$}
\put(-365,127){$0$}
\put(-372,113){$-2$}
\put(-372,98){$-4$}
\put(-372,83){$-6$}
\put(-372,68){$-8$}
\put(-377,54){$-10$}
\put(-377,39){$-12$}
\put(-300,11){$k_x$}
\put(-357,24){$0$}
\put(-335,24){$\pi/4$}
\put(-308,24){$\pi/2$}
\put(-282,24){$3\pi/4$}
\put(-249,24){$\pi$}
\put(-185,11){$k_x$}
\put(-239,24){$0$}
\put(-217,24){$\pi/4$}
\put(-190,24){$\pi/2$}
\put(-165,24){$3\pi/4$}
\put(-131,24){$\pi$}
\put(-70,11){$k_x$}
\put(-122,24){$0$}
\put(-100,24){$\pi/4$}
\put(-73,24){$\pi/2$}
\put(-48,24){$3\pi/4$}
\put(-13,24){$\pi$}
\vspace{-5mm}
\caption{Propagation (top) and dissipation (bottom) curves of the nines modes of the BGK-D2Q9 lattice with $\tau=10^{-5}$, $N=4^*$, $k_y=0$ and three values of the horizontal mean flow: $\overline{\mathrm{Ma}}=0.2$ (left), $\overline{\mathrm{Ma}}=0.4$ (middle), $\overline{\mathrm{Ma}}=0.6$ (right). Modes carrying more than $\eta=90\%$ of a physical wave are identified:~\markershear: shear, \markeracp:~downstream acoustics, \markeracm:~upstream acoustics, \markernum:~non-identified wave, \tiny{+}\,\normalsize{: non-observable mode. Navier-Stokes reference curves are displayed as: \reddashedline:~shear, \bluedashedline:~downstream acoustics, \greendashedline:~upstream acoustics.}\label{fig:D2Q9_BGK_Mach_spectrums}}
\end{minipage}
\vspace{5mm}
%\end{figure*}
%\twocolumngrid
%\onecolumngrid
%\begin{figure*}[ht]
\hspace{-1cm}
\begin{minipage}{1.\textwidth}
\centering
\includegraphics[scale=1.]{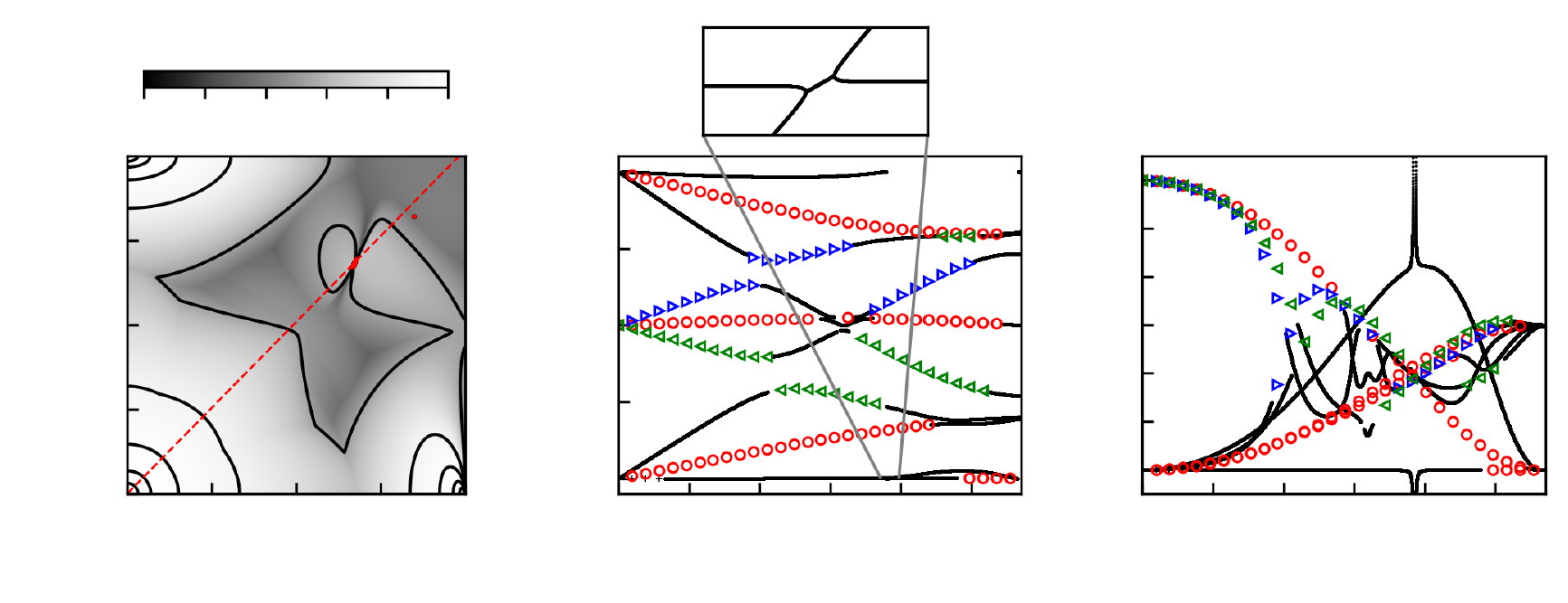}
\put(-462,24){$0$}
\put(-441,24){$\pi/4$}
\put(-414,24){$\pi/2$}
\put(-390,24){$3\pi/4$}
\put(-355,24){$\pi$}
\put(-410,11){$k_x$}
\put(-470,32){$0$}
\put(-480,59){$\pi/4$}
\put(-480,86){$\pi/2$}
\put(-484.5,113){$3\pi/4$}
\put(-470,141){$\pi$}
\put(-463 ,151){$\footnotesize{-10}$}
\put(-442 ,151){$\footnotesize{-8}$}
\put(-423 ,151){$\footnotesize{-6}$}
\put(-404 ,151){$\footnotesize{-4}$}
\put(-384 ,151){$\footnotesize{-2}$}
\put(-360 ,151){$\footnotesize{0}$}
\put(-423 ,177){$\max(\omega_i/\nu)$}
\put(-240,11){$k$}
\put(-306,24){$0$}
\put(-287,24){$\pi/4$}
\put(-265,24){$\pi/2$}
\put(-244,24){$3\pi/4$}
\put(-215,24){$\pi$}
\put(-200,24){$5\pi/4$}
\put(-333,90){$\omega_r$}
\put(-314,135){$\pi$}
\put(-324,110){$\pi/2$}
\put(-314,86){$0$}
\put(-331,62){$-\pi/2$}
\put(-321,38){$-\pi$}
\put(-70,11){$k$}
\put(-139,24){$0$}
\put(-120,24){$\pi/4$}
\put(-98,24){$\pi/2$}
\put(-77,24){$3\pi/4$}
\put(-48,24){$\pi$}
\put(-33,24){$5\pi/4$}
\put(-146,132){$0$}
\put(-153,117){$-2$}
\put(-153,102){$-4$}
\put(-153,86){$-6$}
\put(-153,71){$-8$}
\put(-157,55){$-10$}
\put(-157,40){$-12$}
\put(-166,90){$\frac{\omega_i}{\nu}$}
\end{minipage}\\
\begin{minipage}{1.\textwidth}
\begin{minipage}{0.33\textwidth}
\subcaption{Maximal amplification rate for any wavenumber vector $\boldsymbol{k}$.\label{fig:D2Q9_BGK_maxwi_map}}
\end{minipage}
\begin{minipage}{0.66\textwidth}
\subcaption{Propagation (left) and dissipation (right) curves for plane waves for which $\theta_k = \arctan(k_y/k_x)=45.5^\circ$ (\textit{cf}. dashed line on Fig.~\ref{fig:D2Q9_BGK_maxwi_map}). Symbols are identical to Fig.~\ref{fig:D2Q9_BGK_Mach_spectrums}. An eigenvalue collision is highlighted as the instability cause. \label{fig:D2Q9_BGK_spectrum_line}}
\end{minipage}
\caption{Linear stability analysis of the BGK-D2Q9 LB scheme with $\tau=10^{-5}$, $N=4^*$ and a horizontal mean flow at $\overline{\mathrm{Ma}}=0.2$.}
\end{minipage}
\end{figure*}
%\twocolumngrid

\subsubsection{BGK collision model}

Linear stability analyses of the BGK collision model have been extensively studied in the litterature, and the interested reader may refer to previous work for more in-depth studies~\cite{Sterling1996, Worthing1997, Marie2009, Cleon2014, Wissocq2019}. Thus, this section mainly focuses on key points that should be kept in mind when studying regularized models.

Fig.~\ref{fig:D2Q9_BGK_Mach_spectrums} displays the propagation [$\omega_r=f(k_x)$] and dissipation [$\omega_i/\nu=f(k_x)$] curves of the D2Q9 lattice for horizontal plane waves ($k_y=0$) with three horizontal mean flows: $\overline{\mathrm{Ma}}=0.2$, $\overline{\mathrm{Ma}}=0.4$ and $\overline{\mathrm{Ma}}=0.6$, where $\overline{\mathrm{Ma}}=\overline{u}/c_s$ is the mean flow Mach number. The dimensionless relaxation time is set to $\tau=10^{-5}$, which is a commonly encountered value for air flow simulations when an acoustic scaling, linking space and time steps, is adopted~\cite{Kruger2017}.

Several observations are worth noting. First, nine modes can be identified: three of them carry no macroscopic information whatever the wavenumber $k_x$ (identified with `\tiny{+}\normalsize'), while six modes are observable. Among the latter, three modes carry a shear information (\textit{i.e} a transverse velocity only, identified with `\markershear'). The other three modes either carry upstream (`\markeracm') and downstream (`\markeracp') acoustics, or a non-identified macroscopic information (`\markernum'), which is necessarily a linear superposition of physical waves (shear and acoustics). More dedicated studies indicate that the latter mode is nothing more than a linear combination of acoustic waves, referred to as `spurious acoustics'~\cite{ASTOUL_ARXIV_2020_11863}. Furthermore, a modal interaction exhibited in~\cite{Wissocq2019} as a \textit{curve veering}, or \textit{avoided crossings} phenomenon is evidenced on the $\overline{\mathrm{Ma}}=0.6$-case: two eigencurves repel each other and a swap of the continuous mode carrying the downstream acoustic information is noticed. All in all, for the well-resolved wavelengths -- at least 8 points per wavelength ($k_x \leq \pi/4$) --, the linear behavior of any mode carrying physical waves is consistent with the Navier-Stokes expectations both in propagation and dissipation, and no instability region ($\omega_i >0$) is noticed in the horizontal direction for these cases. 

In order to investigate more precisely the linear stability of the case $\overline{\mathrm{Ma}}=0.2$ (in the horizontal direction), a spectral map of the maximal amplification rate $\omega_i/\nu$ for any value of $k_x$ and $k_y$ is displayed on Fig.~\ref{fig:D2Q9_BGK_maxwi_map}. A very thin instability zone [$\max(\omega_i)/\nu >0$] can be observed in the region $k_x \approx 2.11$, $k_y \approx 2.13$. Looking at the propagation and dissipation curves of the waves travelling in the corresponding direction (\textit{cf.} Fig.~\ref{fig:D2Q9_BGK_spectrum_line}) allows highlighting another kind of modal interaction, referred to as \textit{eigenvalue collision} in~\cite{Wissocq2019}. The latter consists in a local degeneracy of two eigencurves carrying a macroscopic information, leading to a severe instability peak. As a consequence, any LB simulation run under these conditions is expected to become unstable because of a strong amplification of the corresponding wavenumbers.

%\onecolumngrid
\begin{figure*}[h!]
\begin{minipage}{1.\textwidth}
\centering
\includegraphics[scale=0.8]{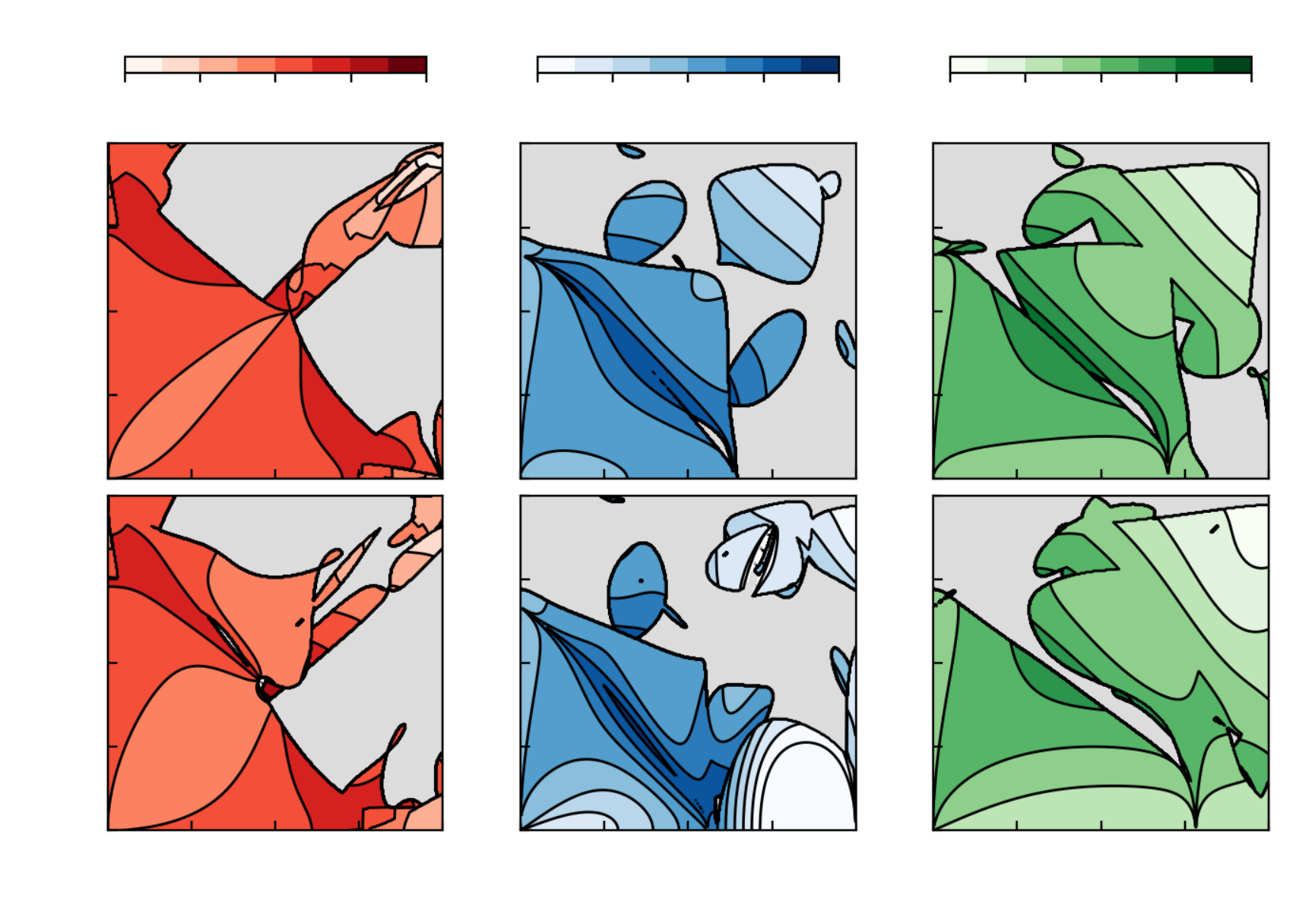}
\put(-310,9){$k_x$}
\put(-364,20){$0$}
\put(-343,20){$\pi/4$}
\put(-317,20){$\pi/2$}
\put(-294,20){$3\pi/4$}
\put(-262,20){$\pi$}
\put(-185,9){$k_x$}
\put(-238,20){$0$}
\put(-217,20){$\pi/4$}
\put(-192,20){$\pi/2$}
\put(-168,20){$3\pi/4$}
\put(-137,20){$\pi$}
\put(-55,9){$k_x$}
\put(-113,20){$0$}
\put(-93,20){$\pi/4$}
\put(-67,20){$\pi/2$}
\put(-44,20){$3\pi/4$}
\put(-11,20){$\pi$}
\put(-400,77){$k_y$}
\put(-370,26){$0$}
\put(-380,52){$\pi/4$}
\put(-380,77){$\pi/2$}
\put(-385,102){$3\pi/4$}
\put(-370,128){$\pi$}
\put(-400,184){$k_y$}
\put(-370,134){$0$}
\put(-380,158){$\pi/4$}
\put(-380,184){$\pi/2$}
\put(-385,209){$3\pi/4$}
\put(-370,235){$\pi$}
%\put(-400,264){$k_y$}
%\put(-370,214){$0$}
%\put(-380,238){$\pi/4$}
%\put(-380,264){$\pi/2$}
%\put(-385,289){$3\pi/4$}
%\put(-370,315){$\pi$}
\put(-366,248){\scriptsize{$0.25$}}
\put(-339,248){\scriptsize{$0.5$}}
\put(-313,248){\scriptsize{$1$}}
\put(-290,248){\scriptsize{$2$}}
\put(-267,248){\scriptsize{$4$}}
\put(-239,248){\scriptsize{$0.25$}}
\put(-213,248){\scriptsize{$0.5$}}
\put(-187,248){\scriptsize{$1$}}
\put(-164,248){\scriptsize{$2$}}
\put(-141,248){\scriptsize{$4$}}
\put(-113,248){\scriptsize{$0.25$}}
\put(-87,248){\scriptsize{$0.5$}}
\put(-61,248){\scriptsize{$1$}}
\put(-38,248){\scriptsize{$2$}}
\put(-15,248){\scriptsize{$4$}}
\put(-324,271){$\nu_e/\nu$}
\put(-198,271){$\nu_e/\nu$}
\put(-72,271){$\nu_e/\nu$}
\caption{Effective viscosity $\nu_e/\nu$ of the modes carrying physical waves with D2Q9 lattice, BGK collision, $\tau=10^{-5}$, $N=4^*$. Top: $\overline{\mathrm{Ma}}=0.2$, bottom: $\overline{\mathrm{Ma}}=0.6$. Left: shear, middle: downstream acoustics, right: upstream acoustics. Grey color indicate zones where no physical wave could be identified with $\eta=90\%$.\label{fig:OmegaI_map_BGK}}
\end{minipage}
\end{figure*}
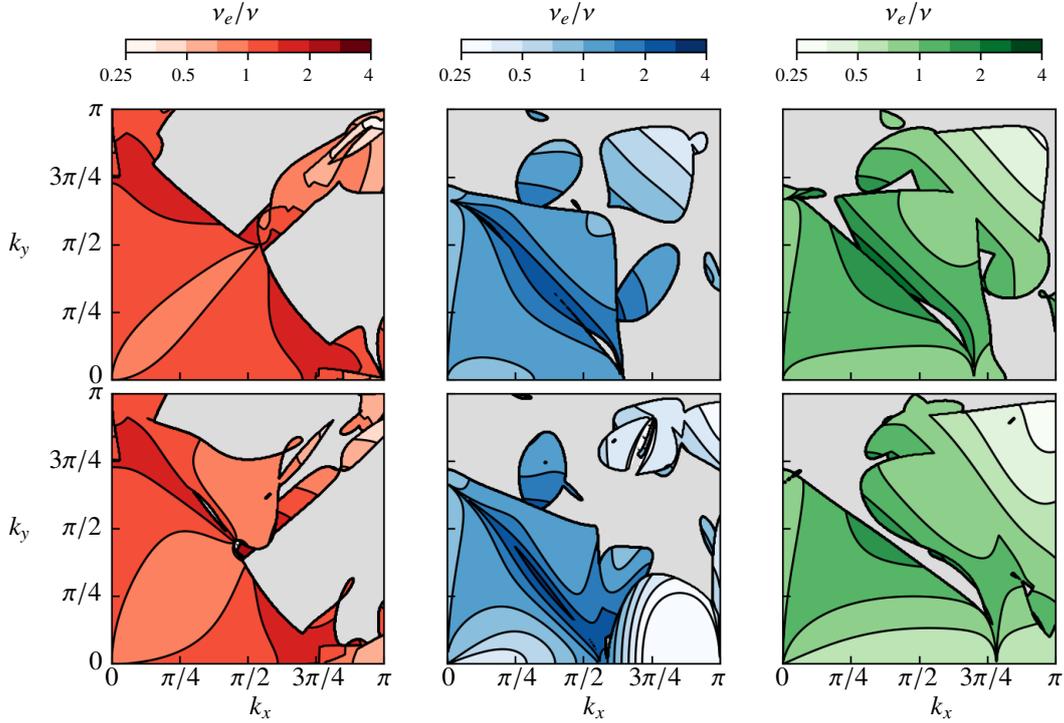
%\twocolumngrid

To complete these analyses, spectral maps of the effective viscosity of any mode carrying a given physical wave (shear, downstream and upstream acoustics) are displayed on Fig.~\ref{fig:OmegaI_map_BGK} for cases $\overline{\mathrm{Ma}}=0.2$ and $\overline{\mathrm{Ma}}=0.6$. When several modes eventually carry a similar macroscopic information (\textit{e.g.} the three modes carrying shear on Fig.~\ref{fig:D2Q9_BGK_Mach_spectrums}), only the one of maximal amplification rate ($\omega_i$) is displayed. In any case, the effective viscosity is computed as $\nu_e= -\omega_i / ||\boldsymbol{k}||^2$ and further dimensionalized by the expected viscosity $\nu$. For $\overline{\mathrm{Ma}}=0.2$ a rather isotropic behavior is observed for any well-resolved wavelength ($|| \boldsymbol{k}|| < \pi/4$). However, when increasing the Mach number of the mean flow to $\overline{\mathrm{Ma}}=0.6$, the dissipation rate becomes anisotropic, especially regarding the acoustics and even in the most-resolved wavelengths. This unphysical observation can be attributed to the $O(\mathrm{Ma}^3)$ error induced by the truncation of the equilibrium distribution function, which is known to be responsible for an anti-dissipative behavior~\cite{Marie2009, Dellar2014}, yielding $\nu_e/\nu <1$. Yet, no instability of any mode carrying more than $90\%$ of a physical wave can be identified on these figures.

To summarize these investigations, all the results carried in the current work, as well as previous analyses of the BGK collision model~\cite{Wissocq2019_these, PAM2019}, exhibit two main sources of instability for the D2Q9 lattice:
\begin{itemize}
    \item for $\overline{\mathrm{Ma}} > \sqrt{3}-1 \approx 0.73$, the negative dissipation due to the $O(\mathrm{Ma}^3)$ error is a cause of instability. 
    \item for $\overline{\mathrm{Ma}} < \sqrt{3}-1 \approx 0.73$, the only linear instabilities are due to destructive modal interactions occurring in the form of \textit{eigenvalue collision}.
\end{itemize}
The first error is inherent to the discretization of the velocity space and the corresponding equilibrium distribution truncation. Note that the critical value $\sqrt{3}-1$ is, in theory, only valid in the low-Knudsen number limit~\cite{PAM2019}, which is the range of application of the Navier-Stokes equations. The second cause is a purely numerical instability, consequent of the space/time discretization of the LB scheme.

%the only linear instabilities observed with the BGK collision model, provided $\overline{\mathrm{Ma}} < 0.73$, are due to destructive modal interactions occurring in the form of \textit{eigenvalue collision}. On the other hand, when $\overline{\mathrm{Ma}} > 0.73$, previous work has shown that the cubic Mach error can also be responsible for instabilities~\textcolor{red}{CITE}.
%\clearpage

\subsubsection{Regularized collision models}

Similar analyses are now performed on the regularized collision models. Three of them are of particular interest in this section: the PR, RR$3^*$ and RR$4^*$ models. %Since FR$3^*$ and FR$4^*$ models exhibit very similar linear properties as the FR2 one, they are not shown in this section.

%\onecolumngrid
\begin{figure*}[ht]
\begin{minipage}{1.\textwidth}
\centering
\includegraphics[scale=1.]{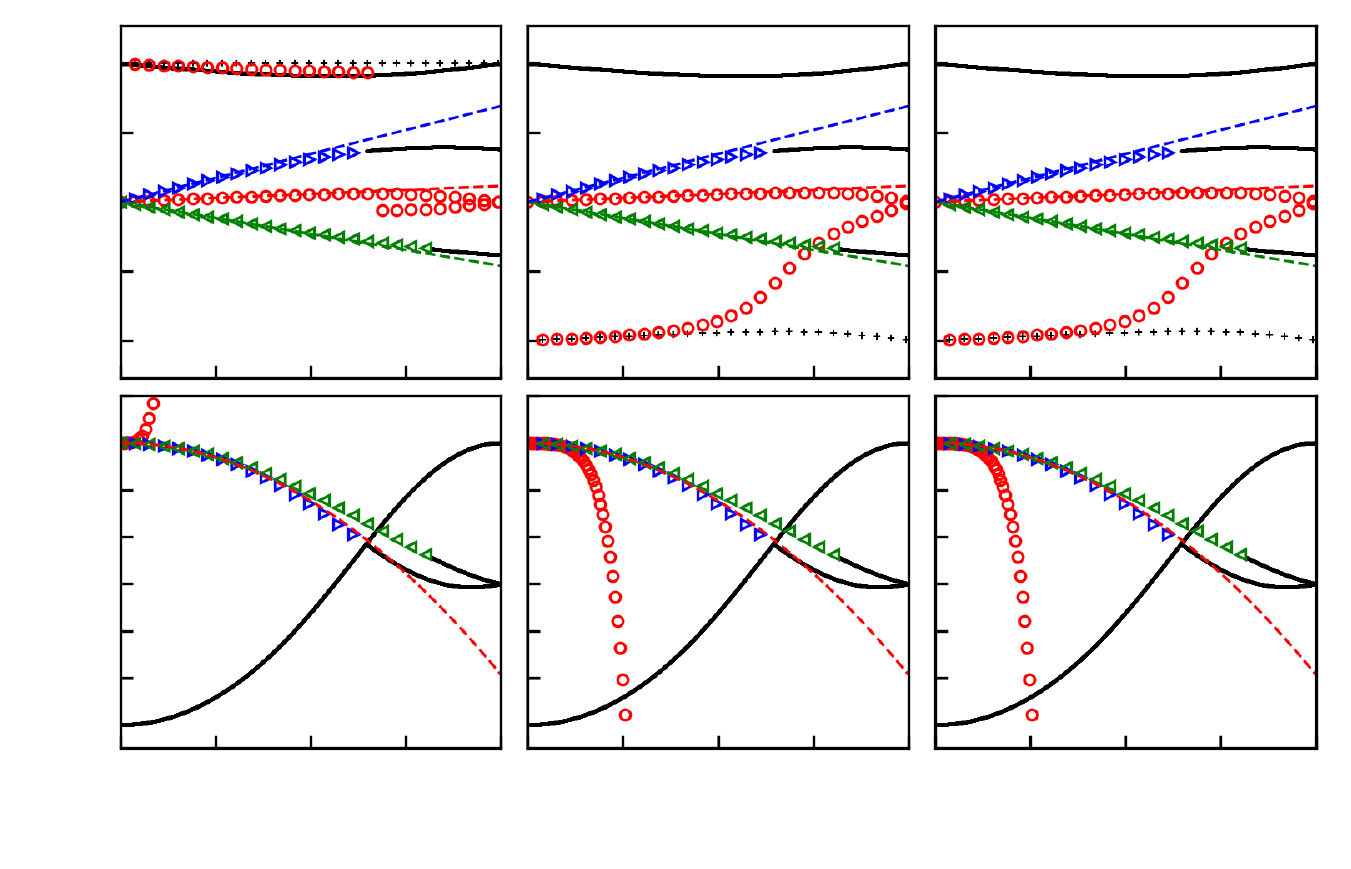}
\put(-400,189){$\omega_r$}
\put(-365,230){$\pi$}
\put(-375,210){$\pi/2$}
\put(-365,190){$0$}
\put(-382,170){$-\pi/2$}
\put(-372,150){$-\pi$}
\put(-400,83){$\omega_i/\nu$}
\put(-365,120){$0$}
\put(-372,107){$-2$}
\put(-372,94){$-4$}
\put(-372,80){$-6$}
\put(-372,66){$-8$}
\put(-377,52){$-10$}
\put(-377,39){$-12$}
%\put(-420,11){$k_x$}
%\put(-476,24){$0$}
%\put(-455,24){$\pi/4$}
%\put(-428,24){$\pi/2$}
%\put(-402,24){$3\pi/4$}
%\put(-367,24){$\pi$}
\put(-300,11){$k_x$}
\put(-357,24){$0$}
\put(-335,24){$\pi/4$}
\put(-308,24){$\pi/2$}
\put(-282,24){$3\pi/4$}
\put(-249,24){$\pi$}
\put(-185,11){$k_x$}
\put(-239,24){$0$}
\put(-217,24){$\pi/4$}
\put(-190,24){$\pi/2$}
\put(-165,24){$3\pi/4$}
\put(-131,24){$\pi$}
\put(-70,11){$k_x$}
\put(-122,24){$0$}
\put(-100,24){$\pi/4$}
\put(-73,24){$\pi/2$}
\put(-48,24){$3\pi/4$}
\put(-13,24){$\pi$}
\caption{Propagation (top) and dissipation (bottom) curves of the D2Q9 lattice with $\tau=10^{-5}$, $N=4^*$, $\overline{\mathrm{Ma}}=0.2$, $k_y=0$ and four regularized collision models, from left to right: PR, RR$3^*$ and RR$4^*$. Modes carrying more than $\eta=90\%$ of a physical wave are identified: \markershear:~shear, \markeracp:~downstream acoustics, \markeracm:~upstream acoustics, \markernum:~non-identified wave, \footnotesize{+}\,\normalsize{:~non-observable mode. Navier-Stokes reference curves are displayed as: \reddashedline:~shear, \bluedashedline:~downstream acoustics, \greendashedline:~upstream acoustics.}\label{fig:D2Q9_regul_spectrums}}
\end{minipage}
\end{figure*}
%\twocolumngrid

Fig.~\ref{fig:D2Q9_regul_spectrums} displays the propagation and dissipation curves for $\tau=10^{-5}$ and a horizontal mean flow at $\overline{\mathrm{Ma}}=0.2$. Only monochromatic plane waves travelling in the horizontal direction ($k_y=0$) are considered here. To avoid a numerical noise in the solutions due to a lack of accuracy in the eigenvalue problem resolution, modes for which $|e^{\mathrm{i} \omega}|<10^{-15}$ have not been plotted on the figure. A first observation can be made: instead of the nine modes observed with the BGK collision model (consistent with the nine velocities of the lattice), six modes are present with the PR, RR$3^*$ and RR$4^*$ models. A very interesting property of the regularization can immediately be suggested: by reducing the number of modes, the destructive interactions occurring with the BGK collision model, and responsible for strong instability issues, are likely to be less frequent. This \textit{mode filtering} property will be particularly investigated in Sec.~\ref{sec:discussion}.

Let us now look closer at the spectral curves for each case of Fig.~\ref{fig:D2Q9_regul_spectrums}. In the PR-case, one mode is non-observable whatever the wavenumber (identified with `\tiny{+}\normalsize'), two modes carry the acoustic information, at least for the more resolved wavelengths (`\markeracp' and `\markeracm') and two modes carry a shear information (`\markershear'). Note that one of the latter has such a large attenuation rate ($\omega_i/\nu\approx-10^{5}$) that it cannot be observed on the dissipation curve. Surprisingly, the other one has a positive amplification rate for any wavenumber $k_x \in [0, \pi]$. As a consequence, in a LB computation, a shear wave is expected to grow with time, leading to an unavoidable instability. A very important point is that, unlike the BGK collision model, this instability does not seem to be related to any \textit{eigenvalue collision}. This mode carrying shear seems indeed to be intrinsically unstable. This specificity will be further discussed in Sec.~\ref{sec:discussion}.

%\clearpage
%\onecolumngrid
\begin{figure*}[h!]
\begin{minipage}{1.\textwidth}
\hspace{-8mm}
\centering
\includegraphics[scale=1.0]{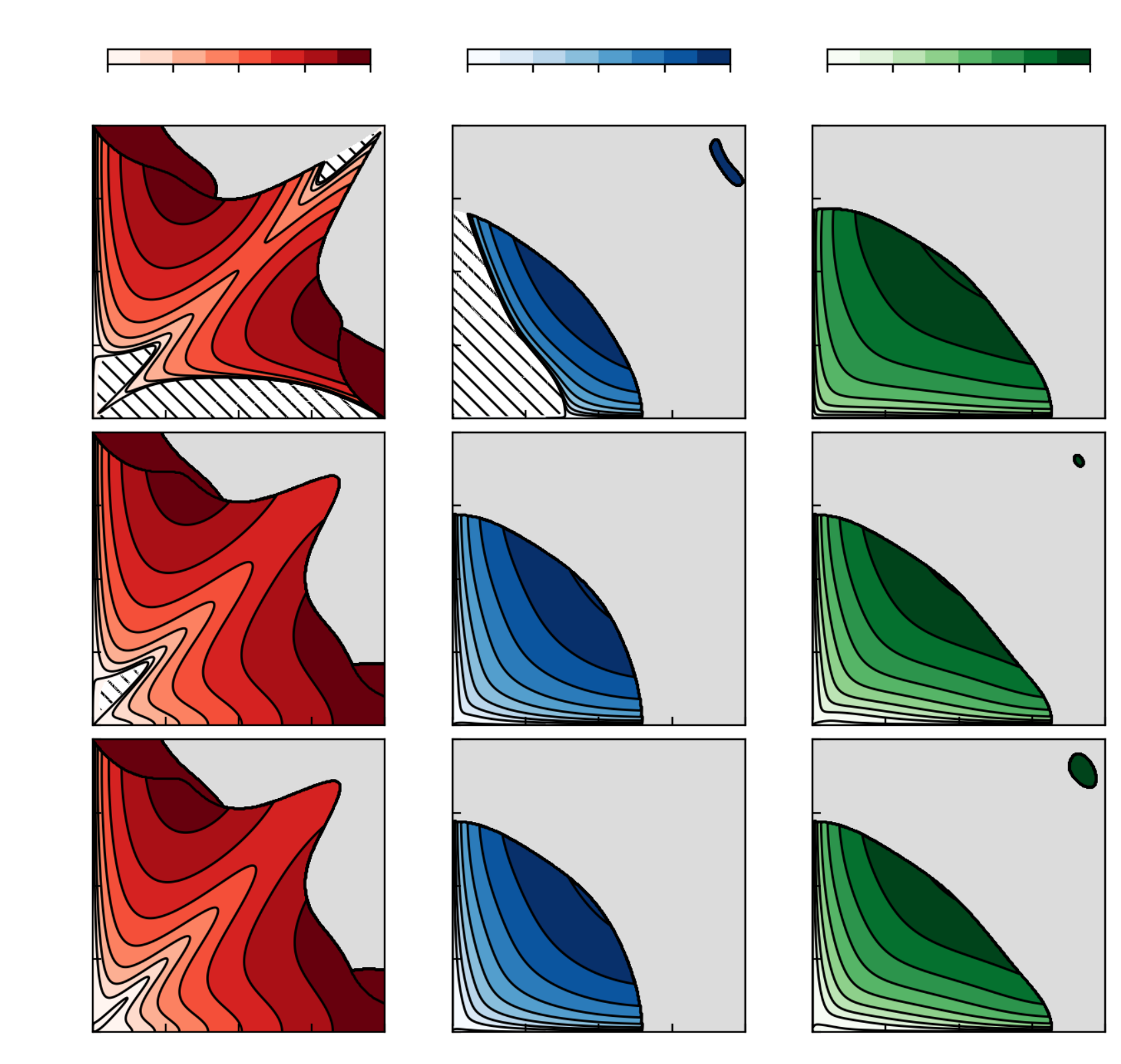}
\put(-397,450){$\nu_e/\nu$}
\put(-447,424){\footnotesize{$10^{0}$}}
\put(-418,424){\footnotesize{$10^{1}$}}
\put(-390,424){\footnotesize{$10^{2}$}}
\put(-362,424){\footnotesize{$10^{3}$}}
\put(-334,424){\footnotesize{$10^{4}$}}
\put(-241,450){$\nu_e/\nu$}
\put(-290,424){\footnotesize{$10^{0}$}}
\put(-262,424){\footnotesize{$10^{1}$}}
\put(-234,424){\footnotesize{$10^{2}$}}
\put(-205,424){\footnotesize{$10^{3}$}}
\put(-176,424){\footnotesize{$10^{4}$}}
\put(-84,450){$\nu_e/\nu$}
\put(-134,424){\footnotesize{$10^{0}$}}
\put(-106,424){\footnotesize{$10^{1}$}}
\put(-77,424){\footnotesize{$10^{2}$}}
\put(-49,424){\footnotesize{$10^{3}$}}
\put(-20,424){\footnotesize{$10^{4}$}}
\put(-481,350){$k_y$}
\put(-457,407){$\pi$}
\put(-472,375){$3\pi/4$}
\put(-467,343){$\pi/2$}
\put(-467,311){$\pi/4$}
\put(-457,279.5){$0$}
\put(-481,216){$k_y$}
\put(-457,273){$\pi$}
\put(-472,241){$3\pi/4$}
\put(-467,209){$\pi/2$}
\put(-467,177){$\pi/4$}
\put(-457,145){$0$}
\put(-481,80){$k_y$}
\put(-457,139){$\pi$}
\put(-472,107){$3\pi/4$}
\put(-467,75){$\pi/2$}
\put(-467,43){$\pi/4$}
\put(-457,12){$0$}
\put(-392,-7){$k_x$}
\put(-450.5,5){$0$}
\put(-423.5,5){$\pi/4$}
\put(-392,5){$\pi/2$}
\put(-362,5){$3\pi/4$}
\put(-324,5){$\pi$}
\put(-235,-7){$k_x$}
\put(-294,5){$0$}
\put(-267,5){$\pi/4$}
\put(-235,5){$\pi/2$}
\put(-205,5){$3\pi/4$}
\put(-167,5){$\pi$}
\put(-79,-7){$k_x$}
\put(-137,5){$0$}
\put(-110,5){$\pi/4$}
\put(-78,5){$\pi/2$}
\put(-49,5){$3\pi/4$}
\put(-10,5){$\pi$}
\caption{Dissipation properties of the modes carrying physical waves with the D2Q9 lattice, $\tau=10^{-5}$, $N=4^*$ and a horizontal mean flow at $\overline{\mathrm{Ma}}=0.2$. Hatched areas indicate zones where $\omega_i>0$ (unstable wavenumbers). From top to bottom: PR, RR$3^*$ and RR$4^*$ collision models. Left: shear, middle: downstream acoustics, right: upstream acoustics. Grey color indicate zones where no physical wave could be identified with $\eta=90\%$. \label{fig:D2Q9_dissipation_isotropy_regul}}
\end{minipage}
\end{figure*}
%\twocolumngrid
%\clearpage

RR$3^*$ and RR$4^*$ models are perfectly equivalent in the horizontal direction. The physical content of the six modes is the same as with the PR model: one mode is non-observable, two modes carry shear (one of which is very damped) and two modes carry the acoustics. However, unlike the PR model, the shear mode of larger amplification rate is not unstable. On the contrary, it is rather very attenuated compared to the Navier-Stokes expectations. For instance, for $k_x = \pi/4$, the effective viscosity of this mode is about $15\nu$. Acoustic modes travelling in the horizontal direction, for their part, do not seem to suffer from an attenuation compared to the BGK case of Fig.~\ref{fig:D2Q9_BGK_Mach_spectrums}.

%Regarding the FR2 model, three modes only remain, which makes their identification much easier. One of them carries the shear wave whatever the wavenumber, while the two other ones carry the acoustics, at least for $k_x < 2\pi/3$. A very surprising dissipative behavior can be noticed for the three modes. The shear one is even more attenuated than with the RR$3^*$ and RR$4^*$ models. The acoustic ones are unexpectedly amplified for the more resolved wavelengths ($k_x<\pi/8$), and significantly attenuated for larger wavenumbers $k_x$. All in all, whether for acoustics or shear, the effective viscosity is in the order of $1000\nu$ for plane waves with $k_x=\pi/4$, \textit{i.e.} 8 points per wavelengh. While the propagation properties of this model seem in good agreement with the Navier-Stokes expectations, the dissipation is completely unphysical, and the numerical scheme is linearly unstable without noticing any \textit{eigenvalue collision} phenomenon. Possible reasons for these wrong dissipative properties will be investigated in Sec.~\ref{sec:discussion}.\newline

A better overview of the dissipative and isotropy property of every model is displayed on Fig.~\ref{fig:D2Q9_dissipation_isotropy_regul}, where $\nu_e/\nu$ is plotted over all the possible wavenumber vectors $\boldsymbol{k}$, for each identified physical information. Similar flow conditions as previously are simulated ($\tau=10^{-5}$, $\overline{\mathrm{Ma}}=0.2$). It immediately highlights an overall more dissipative behavior of all the considered regularized models, for any physical wave and in every direction, as well as some unstable regions.

More specifically, regarding the PR model, a large instability zone of the shear mode is highlighted, as it could be guessed from Fig.~\ref{fig:D2Q9_regul_spectrums}. While the acoustic waves travelling along the horizontal direction have a rather correct dissipation rate (\textit{cf.} Fig.~\ref{fig:D2Q9_regul_spectrums}), the downstream ones are unstable in all other directions. Moreover, a strong anisotropic behavior makes the upstream one significantly attenuated in the diagonal directions. 

The RR$3^*$ model is actually unstable under these flow conditions, which could not be guessed from Fig.~\ref{fig:D2Q9_regul_spectrums}. This is due to a small instability zone exhibited for the shear waves in the diagonal direction. The two acoustic waves remain stable, but they suffer from the same kind of anisotropy as the PR model. This is also the case of the RR$4^*$ model, which has very similar properties as the RR$3^*$ one, while remaining stable. %Finally, regarding the FR2 model, the aforementioned over-dissipation is exhibited in every direction. In addition to the unstable acoustics, an instability zone can be observed on the shear modes travelling in the diagonal directions. FR$3^*$ and FR$4^*$ turn out to have completely equivalent properties, this is why they are not represented here.

To summarize these observations, %the physical nature of the dissipation properties can undoubtedly be questioned for all the regularized models studied here. These properties 
strong anisotropic deviations are observed on the spectral dissipation of all regularized models. The latter do not seem to have a physical meaning, but on the contrary, they might be related to the numerical discretization of the regularized schemes. To investigate this aspect, the numerical properties of regularized approaches will be numerically emphasized in Sec.~\ref{sec:numerical_validation}, then possible explanations will be provided in Sec.~\ref{sec:discussion}. 

\subsection{D2V17 lattice}

%\onecolumngrid
\begin{figure*}[htb]
\begin{minipage}{1.\textwidth}
\centering
\includegraphics[scale=1.]{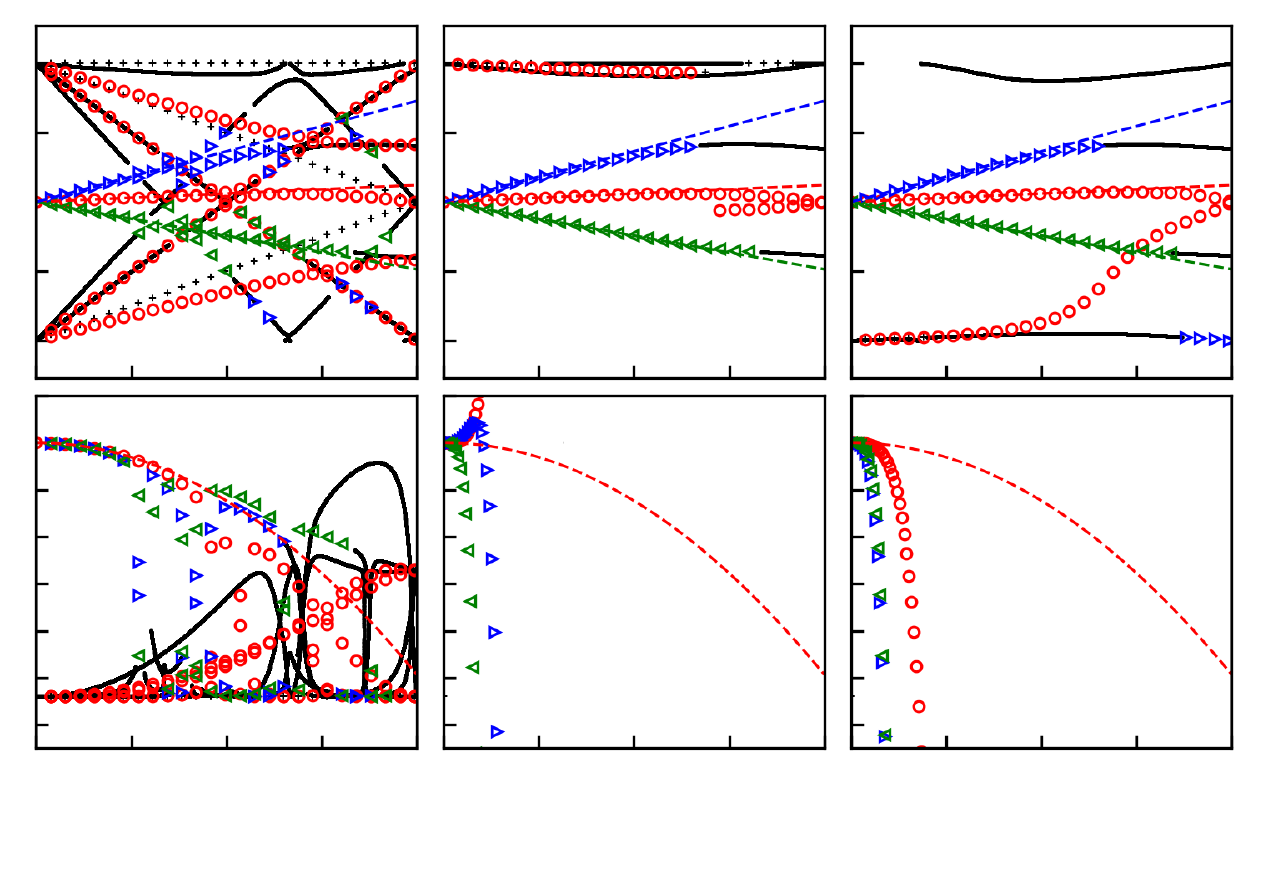}
\put(-388,189){$\omega_r$}
\put(-364,230){$\pi$}
\put(-374,210){$\pi/2$}
\put(-364,190){$0$}
\put(-381,170){$-\pi/2$}
\put(-371,150){$-\pi$}
\put(-399,83){$\omega_i/\nu$}
\put(-364,120){$0$}
\put(-371,107){$-2$}
\put(-371,94){$-4$}
\put(-371,80){$-6$}
\put(-371,66){$-8$}
\put(-376,52){$-10$}
\put(-376,39){$-12$}
%\put(-420,11){$k_x$}
%\put(-476,24){$0$}
%\put(-455,24){$\pi/4$}
%\put(-428,24){$\pi/2$}
%\put(-402,24){$3\pi/4$}
%\put(-367,24){$\pi$}
\put(-300,11){$k_x$}
\put(-357,24){$0$}
\put(-335,24){$\pi/4$}
\put(-308,24){$\pi/2$}
\put(-282,24){$3\pi/4$}
\put(-249,24){$\pi$}
\put(-185,11){$k_x$}
\put(-239,24){$0$}
\put(-217,24){$\pi/4$}
\put(-190,24){$\pi/2$}
\put(-165,24){$3\pi/4$}
\put(-131,24){$\pi$}
\put(-70,11){$k_x$}
\put(-122,24){$0$}
\put(-100,24){$\pi/4$}
\put(-73,24){$\pi/2$}
\put(-48,24){$3\pi/4$}
\put(-13,24){$\pi$}
\caption{Propagation (top) and dissipation (bottom) curves of the D2V17 lattice with $\tau=10^{-5}$, $N=3$, $\overline{\mathrm{Ma}}=0.2$ and four regularized collision models, from left to right: BGK, PR and RR3. Modes carrying more than $\eta=90\%$ of a physical wave are identified: \markershear:~shear, \markeracp:~downstream acoustics, \markeracm:~upstream acoustics, \markernum:~non-identified wave, \footnotesize{+}\,\normalsize{:~non-observable mode. Navier-Stokes reference curves are displayed as: \reddashedline:~shear, \bluedashedline:~downstream acoustics, \greendashedline:~upstream acoustics.}\label{fig:D2V17_regul_spectrums}}
\end{minipage}
\end{figure*}
%\twocolumngrid

To complete these analyses, similar investigations are now performed on the D2V17 lattice whose features are recalled in \ref{app:lattices}. The order of quadrature of this lattice is $Q=7$, so that the equilibrium distribution function can be expanded up to $N=3$, which allows recovering the athermal NS behavior without any Mach error in the momentum equation~\cite{SHAN2006}.  For the same reason, the order of regularized schemes cannot exceed $N_r=3$. The choice $N=3$ is adopted for all the analyses of this section.

Fig.~\ref{fig:D2V17_regul_spectrums} displays propagation and dissipation curves of the D2V17 lattice for $\tau=10^{-5}$, a horizontal mean flow at $\overline{\mathrm{Ma}}=0.2$ and three collision models: the BGK, PR and RR3 collision models.

With the BGK collision model, 17 modes can be identified, which is in agreement with the $(17 \times 17)$-shape matrix  of the linear system. Due to their large number, modal interactions in the form of \textit{curve veering} phenomena frequently occur. Yet, no positive amplification rate is captured for the waves travelling in the horizontal direction. Further analyses over the full set of possible wavenumber vectors $\boldsymbol{k}$, not shown here, highlight severe instability zones for non-horizontal plane waves due to \textit{eigenvalue collision} phenomena.

PR and RR3 models have a rather similar behavior with only six modes remaining, exactly like with the D2Q9 lattice (\textit{cf.} Fig.~\ref{fig:D2Q9_regul_spectrums}). Two modes carry a shear information, two modes carry the acoustics (at least for $k_x < 2\pi/3$), and the last two modes advect a non-identified macroscopic information. Note that contrary to the RR3 model, the PR one is unstable because of a positive amplification rate of the shear mode whatever the wavenumber, and a small instability zone of the downstream acoustic mode.

%Exactly like in the D2Q9 case, only three modes remain with the FR2 model. Compared to the BGK model, the important reduction in the number of modes, especially those carrying a non-physical information, suggests an enhanced numerical stability by avoiding \textit{eigenvalue collisions}. Unlike the D2Q9 case, the FR2 model is stable in the horizontal direction with the D2V17 lattice, but still very dissipative: effective viscosity of waves at $k_x=\pi/4$ is about $3000 \nu$. Spectral properties of the FR3 model are noticed to be very similar to that of the FR2, this is why they are not shown here. \newline

%\onecolumngrid
\begin{figure*}[h!]
\begin{minipage}{1.\textwidth}
\hspace{-8mm}
\centering
\includegraphics[scale=1.]{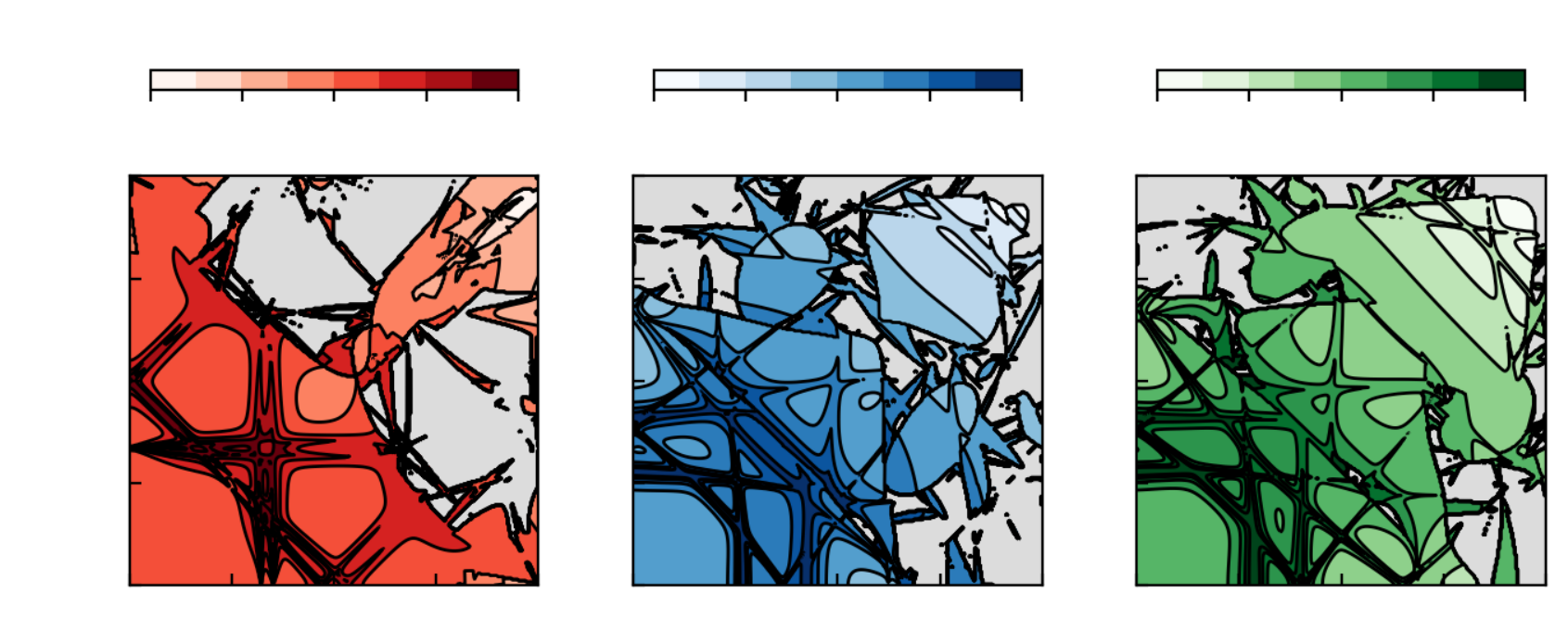}
\put(-397,182){$\nu_e/\nu$}
\put(-450,156){\footnotesize{$0.25$}}
\put(-419,156){\footnotesize{$0.5$}}
\put(-387,156){\footnotesize{$1$}}
\put(-358,156){\footnotesize{$2$}}
\put(-331,156){\footnotesize{$4$}}
\put(-241,182){$\nu_e/\nu$}
\put(-293,156){\footnotesize{$0.25$}}
\put(-262,156){\footnotesize{$0.5$}}
\put(-230,156){\footnotesize{$1$}}
\put(-201,156){\footnotesize{$2$}}
\put(-172,156){\footnotesize{$4$}}
\put(-84,182){$\nu_e/\nu$}
\put(-137,156){\footnotesize{$0.25$}}
\put(-105,156){\footnotesize{$0.5$}}
\put(-73,156){\footnotesize{$1$}}
\put(-45,156){\footnotesize{$2$}}
\put(-16,156){\footnotesize{$4$}}
\put(-481,80){$k_y$}
\put(-457,139){$\pi$}
\put(-472,107){$3\pi/4$}
\put(-467,75){$\pi/2$}
\put(-467,43){$\pi/4$}
\put(-457,12){$0$}
\put(-392,-7){$k_x$}
\put(-450.5,5){$0$}
\put(-423.5,5){$\pi/4$}
\put(-392,5){$\pi/2$}
\put(-362,5){$3\pi/4$}
\put(-324,5){$\pi$}
\put(-235,-7){$k_x$}
\put(-294,5){$0$}
\put(-267,5){$\pi/4$}
\put(-235,5){$\pi/2$}
\put(-205,5){$3\pi/4$}
\put(-167,5){$\pi$}
\put(-79,-7){$k_x$}
\put(-137,5){$0$}
\put(-110,5){$\pi/4$}
\put(-78,5){$\pi/2$}
\put(-49,5){$3\pi/4$}
\put(-10,5){$\pi$}
\subcaption{BGK collision model}
\hspace{-8mm}
\includegraphics[scale=1.]{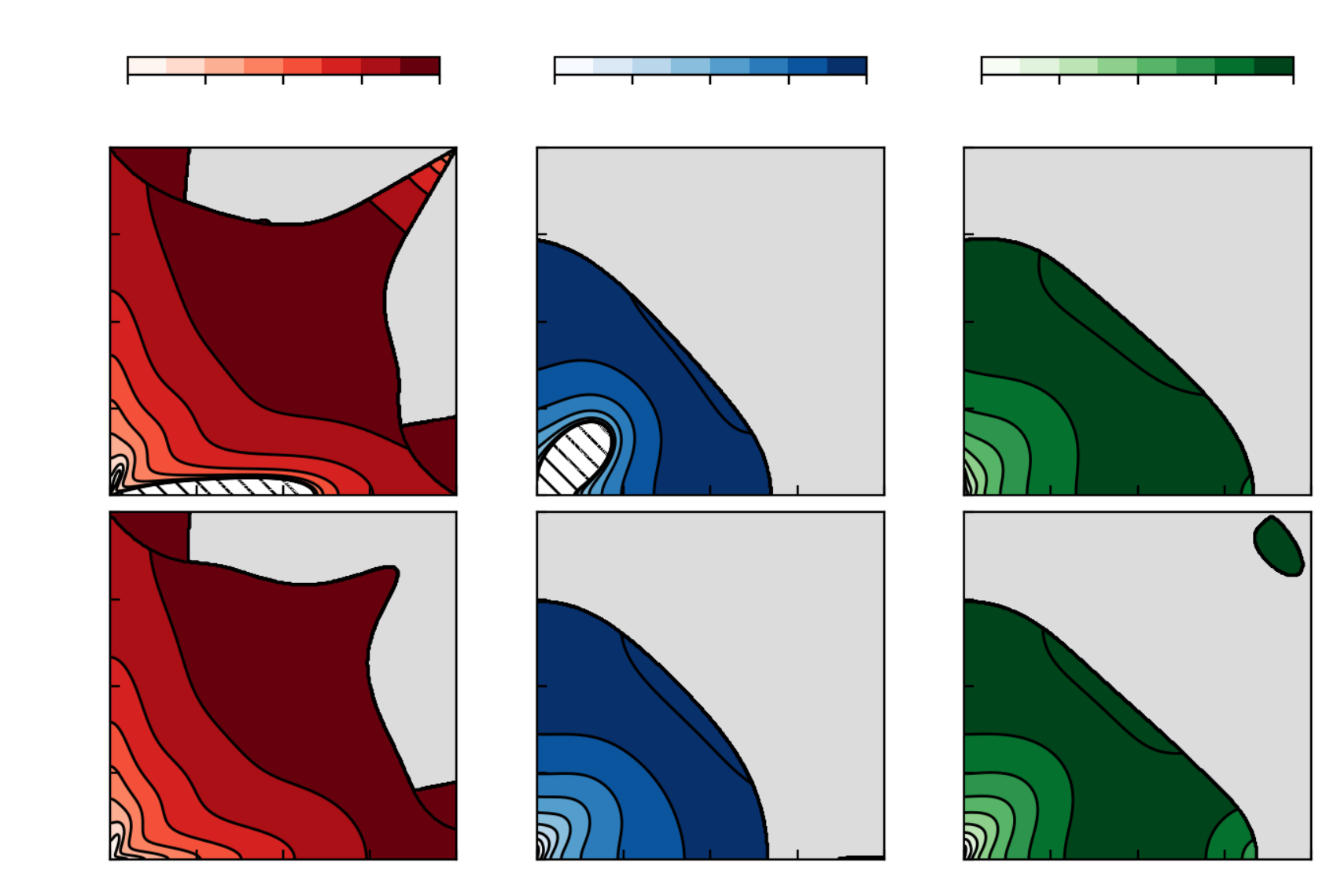}
\put(-397,316){$\nu_e/\nu$}
\put(-447,290){\footnotesize{$10^{0}$}}
\put(-418,290){\footnotesize{$10^{1}$}}
\put(-390,290){\footnotesize{$10^{2}$}}
\put(-362,290){\footnotesize{$10^{3}$}}
\put(-334,290){\footnotesize{$10^{4}$}}
\put(-241,316){$\nu_e/\nu$}
\put(-290,290){\footnotesize{$10^{0}$}}
\put(-262,290){\footnotesize{$10^{1}$}}
\put(-234,290){\footnotesize{$10^{2}$}}
\put(-205,290){\footnotesize{$10^{3}$}}
\put(-176,290){\footnotesize{$10^{4}$}}
\put(-84,316){$\nu_e/\nu$}
\put(-134,290){\footnotesize{$10^{0}$}}
\put(-106,290){\footnotesize{$10^{1}$}}
\put(-77,290){\footnotesize{$10^{2}$}}
\put(-49,290){\footnotesize{$10^{3}$}}
\put(-20,290){\footnotesize{$10^{4}$}}
\put(-481,216){$k_y$}
\put(-457,273){$\pi$}
\put(-472,241){$3\pi/4$}
\put(-467,209){$\pi/2$}
\put(-467,177){$\pi/4$}
\put(-457,145){$0$}
\put(-481,80){$k_y$}
\put(-457,139){$\pi$}
\put(-472,107){$3\pi/4$}
\put(-467,75){$\pi/2$}
\put(-467,43){$\pi/4$}
\put(-457,12){$0$}
\put(-392,-7){$k_x$}
\put(-450.5,5){$0$}
\put(-423.5,5){$\pi/4$}
\put(-392,5){$\pi/2$}
\put(-362,5){$3\pi/4$}
\put(-324,5){$\pi$}
\put(-235,-7){$k_x$}
\put(-294,5){$0$}
\put(-267,5){$\pi/4$}
\put(-235,5){$\pi/2$}
\put(-205,5){$3\pi/4$}
\put(-167,5){$\pi$}
\put(-79,-7){$k_x$}
\put(-137,5){$0$}
\put(-110,5){$\pi/4$}
\put(-78,5){$\pi/2$}
\put(-49,5){$3\pi/4$}
\put(-10,5){$\pi$}
\subcaption{Top: PR, bottom: RR3 collision models}
\end{minipage}
\caption{Dissipation properties of the modes carrying physical waves with the D2V17 lattice, $\tau=10^{-5}$, $N=3$ and a horizontal mean flow at $\overline{\mathrm{Ma}}=0.2$. Hatched areas indicate zones where $\omega_i>0$ (unstable wavenumbers). Left: shear, middle: downstream acoustics, right: upstream acoustics. Grey color indicate zones where no physical wave could be identified with $\eta=90\%$. \label{fig:D2V17_dissipation_isotropy_regul}}
\end{figure*}
%\twocolumngrid
%\clearpage

Ratio of effective viscosities $\nu_e/\nu$ are displayed on Fig.~\ref{fig:D2V17_dissipation_isotropy_regul} over all the possible wavenumber vectors $\boldsymbol{k}$, for each identified macroscopic information (shear or acoustics) and for the three following collision models: BGK, PR and RR3. Several observations are worth noting:
\begin{itemize}
    \item The BGK collision model has a rather isotropic behavior, at least for the most resolved wavelengths. Its dissipation rate is in the order of magnitude of the Navier-Stokes expectations, and no positive amplification of the identified physical waves is noticed. It confirms the fact that eigenvalue collisions of non-physical modes are responsible for the numerical instabilities.
    \item As already noticed with the D2Q9 lattice, the regularized models lead to an over-dissipation of under-resolved modes whatever the direction considered.
    \item Unlike the RR3 model, the PR one is unstable, which cannot be related to any eigenvalue collision phenomenon.
    \item Regularized models suffer from a severe anisotropic dissipation, especially for the shear wave.
%    \item Even if it suffers from an over-dissipation, the FR2 models seems to have correct isotropy properties.
\end{itemize}

\subsection{Linear stability domains}

A better insight of the stability properties of each collision model can be obtained by computing the maximal reachable Mach number under stable flow conditions. To this extent, numerical stability analyses are performed for $k_x \in [-\pi, \pi]$, $k_y \in [0, \pi]$ and mean flow angles $\overline{\theta} \in [0^\circ, 45^\circ]$ with a step $\Delta \overline{\theta} = 1^\circ$. Thanks to the lattice symmetry properties, it is not necessary to investigate mean flows for which $\overline{\theta} > 45^\circ$. Regarding the value of the wavenumber step $\Delta k$, it has to be fine enough so as to well capture the instabilities of a given model, especially with the BGK operator, where instability peaks are likely to be sharp. It is here set to $\Delta k=0.005$, which, as shown in \ref{app:convergence_D2Q9_BGK}, is sufficient to obtain a convergence in the linear stability results of the BGK model. The mean flow Mach number $\overline{\mathrm{Ma}}=\overline{u}/c_s$ is progressively increased by a step $0.001$ until a critical value $\overline{\mathrm{Ma}}^c$ is reached.

\begin{figure*}[ht]
\begin{minipage}{1.\textwidth}
\centering
\hspace{-8mm}
\includegraphics[scale=0.95]{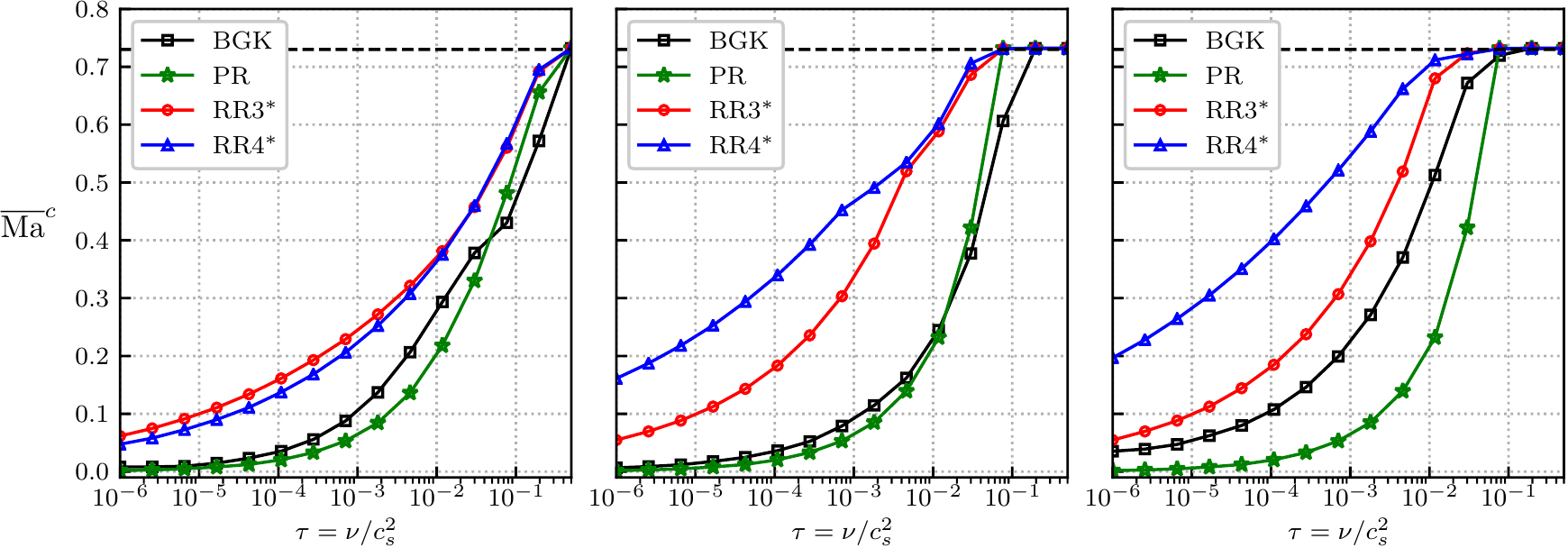}
\begin{minipage}{1.\textwidth}
\hspace{1.5cm}
\begin{minipage}{0.2\textwidth}
\subcaption{$N=2$}
\end{minipage}
\begin{minipage}{0.4\textwidth}
\subcaption{$N=3^*$}
\end{minipage}
\begin{minipage}{0.2\textwidth}
\subcaption{$N=4^*$}
\end{minipage}
\end{minipage}
\caption{Critical mean flow Mach number $\overline{\mathrm{Ma}}^c$ as function of the dimensionless relaxation time $\tau$ for BGK and regularized collision models, with the D2Q9 lattice and several equilibrium distribution orders $N$.  The dashed line represents the theoretical limit of lattice Boltzmann models with a second-order equilibrium: $\overline{\mathrm{Ma}}^c = \sqrt{3}-1 \approx 0.73$~\cite{PAM2019}. \label{fig:critical_Mach_D2Q9}}
\end{minipage}
\end{figure*}

Critical Mach numbers obtained with the D2Q9 lattice are displayed on Fig.~\ref{fig:critical_Mach_D2Q9} as function of the dimensionless relaxation time $\tau$. %Again, FR$3^*$ and FR$4^*$ models are not represented since they share similar properties as the FR2 one. 
A monotonous stability increase is evidenced as $\tau$ increases, whatever the adopted collision model. It has two important consequences:
\begin{itemize}
    \item For a given Mach number, stability can be recovered by increasing the kinematic viscosity $\nu$, \textit{a fortiori} by decreasing the Reynolds number of the simulation.
    \item With an acoustic scaling~\cite{Kruger2017}, for which $\Delta x/\Delta t = c_0/c_s$ where $c_0$ is the physical (dimensional) sound speed, the dimensionless viscosity $\nu$ is related to the dimensional one $\nu^*$ as
\begin{align}
    \nu = \nu^* \frac{\Delta t}{\Delta x^2} = \frac{\nu^* c_s}{c_0 \Delta x}.
\end{align} 
        Hence, $\nu$ can be increased for a given $\nu^*$ by refining the local mesh size without affecting the simulated Reynolds number, which therefore helps increasing numerical stability.
\end{itemize}

Surprisingly, PR models are overall linearly less stable than the standard BGK model, which is probably due to their unexpected dissipation properties exhibited in the previous section. This result may seem in disagreement with previous simulations showing better stability of regularized models%~\cite{Zhang2006, Sengissen2015}
~\cite{Latt2006,Zhang2006,NIU_PRE_76_2007,Montessori2014a,Mattila2015,BA_PRE_97_2018,LI_CF_190_2019}. In fact, one should be cautious with the notions of ``more stable'' or ``less stable'' configurations. In the current context, only maximal reachable Mach numbers ensuring stable numerical simulations are investigated. The present analyses do not focus on the value of the amplification rate, which might be much lower for the PR model than for the BGK one. For instance, with $\mathrm{N=4}$, $\tau=10^{-5}$ and a horizontal mean flow at $\overline{\mathrm{Ma}}=0.2$, one has $\max(\omega_i)/\tau \approx 150$ %for with the PR scheme and $\max(\omega_i)/\tau \approx 150$ for the BGK one.
and $2000$ for the PR and the BGK schemes respectively. Hence, for the former, it is quite conceivable that instabilities require so many iterations to develop that it is not visible in a real computation. It would then be considered more stable in practice than the BGK model for a given Mach number, even if both models are unstable in theory. %This question will be further raised in Sec.~\ref{sec:numerical_validation}.

For their part, recursive regularized schemes %succeed in considerably increasing 
systematically increase the maximal stable Mach number whatever the relaxation time. Most significant gains are observed in the zero-viscosity limit. This result is in agreement with previous numerical simulations which underlined a better stability of these models~\cite{Malaspinas2015,BROGI_JASA_142_2017, Coreixas2017, Coreixas2018,Wissocq2019_these,COREIXAS_RSTA_378_2020}. Among all the investigated models, the RR$4^*$ one with an equilibrium distribution function expanded up to $N=4^*$ leads to the best increase in stability range.

Note finally that, whatever the adopted collision model, a ceiling cannot be exceeded at $\overline{\mathrm{Ma}}=0.73$. This particular limitation is a consequence of the $O(\mathrm{Ma}^3)$ error inherent to the D2Q9 lattice~\cite{Wissocq2019_these, PAM2019}. This is also in %perfect 
agreement with previous analyses dedicated to different collision models and numerical schemes~\cite{COREIXAS_RSTA_378_2020,Wilde2019}.

%\onecolumngrid
%\twocolumngrid

%\onecolumngrid
\begin{figure*}[ht]
\begin{minipage}{0.95\textwidth}
\centering
\hspace{-8mm}
\includegraphics[scale=1.]{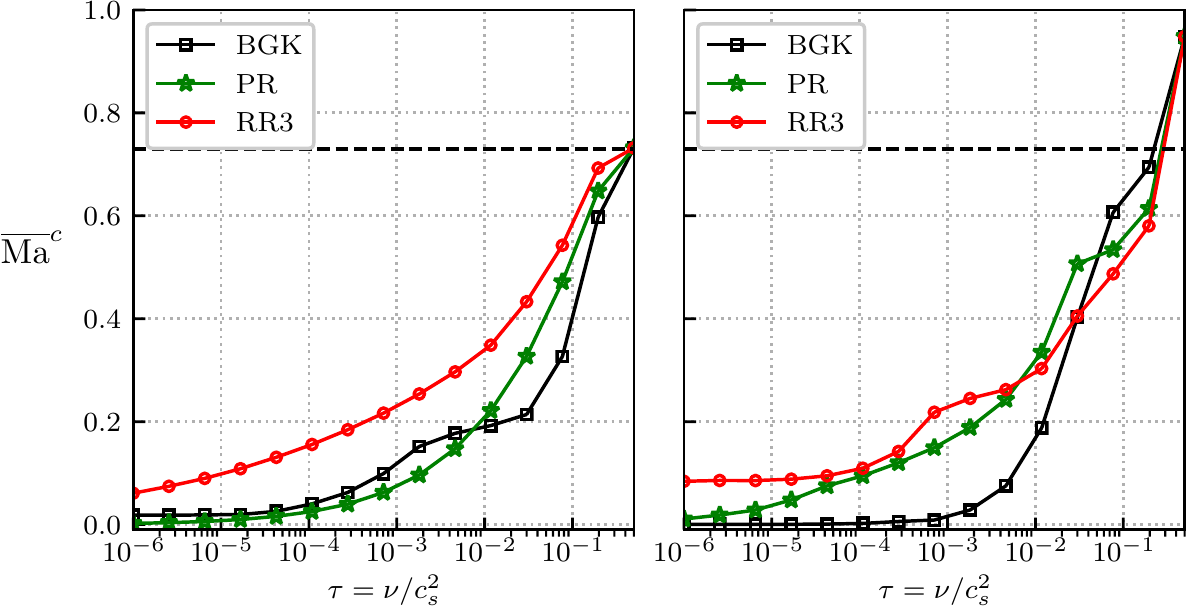}
\begin{minipage}{1.\textwidth}
\hspace{3.5cm}
\begin{minipage}{0.25\textwidth}
\subcaption{$N=2$}
\end{minipage}
\begin{minipage}{0.4\textwidth}
\subcaption{$N=3$}
\end{minipage}
\end{minipage} 
\caption{Critical mean flow Mach number $\overline{\mathrm{Ma}}^c$ as function of the dimensionless relaxation time $\tau$ for BGK and regularized collision models, with the D2V17 lattice and several equilibrium distribution orders $N$. The dashed line represents the theoretical limit of lattice Boltzmann models with a second-order equilibrium: $\overline{\mathrm{Ma}}^c = \sqrt{3}-1 \approx 0.73$~\cite{PAM2019}.\label{fig:critical_Mach_D2V17}}
\end{minipage}
\end{figure*}
%\twocolumngrid 
%\clearpage

To conclude this investigation, similar linear stability analyses are applied to the D2V17 lattice. Results are displayed on Fig.~\ref{fig:critical_Mach_D2V17} for second- and third-order equilibrium distribution functions. With $N=2$, similar observations as with the D2Q9 lattice can be drawn: the recursive regularized model is the most stable one and the maximal Mach number cannot exceed $\overline{\mathrm{Ma}}=0.73$. With $N=3$, stability gains can be effectively achieved for the largest values of $\tau$. The stability range can indeed exceed $\overline{\mathrm{Ma}}=0.73$, which is due to a correctly recovered momentum equation without any Mach error in the shear stress tensor. However, in the low-viscosity region, no significant effect of the equilibrium order can be noticed.

%, in particular:
%\begin{itemize}
%	\item in the low-viscosity region, the stability of the RR3 model is significantly improved,
%	\item for the largest values of $\tau$, the BGK model turns out to be the most stable one.
%\end{itemize}
%%On the contrary, with $N=3$, impressive stability gains can be achieved, especially with the fully reconstructed regularized models. Notably, in the zero viscosity limit, linear stability is ensured while $\overline{\mathrm{Ma}} < 0.67$. This result seems very attractive, even if this stability gain may be obtained at the cost of an over-dissipative behavior, as shown in the previous section. 
%Also note that with $N=3$, the stability range can exceed $\overline{\mathrm{Ma}}=0.73$, which is due to a correctly recovered momentum equation without any Mach error in the shear stress tensor.

%\clearpage
\section{Numerical validation}
\label{sec:numerical_validation}

Sec.~\ref{sec:LSA_results} highlighted the surprising dissipative behavior of regularized collision models. Unlike the standard BGK model, an anisotropic dissipation rate has been exhibited even for relatively well resolved wavelengths, as well as an over-dissipation of some wavenumbers and instabilities that could not be related to \textit{eigenvalue collision} phenomena. For these reasons, the numerical properties of the regularized collisions seem very different from that of the BGK one. The present section aims at performing a numerical validation of the linear results of the previous section. To this extent, several plane monochromatic waves will be simulated in a real LB solver with both the D2Q9 and D2V17 lattices. Two kinds of waves will be considered in this section: shear waves and downstream acoustic ones. To be consistent with the previous analyses, they will be superimposed to a horizontal mean flow at $\overline{\mathrm{Ma}}=0.2$ with a mean dimensionless density $\overline{\rho}=1$. The relaxation time of the collision models will be set to $\tau=10^{-5}$. A partial fourth-order equilibrium ($N=4^*$) will be adopted with the D2Q9 lattice, and a third-order one ($N=3$) with the D2V17 lattice.

\subsection{Shear waves}

Numerical simulations of shear waves are performed in this section. A horizontal one is first considered thanks to the following initialization of macroscopic fields:
\begin{align}
    \rho_0 (x, y) &= \overline{\rho}, \\
    u_{0_x}(x,y) &= \overline{\mathrm{Ma}}\, c_s, \\
    u_{0_y}(x,y) &= \epsilon\, \overline{\mathrm{Ma}}\, c_s \, \cos(k_x x),
\end{align}
with $\epsilon=0.001$ and $k_x=2\pi/8$, so as to simulate a sine wave with eight voxels per wavelength. Distribution functions are initialized at the corresponding equilibrium $f^{eq,N}(\rho_0, \boldsymbol{u_0})$ on the D2Q9 and D2V17 lattices, with the BGK and regularized collision models. Computations are performed on a 2D numerical domain of $(80 \times 2)$ voxels with fully periodic boundary conditions, so as to simulate ten periods of the shear wave. Note that a decay of the transverse velocity is expected by the Navier-Stokes equations as
\begin{align}
    u_y \sim e^{-\nu k_x^2 t}.
\end{align}
On the contrary, with each of the investigated collision models, an \textit{effective} viscosity $\nu_e$ will be involved instead of the physical viscosity $\nu$. Note that $\nu_e$ can be negative, as some models are expected to be unstable.

\begin{figure}[ht]
\centering
\hspace{-7mm}
\includegraphics[scale=0.9]{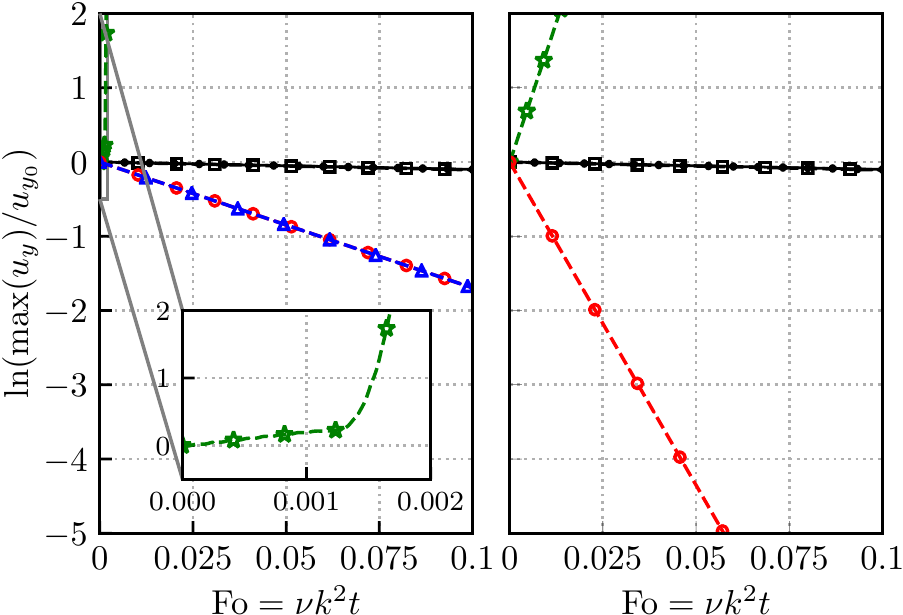}
\begin{minipage}{0.23\textwidth}
	\vspace{2mm}
    \subcaption{D2Q9 lattice}
\end{minipage}
\begin{minipage}{0.15\textwidth}
	\vspace{2mm}
    \subcaption{D2V17 lattice}
\end{minipage}
\caption{Decay/growth of the vertical velocity of a horizontal  shear wave as function of the Fourier number. \blackdashedlinesquare:~BGK, \greendashedlinestar:~PR, \reddashedlinecircle:~RR3, \bluedashedlinetriangle:~RR4, %\orangedashedlinetriangle:~FR2, \slatebluedashedlinetriangle:~FR3, \purpledashedlinediamond:~FR4, 
\blacklinefullcircle:~NS.\label{fig:horizontal_shear_max_uy_time}}
\end{figure}

Fig.~\ref{fig:horizontal_shear_max_uy_time} displays the logarithm of the maximal vertical velocity $\max(u_y)$ normalized by $u_{0_y}$, as function of a dimensional time expressed as a Fourier number $\mathrm{Fo}=\nu k^2 t$. Plotting these quantities allows an easy identification of the ratio $\nu_e/\nu$ as the slope of a given curve. The following conclusions can be drawn:
\begin{itemize}
    \item with both lattices, the dissipation of the BGK collision model is close to the NS expectations,
    \item an amplification of the shear wave is observed with the PR model for both lattices, 
    \item with the D2Q9 lattice, RR$3^*$ and RR$4^*$ models have the same dissipation rate, which is over-estimated compared to the theory,
    \item the same behavior, although more amplified, is found with the RR3 model on the D2V17 lattice. 
%    \item the full reconstructed regularizations are even much more dissipative.
\end{itemize}
All these conclusions are in perfect agreement with the linear stability analyses of Sec.~\ref{sec:LSA_results}.

\begin{figure}[t]
\centering
\hspace{-6mm}
\includegraphics[scale=0.9]{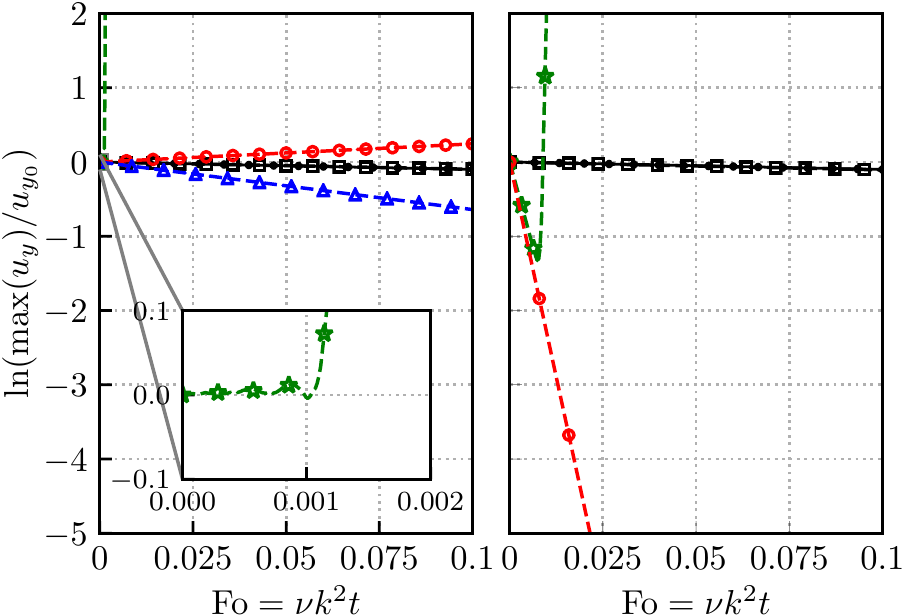}
\begin{minipage}{0.23\textwidth}
	\vspace{2mm}
    \subcaption{D2Q9 lattice}
\end{minipage}
\begin{minipage}{0.15\textwidth}
	\vspace{2mm}
    \subcaption{D2V17 lattice}
\end{minipage}
\caption{Decay/growth of the vertical velocity of an inclined shear wave as function of the Fourier number. \blackdashedlinesquare:~BGK, \greendashedlinestar:~PR, \reddashedlinecircle:~RR3, \bluedashedlinetriangle:~RR4, %\orangedashedlinetriangle:~FR2, \slatebluedashedlinetriangle:~FR3, \purpledashedlinediamond:~FR4, 
\blacklinefullcircle:~NS.\label{fig:diagonal_shear_max_uy_time}}
\end{figure}

Similarly, an inclined shear wave is simulated in order to assess the dissipation properties of non-horizontal waves exhibited in the previous section. To that end, the velocity field is now initialized as %$\boldsymbol{u}_0 = \overline{\boldsymbol{u}_0} + \boldsymbol{u'_0}$ with
\begin{align}
	u_{0_x} &= \overline{\mathrm{Ma}}\, c_s - \epsilon \, \overline{u_x} \, \sin(\theta_k) \cos(k_x x + k_y y), \\
	u_{0_y} &= \epsilon \, \overline{u_x} \, \cos(\theta_k) \cos(k_x x + k_y y),
%	\overline{u_{0}}_x &= \overline{\mathrm{Ma}}\, c_s, \qquad \overline{u_{0}}_y = 0, \\
%    u'_{0_x} &= - \epsilon \, \overline{u_x} \, \sin(\theta_k) \cos(k_x x + k_y y), \\
%    u'_{0_y} &= \epsilon \, \overline{u_x} \, \cos(\theta_k) \cos(k_x x + k_y y),
\end{align}
with $\theta_k=\mathrm{atan2}(k_y,k_x)$. In the following, $k_x=2 \pi/16$ and $k_y=2 \pi/12$ are adopted to deliberately capture the instability of the D2Q9-RR$3^*$ model (\textit{cf.} Fig.~\ref{fig:D2Q9_dissipation_isotropy_regul}). A periodic domain of size $(160 \times 120)$ voxels is used. Linear growths and decays of this inclined wave are displayed on Fig.~\ref{fig:diagonal_shear_max_uy_time}. A perfect agreement with the linear analyses of Sec.~\ref{sec:LSA_results} are obtained. In particular, with the D2Q9 lattice, both PR and RR$3^*$ models are unstable, while the RR$4^*$ one remains stable. An interesting behavior of the D2V17-PR model can be noticed: even though the excited wave is stable (resulting in an initial decay of the velocity amplitude), a severe instability occurs. This is due to the positive amplification rate of horizontal shear waves, which can still exist in the simulation even if not initially triggered.

Note that, under these mean flow conditions, the BGK collision model is expected to be unstable with both the D2Q9 and D2V17 lattices, as displayed on Figs.~\ref{fig:critical_Mach_D2Q9}-\ref{fig:critical_Mach_D2V17}. However, no instability is observed in the test cases considered in the present section. This is due to two distinct phenomena, both illustrated in \ref{app:convergence_D2Q9_BGK}. First, with a periodic domain of size $(N_x, N_y)$, only discrete values of the wavenumbers $k_x \in \{j\pi/N_x, \, j \in \llbracket 0, N_x \rrbracket \}$, $k_y \in \{j\pi/N_y, \, j \in \llbracket 0, N_y \rrbracket \}$ are considered, so that $\Delta k_x = \pi/N_x$ and $\Delta k_y=\pi/N_y$~\cite{HIRSCH_Book_2007}. Hence, the thin instability peaks of the BGK model may not be triggered in such simulations. The second reason lies in the fact that the $x$-aligned mean flow is not the most critical configuration, as displayed on Fig.~\ref{fig:convergence_Mac_BGK_D2Q9_Horizontalflow}.

%However, this instability is not triggered since the wavenumber vectors $\boldsymbol{k}$ corresponding to positive amplification rates are not resolved in the above simulations, due to the size of the periodic domain. Indeed, only discrete values of the wavenumbers $k_x \in \{j\pi/N_x, \, j \in \llbracket 0, N_x \rrbracket \}$, $k_y \in \{j\pi/N_y, \, j \in \llbracket 0, N_y \rrbracket \}$ are considered in such simulations~\cite{HIRSCH_Book_2007}, which does not allows for triggering the very thin instability peaks observed with the BGK model.

\begin{table*}[t]
\centering
\begin{tabular}{c c >{\centering}p{15mm} >{\centering}p{15mm} c >{\centering}p{16mm} c}\toprule
  && \multicolumn{2}{c}{Horizontal} && \multicolumn{2}{c}{Inclined} \\\cmidrule{3-4}\cmidrule{6-7}
  && \multicolumn{2}{c}{$N_{ppx}=8, N_{ppy}=\infty$} && \multicolumn{2}{c}{$N_{ppx}=16, N_{ppy}=12$} \\
  \midrule
  Collision && D2Q9 & D2V17 && D2Q9 & D2V17 \\ \cmidrule{1-1}\cmidrule{3-4}\cmidrule{6-7}
  BGK && 1.15 & 1.07 && 0.98 & 1.06 \\
  PR && -195 & -150 && -2.1 & 180 \\
  RR3 && 17 & 87 && -2.5 & 230 \\
  RR4 && 17 & X && 6.4 & X \\\bottomrule
\end{tabular}
\caption{Estimated ratio $\nu_e/\nu$ for horizontal and inclined shear waves. $N_{ppx}=2\pi/k_x$, $N_{ppy}=2\pi/k_y$.\label{tab:shear_wave_ratios}}
\end{table*}

More quantitatively, estimated ratios of the effective viscosity on the real kinematic one $\nu_e/\nu$ are compiled in Table~\ref{tab:shear_wave_ratios}. Negative values correspond to an anti-dissipative behavior, \textit{i.e.} an instability. Their order of magnitude is in perfect agreement with the linear stability results of Figs.~\ref{fig:D2Q9_dissipation_isotropy_regul}-\ref{fig:D2V17_dissipation_isotropy_regul}.

%\begin{table}[h]
%\begin{tabular}{>{\centering}b{14mm}|>{\centering}p{15mm}|>{\centering}p{15mm}|>{\centering}p{16mm}|c}
%  & \multicolumn{2}{c|}{Horizontal} & \multicolumn{2}{c}{Inclined} \\
%  & \multicolumn{2}{c|}{$N_{ppx}=8, N_{ppy}=\infty$} & \multicolumn{2}{c}{$N_{ppx}=16, N_{ppy}=12$} \\
%  \hline
%  Collision & D2Q9 & D2V17 & D2Q9 & D2V17 \\
%  \hline
%  BGK & 1.15 & 1.07 & 0.98 & 1.06 \\
%  \hline
%  PR & -195 & -150 & -2.1 & 180 \\
%  \hline
%  RR3 & 17 & 87 & -2.5 & 230 \\
%  RR4 & 17 & X & 6.4 & X\\
%%  \hline
%%  FR2 & 7500 & 7400 & 1450 & 3550 \\
%%  FR3 & 7200 & 7000 & 1450 & 3550\\
%%  FR4 & 7200 & X & 1440 & X \\
%\end{tabular}
%\caption{Estimated ratio $\nu_e/\nu$ for horizontal and inclined shear waves. $N_{ppx}=2\pi/k_x$, $N_{ppy}=2\pi/k_y$.\label{tab:shear_wave_ratios}}
%\end{table}

\subsection{Acoustic waves}

\begin{figure}[ht]
\centering
\hspace{-5mm}
\includegraphics[scale=0.9]{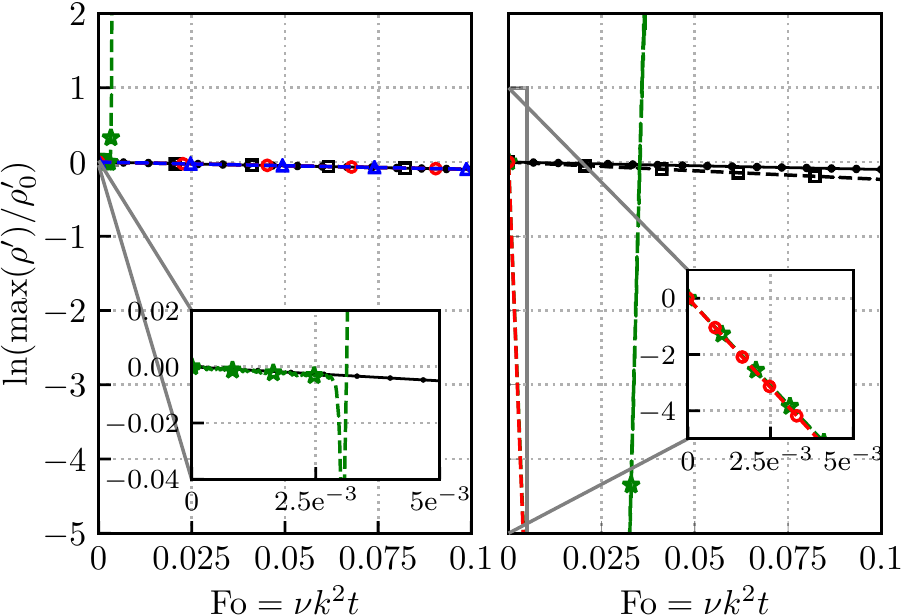}
\begin{minipage}{0.23\textwidth}
	\vspace{2mm}
    \subcaption{D2Q9 lattice}
\end{minipage}
\begin{minipage}{0.15\textwidth}
	\vspace{2mm}
    \subcaption{D2V17 lattice}
\end{minipage}
\caption{Decay/growth of the density amplitude of a horizontal acoustic wave as function of the Fourier number. \blackdashedlinesquare:~BGK, \greendashedlinestar:~PR, \reddashedlinecircle:~RR3, \bluedashedlinetriangle:~RR4, %\orangedashedlinetriangle:~FR2, \slatebluedashedlinetriangle:~FR3, \purpledashedlinediamond:~FR4, 
\blacklinefullcircle:~NS.\label{fig:horizontal_ac_max_rho_time}}
\end{figure}

Similar simulations are now performed with downstream acoustic waves. A horizontal one can first be considered thanks to the following initialization of macroscopic variables:
\begin{align}
    \rho_0 &= \overline{\rho} + \rho_0', \qquad \rho_0'= \epsilon \, \overline{\rho} \, \cos(k_x x), \\
    u_{0_x} &= \overline{\mathrm{Ma}}\, c_s +\rho'_0 c_s/\overline{\rho}, \qquad u_{0_y} = 0,
\end{align}
with $\epsilon=0.001$ and $k_x=2\pi/8$. A periodic domain of size $(80 \times 2)$ voxels is adopted for this case. Linear growth and decay of this wave with the BGK and regularized collision models are displayed on Fig.~\ref{fig:horizontal_ac_max_rho_time} for the D2Q9 and D2V17 lattices. Once again, a very good agreement with the linear stability analyses of Sec.~\ref{sec:LSA_results} are obtained. In particular, with the D2Q9-PR model, the acoustic wave is correctly attenuated during the first instants, even if a strong amplification suddenly occurs, due to the instability of an unexpected shear mode for this pure acoustic test case. D2Q9-RR models behave close to the BGK one in terms of dissipation. With the D2V17 lattice, the acoustic wave is stable with every model, even if a sudden amplification occurs with the PR one, which is, here again, due to the instability of the shear wave. Also note that the acoustic wave is more dissipated than expected by the NS equations, even with the BGK collision model. This is due to the modal interaction highlighted on Fig.~\ref{fig:D2V17_regul_spectrums} (left), responsible for a sudden attenuation of the acoustic wave for $k_x \approx 2\pi/8$. 

\begin{figure}[ht]
\centering
\hspace{-7mm}
\includegraphics[scale=0.9]{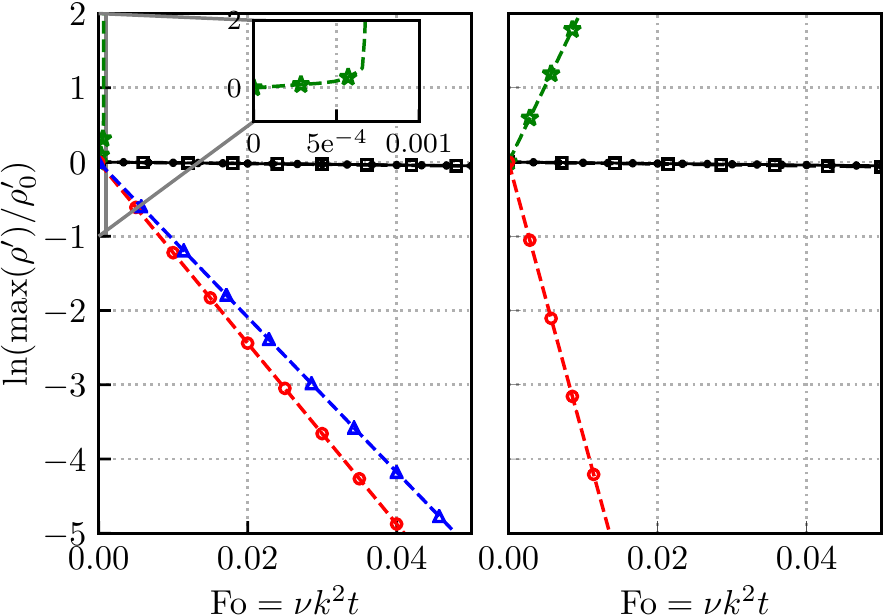}
\begin{minipage}{0.23\textwidth}
	\vspace{2mm}
    \subcaption{D2Q9 lattice}
\end{minipage}
\begin{minipage}{0.15\textwidth}
	\vspace{2mm}
    \subcaption{D2V17 lattice}
\end{minipage}
\caption{Decay/growth of the density amplitude of an inclined  acoustic wave as function of the Fourier number. \blackdashedlinesquare:~BGK, \greendashedlinestar:~PR, \reddashedlinecircle:~RR3, \bluedashedlinetriangle:~RR4, 
%\orangedashedlinetriangle:~FR2, \slatebluedashedlinetriangle:~FR3, \purpledashedlinediamond:~FR4, 
\blacklinefullcircle:~NS.\label{fig:diagonal_ac_max_rho_time}}
\end{figure}

Similarly, an inclined downstream acoustic wave can be initialized as
\begin{align}
    \rho_0 &= \overline{\rho} + \rho_0', \qquad \rho_0'= \epsilon \, \overline{\rho} \, \cos(k_x x+k_y y), \\
    u_{0_x} &= \overline{\mathrm{Ma}}\, c_s +\rho'_0 c_s \cos(\theta_k)/\overline{\rho}, \\
    u_{0_y} &= \rho'_0 c_s \sin(\theta_k)/\overline{\rho},
\end{align}
with $\theta_k = \mathrm{atan2}(k_y, k_x)$. In the following, the wavenumber vector is set as $k_x = 2\pi/16$, $k_y=2\pi/12$ and a periodic domain of size $(160 \times 120)$ voxels is adopted. The linear growths or decays of this wave are displayed on Fig.~\ref{fig:diagonal_ac_max_rho_time}. As expected by the linear analyses, the PR model is unstable with both the D2Q9 and D2V17 lattices, and all other regularized models are over-dissipative. Only the standard BGK model achieves recovering the linear decay expected by the NS equations. 

These results are more precisely quantified in Table~\ref{tab:acoustic_wave_ratios}, where ratios $\nu_e/\nu$ are compiled for each model. Very fair comparisons with the NS dissipative behavior are obtained for the horizontal acoustic wave with the D2Q9 lattice, whatever the adopted collision model. This is in agreement with the linear studies of Mari\'{e} \textit{et al.}~\cite{Marie2009}, who showed that LB methods are less dissipative than $6th$-order optimized NS schemes regarding the acoustics. Note, however, that this trend is not recovered in non-Cartesian directions. In that case, the numerical dissipation of the BGK model is indeed increased, and, as previously noticed for the shear waves, dissipative properties of regularized models for the acoustics are very far from that expected by the NS equations. In addition to the results themselves, these analyses prove the need to perform investigations of the numerical properties in any direction before concluding on the advantages of a given scheme. 

%\begin{table}[h!]
%\begin{tabular}{>{\centering}b{14mm}|>{\centering}p{15mm}|>{\centering}p{15mm}|>{\centering}p{16mm}|c}
%  & \multicolumn{2}{c|}{Horizontal} & \multicolumn{2}{c}{Inclined} \\
%  & \multicolumn{2}{c|}{$N_{ppx}=8, N_{ppy}=\infty$} & \multicolumn{2}{c}{$N_{ppx}=16, N_{ppy}=12$} \\
%  \hline
%  Collision & D2Q9 & D2V17 & D2Q9 & D2V17 \\
%  \hline
%  BGK & 1.00 & 2.1 & 1.05  & 1.13 \\
%  \hline
%  PR & 1.00 & 1120 & -310 & -190 \\
%  \hline
%  RR3 & 1.00 & 1150 & 120 & 330 \\
%  RR4 & 1.00 & X & 105 & X \\
%%  \hline
%%  FR2 & -980 & 2900 & 4900 & 6000 \\
%%  FR3 & -980 & 2200 & 4100 & 5000 \\
%%  FR4 & -980 & X & 4100 & X \\
%\end{tabular}
%\caption{Estimated ratio $\nu_e/\nu$ for horizontal and inclined acoustic waves. $N_{ppx}=2\pi/k_x$, $N_{ppy}=2\pi/k_y$.\label{tab:acoustic_wave_ratios}}
%\end{table}

\begin{table*}[t]
\centering
\begin{tabular}{c c >{\centering}p{15mm} >{\centering}p{15mm} c >{\centering}p{16mm} c}\toprule
  && \multicolumn{2}{c}{Horizontal} && \multicolumn{2}{c}{Inclined} \\\cmidrule{3-4}\cmidrule{6-7}
  && \multicolumn{2}{c}{$N_{ppx}=8, N_{ppy}=\infty$} && \multicolumn{2}{c}{$N_{ppx}=16, N_{ppy}=12$} \\
  \midrule
  Collision && D2Q9 & D2V17 && D2Q9 & D2V17 \\ \cmidrule{1-1}\cmidrule{3-4}\cmidrule{6-7}
  BGK && 1.00 & 2.1 && 1.05  & 1.13 \\
  PR && 1.00 & 1120 && -310 & -190 \\
  RR3 && 1.00 & 1150 && 120 & 330 \\
  RR4 && 1.00 & X && 105 & X \\\bottomrule
\end{tabular}
\caption{Estimated ratio $\nu_e/\nu$ for horizontal and inclined acoustic waves. $N_{ppx}=2\pi/k_x$, $N_{ppy}=2\pi/k_y$.\label{tab:acoustic_wave_ratios}}
\end{table*}

\section{Discussion: numerical properties of regularized models}
\label{sec:discussion}

The analyses of the previous sections have put the light on two paramount properties of the regularized collision models:
\begin{enumerate}
    \item a mode filtering property: some modes carrying a non-physical information have been filtered out of the computation,
    \item an incorrect dissipation rate of the modes carrying shear and acoustics, yielding either an over-dissipation, or an instability of physical waves.
\end{enumerate}
The first point explains how some regularized models achieve to effectively increase the numerical stability of LB simulations. A decrease in the number of modes indeed helps reducing the occurence of \textit{eigenvalue collisions}, which is the main source of instability with the BGK collision model. It is worth noting that the number of remaining modes does not depend on the lattice considered: whatever the lattice, projected (PR) and recursive (RR) regularizations reduce their number to six modes, as noticed in Sec.~\ref{sec:LSA_results}.

On the other hand, the second property turns out to be problematic both for accuracy, because of an over-dissipation of relatively well resolved waves%in the order of $1000\nu$ of some wavenumbers
, and for numerical stability, since an anti-dissipative behavior could be highlighted. It is important to notice that this last point could not be attributed to any eigenvalue collision phenomenon, which makes the numerical behavior of regularized models very different from that of the BGK one. 

%In particular, the numerical behavior of fully reconstructed models is meaningful. Since there is only three modes remaining (all carrying a physical information) whatever the model, no modal interaction can occur between non-hydrodynamic modes and physical waves. One would therefore expect this model to be much more stable than the BGK one, which is unfortunately not the case, as shown on Fig.~\ref{fig:critical_Mach_D2Q9}. One could argue that the bad dissipative behavior of this model is attributed to an estimation of $\boldsymbol{a}_1^{(2)}$ by second-order finite differences. However, a very similar linear behavior can be obtained by performing analyses with exactly computed gradients -- which is fortunately possible in linear stability analyses. It therefore seems that the numerical error induced by a finite difference estimation of this coefficient is drown by a higher-order error. And such an error would actually concern every regularized collision model.

This section aims at providing some explanations of these two phenomena, specific to regularized collision models. But before focusing on these two properties, the light will be put on the analysis of a so-called ``analytically regularized'' scheme, which might help understand the origins of the observed phenomena.

\subsection{LSA of an analytically regularized model}

In this section, a so-called ``analytically regularized'' (AR) collision model is considered. The latter is based on the observation that a Chapman-Enskog expansion~\cite{CHAPMAN_Book_3rd_1970} allows providing an \textit{analytical} expression for the first-order coefficient $\boldsymbol{a}_1^{(2)}$. In the context of athermal equations, it reads
\begin{align}
    a_{1, \alpha \beta}^{(2), \mathrm{AR}} = -\overline{\tau} \rho c_s^2 \left( \frac{\partial u_\alpha}{\partial x_\beta} + \frac{\partial u_\beta}{\partial x_\alpha} \right).
    \label{eq:coef_a1_analytical}
\end{align}
Based on this expression, an AR collision model can be considered, where the $f_i^{(1)}$ term of Eq.~(\ref{eq:general_form_regul}) is replaced by
\begin{align}
\label{eq:AR_fi1}
	f_i^{(1)} = \frac{w_i}{2c_s^4} \boldsymbol{a}_1^{(2), \mathrm{AR}}: \boldsymbol{\mathcal{H}}_i^{(2)}.
\end{align}
Note that, as for the recursive regularization, higher-order terms in Hermite polynomials could be included in $f_i^{(1)}$. However, without any loss of generality, they will not be considered in this section since these high-order terms are not expected to have any macroscopic constribution at the (athermal) Navier-Stokes level. It is all the more noted that linear analyses including third- and fourth-order Hermite polynomials (not shown below) do not affect the conclusions of this section.

Even if the analytical expression of Eq.~(\ref{eq:coef_a1_analytical}) cannot be used as it stands in a LB solver, for which a discretization of the gradient operator is mandatory, a linear stability analysis can be performed on the corresponding time-advance numerical scheme, without any approximation in the computation of the space derivatives. This is the main interest of the analyses proposed in this section.

The derivation of the matrix representative of the AR scheme is proposed in \ref{app:linear_matrices}. The eigenvalue problem of Eq.~(\ref{eq:LSA_matrix_system}) is recovered with
\begin{align}
    M_{ij}^{\mathrm{AR}} = e^{\mathrm{i} \boldsymbol{k} \cdot \boldsymbol{e_i}} \left[ J_{ij}^{eq,N} + \left( 1-\frac{1}{\overline{\tau}} \right) \frac{w_i}{2c_s^{4}}\, \mathbf{\Lambda}_{1, j}^{(2), \mathrm{AR}}:\boldsymbol{\mathcal{H}}_i^{(2)} \right],
\end{align}
where
\begin{align}
    \left(\mathbf{\Lambda}_{1, j}^{(2), \mathrm{AR}}\right)_{\alpha \beta} = -\mathrm{i} \overline{\tau} c_s^2 \left( (e_{j_\alpha} - \overline{u_\alpha}) k_\beta + (e_{j_\beta} - \overline{u_\beta}) k_\alpha \right).
\end{align}

\begin{figure}[ht]
%\begin{minipage}{0.5\textwidth}
\hspace{-10mm}\includegraphics[scale=0.9]{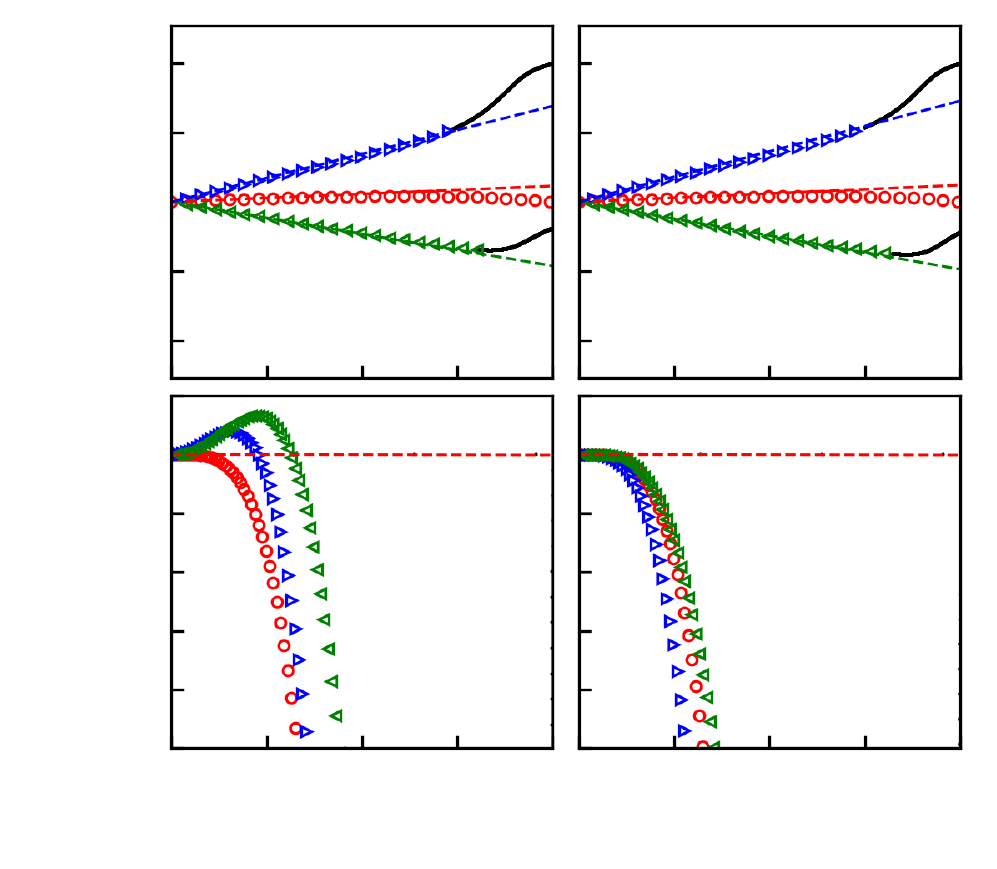}
\put(-168,10){$k_x$}
\put(-217,22){$0$}
\put(-198,22){$\pi/4$}
\put(-173,22){$\pi/2$}
\put(-150,22){$3\pi/4$}
\put(-119,22){$\pi$}
\put(-62,10){$k_x$}
\put(-111,22){$0$}
\put(-92,22){$\pi/4$}
\put(-67,22){$\pi/2$}
\put(-44,22){$3\pi/4$}
\put(-13,22){$\pi$}
\put(-244,27){$-5000$}
\put(-244,42){$-4000$}
\put(-244,58){$-3000$}
\put(-244,73){$-2000$}
\put(-244,88){$-1000$}
\put(-222,104){$0$}
\put(-237,120){$1000$}
\put(-252,104){$\omega_i/\nu$}
\put(-230,135){$-\pi$}
\put(-239.5,152){$-\pi/2$}
\put(-222,170){$0$}
\put(-233,188){$\pi/2$}
\put(-223,207){$\pi$}
\put(-252,170){$\omega_r$}
\caption{Propagation (top) and dissipation (bottom) curves of the AR collision model with $\tau=10^{-5}$, $N=3$, $\overline{\mathrm{Ma}}=0.2$. Left: D2Q9 lattice with $N=4^*$, right: D2V17 lattice with $N=3$. Modes carrying a physical wave are identified as on Fig.~\ref{fig:D2Q9_regul_spectrums}. \label{fig:AR_spectrums} }
%\end{minipage}
\end{figure}
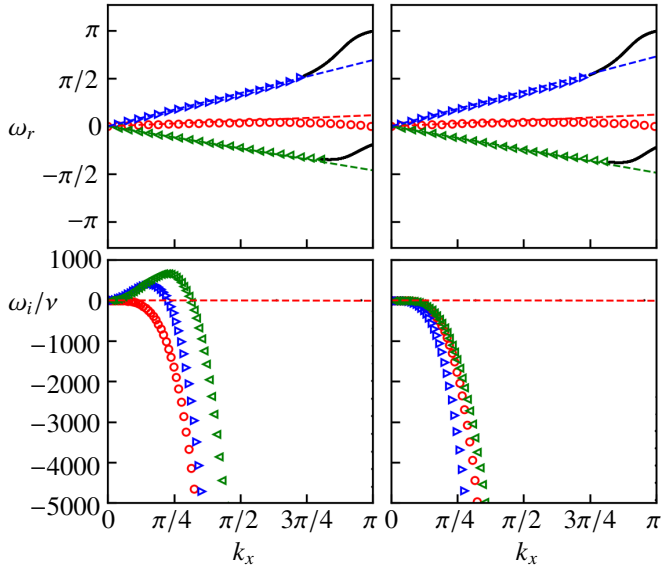

Propagation and dissipation curves along the horizontal direction ($k_y=0$) are displayed on Fig.~\ref{fig:AR_spectrums} for a horizontal mean flow at $\overline{\mathrm{Ma}}=0.2$, $\tau=10^{-5}$ and two lattices: the D2Q9 lattice with $N=4^*$ and the D2V17 lattice with $N=3$. Note that, for the sake of convenience, the scale of the dissipation curves have been adapted for these models.

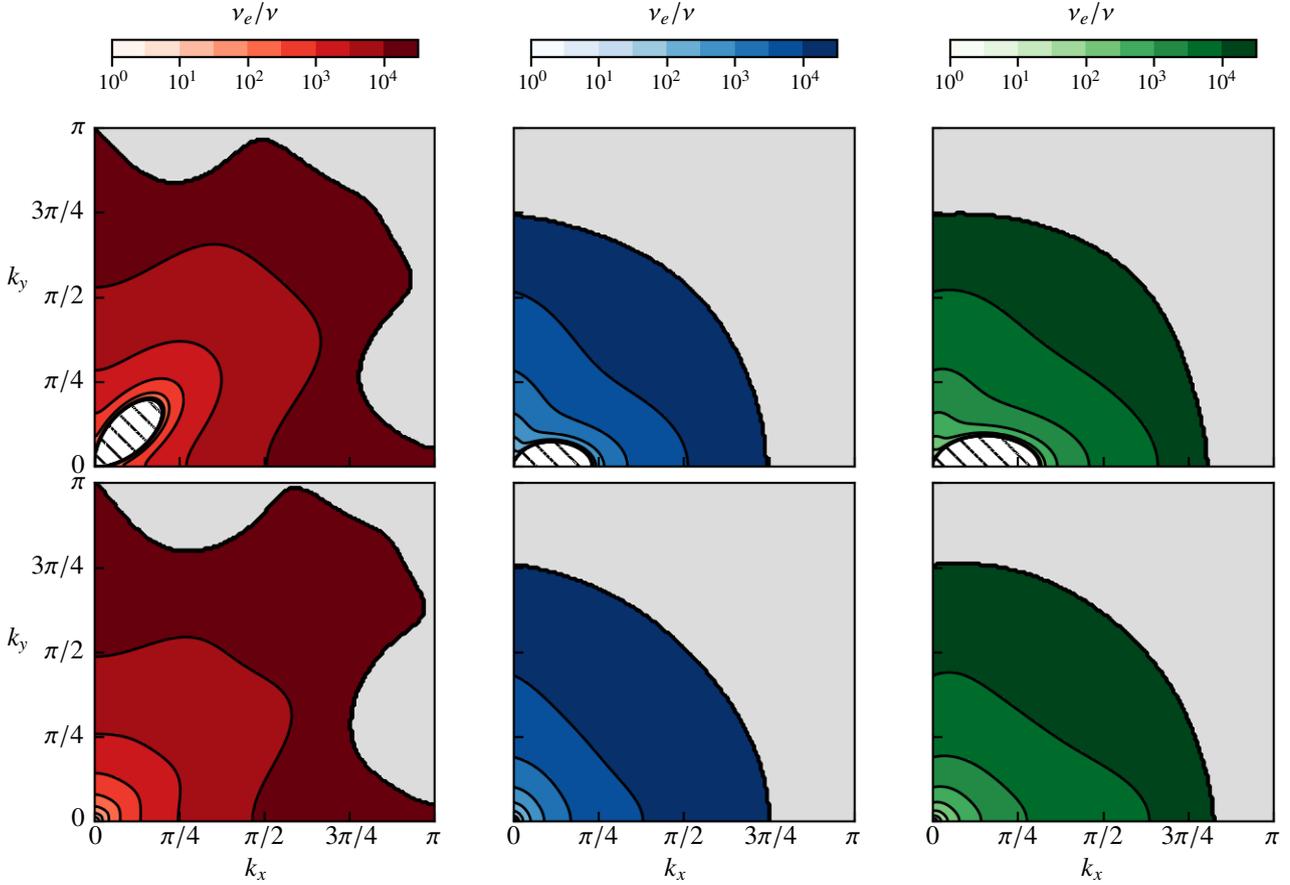
\begin{figure*}[ht]
\begin{minipage}{1.\textwidth}
\centering
\hspace{-1cm}
\includegraphics[scale=1.]{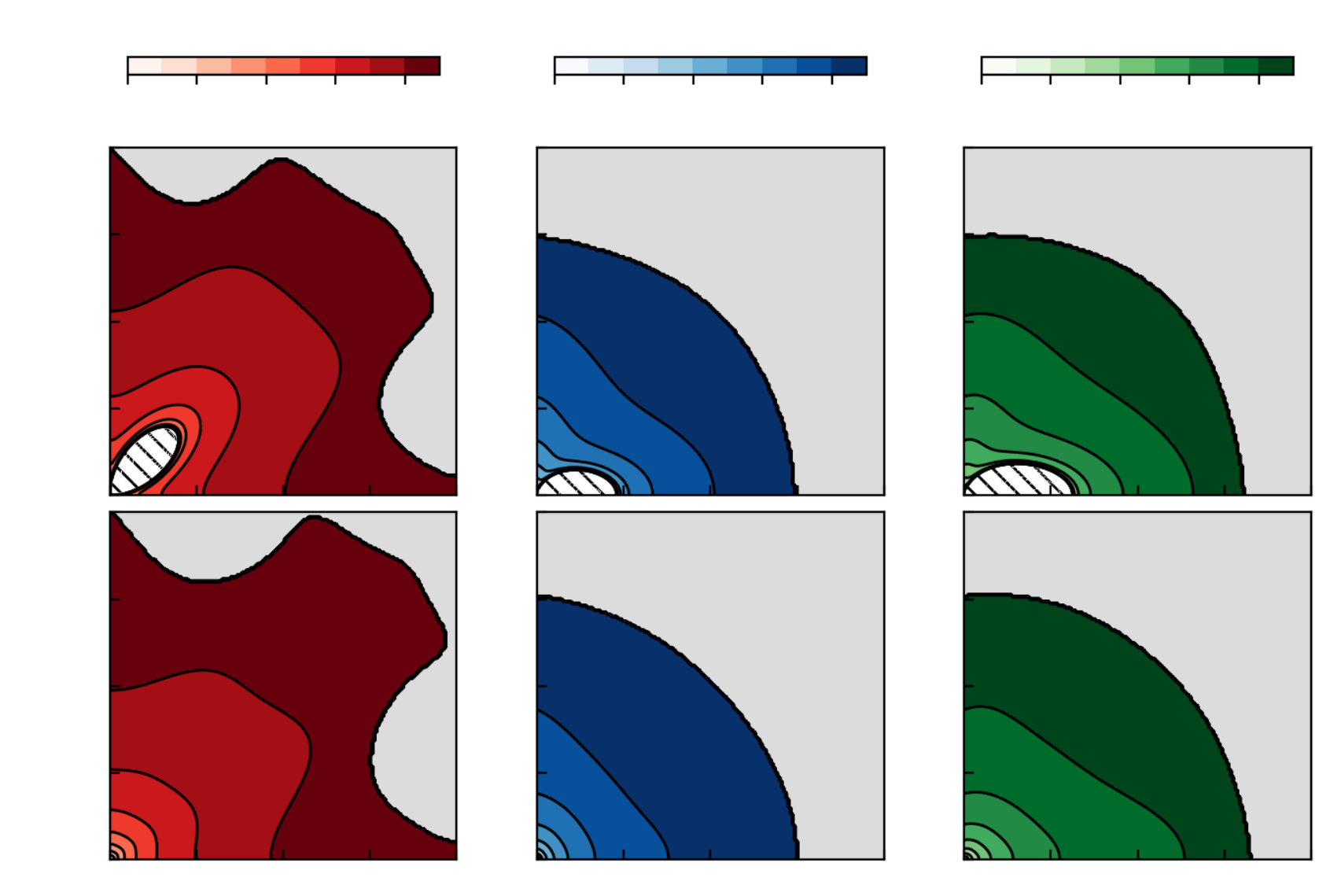}
\put(-397,316){$\nu_e/\nu$}
\put(-447,290){\footnotesize{$10^{0}$}}
\put(-421,290){\footnotesize{$10^{1}$}}
\put(-396,290){\footnotesize{$10^{2}$}}
\put(-371,290){\footnotesize{$10^{3}$}}
\put(-346,290){\footnotesize{$10^{4}$}}
\put(-241,316){$\nu_e/\nu$}
\put(-290,290){\footnotesize{$10^{0}$}}
\put(-265,290){\footnotesize{$10^{1}$}}
\put(-240,290){\footnotesize{$10^{2}$}}
\put(-214,290){\footnotesize{$10^{3}$}}
\put(-188,290){\footnotesize{$10^{4}$}}
\put(-84,316){$\nu_e/\nu$}
\put(-134,290){\footnotesize{$10^{0}$}}
\put(-109,290){\footnotesize{$10^{1}$}}
\put(-83,290){\footnotesize{$10^{2}$}}
\put(-58,290){\footnotesize{$10^{3}$}}
\put(-32 ,290){\footnotesize{$10^{4}$}}
\put(-481,216){$k_y$}
\put(-457,273){$\pi$}
\put(-472,241){$3\pi/4$}
\put(-467,209){$\pi/2$}
\put(-467,177){$\pi/4$}
\put(-457,145){$0$}
\put(-481,80){$k_y$}
\put(-457,139){$\pi$}
\put(-472,107){$3\pi/4$}
\put(-467,75){$\pi/2$}
\put(-467,43){$\pi/4$}
\put(-457,12){$0$}
\put(-392,-7){$k_x$}
\put(-450.5,5){$0$}
\put(-423.5,5){$\pi/4$}
\put(-392,5){$\pi/2$}
\put(-362,5){$3\pi/4$}
\put(-324,5){$\pi$}
\put(-235,-7){$k_x$}
\put(-294,5){$0$}
\put(-267,5){$\pi/4$}
\put(-235,5){$\pi/2$}
\put(-205,5){$3\pi/4$}
\put(-167,5){$\pi$}
\put(-79,-7){$k_x$}
\put(-137,5){$0$}
\put(-110,5){$\pi/4$}
\put(-78,5){$\pi/2$}
\put(-49,5){$3\pi/4$}
\put(-10,5){$\pi$}
\caption{Dissipation properties of the modes carrying physical waves with the AR collision model, $\tau=10^{-5}$, $N=3$ and a horizontal mean flow at $\overline{\mathrm{Ma}}=0.2$. Hatched areas indicate zones where $\omega_i>0$ (unstable wavenumbers). Top: D2Q9 lattice with $N=4^*$, bottom: D2V17 lattice with $N=3$. Left: shear, middle: downstream acoustics, right: upstream acoustics. Grey color indicate zones where no physical wave could be identified with $\eta=90\%$. \label{fig:_dissipation_isotropy_AR}}
\end{minipage}
\end{figure*}

These analyses lead to two major conclusions:
\begin{itemize}
	\item While six modes are present with the PR and RR schemes whatever the lattice, only three modes remain with the AR collision model.
	\item Dissipative properties of both D2Q9 and D2V17 lattice drastically deviate from the NS behavior for $||\boldsymbol{k}||>\pi/8$ (16 points per wavelength), and the D2Q9 lattice is even unstable for this configuration.
\end{itemize}
These incorrect dissipative properties can be more precisely figured out on Fig.~\ref{fig:_dissipation_isotropy_AR}, where any propagation direction of the physical waves is considered. With the D2Q9 lattice with $N=4^*$, all the physical waves (shear and acoustics) have unstable regions, while they remain stable with the D2V17 lattice, even though much more attenuated than expected.

%\onecolumngrid
%\twocolumngrid
%\clearpage

This short study remains useful to better understand the main properties of the regularized models stated above, since they are both recovered in the AR model. In particular, it is important to notice that this AR regularization is \textit{a priori} perfect in the sense of the NS equations. There is indeed no need to enrich, from a physical point of view, the content of the off-equilibrium part of Eq.~(\ref{eq:AR_fi1}), which already contains all the NS physics. Yet, a dissipative issue can be highlighted. It therefore seems that this issue is related to the intrinsic regularization procedure (\textit{i.e.} the fact of re-writing $f_i$ as $f_i
^{eq} + f_i^{(1)},$ with a specific off-equilibrium part, before the collision), rather than to the way coefficients $\boldsymbol{a}_1^{(n)}$ are regularized, since such an error remains whatever the adopted regularized model (PR, RR and AR).

This statement being made, the rest of the section focuses on the origins of the aforementioned properties.

\subsection{Mode filtering property}

Every regularized collision model of interest in the present work can be written as two successive steps: (1) a pre-collision regularization,
\begin{align}
    f_i^{reg} = f_i^{eq, N} + f_i^{(1)},
\end{align} 
followed by (2) a BGK collision,
\begin{align}
    f_i^* = f_i^{reg} - \frac{1}{\overline{\tau}} \left( f_i^{reg} - f_i^{eq, N} \right).
    \label{eq:BGK_collision_part6}
\end{align}
The mode filtering property can be understood by looking carefully at the first step of this scheme. In two dimensions, the computation of the equilibrium distribution function involves three macroscopic quantities $(\rho, u_x, u_y)$, hence three moments of the discrete distributions $f_i$. Regarding $f_i^{(1)}$, it depends on the adopted model:
\begin{itemize}
    \item with the PR and RR models: $f_i^{(1)}$ is a function of six variables: $(\rho, u_x, u_y, a_{1,xx}^{(2)}, a_{1,xy}^{(2)}, a_{1,yy}^{(2)})$, involving six independent moments of the discrete distributions,
    \item with the AR model, since $\boldsymbol{a}_1^{(2)}$ is computed thanks to the knowledge of $(\rho, u_x, u_y)$ only, $f_i^{(1)}$ is a function of these three variables.
\end{itemize}
Pre-collision regularized distribution functions can then be formally written as function of:
\begin{align}
    \mathrm{PR,\ RR}:& \qquad f_i^{reg} (\rho, u_x, u_y, a_{1,xx}^{(2)}, a_{1,xy}^{(2)}, a_{1,yy}^{(2)}), \\
    \mathrm{AR}:& \qquad f_i^{reg} (\rho, u_x, u_y).
\end{align}
This regularization procedure yields a reduction in the rank of the system, which explains the aforementioned decrease in the number of modes: six modes for PR and RR models, three modes for the AR model.

Understanding this property makes it possible to predict the behavior of regularized models in three dimensions. Then, ten modes are expected with the PR and RR models because of the dependency on ten variables: $(\rho, u_x, u_y, u_z, a_{1,xx}^{(2)}, a_{1,xy}^{(2)}, a_{1,xz}^{(2)}, a_{1,yy}^{(2)}, a_{1,yz}^{(2)}, a_{1,zz}^{(2)})$. This larger number of remaining modes may 
%be at the origin of 
possibly lead to a loss of stability for three-dimensional lattices, especially for the BGK collision model, where modal interactions are the main cause of instability. Such investigations will be the purpose of future work.

Finally, note that discussing on the rank of the system is the opportunity to put the light on the particular behavior of any collision model in the case $\tau = 1/2$ (or equivalently $\overline{\tau}=1$). In such a case, Eq.~(\ref{eq:BGK_collision_part6}) becomes $f_i^* = f_i^{eq, N}$, so that the rank of the system is reduced to three in two dimensions (four in three dimensions), yielding a substantial gain in stability induced by a strong mode filtering. It explains the large critical Mach numbers reached on Figs.~\ref{fig:critical_Mach_D2Q9}-\ref{fig:critical_Mach_D2V17}, even with the BGK collision model. 

\subsection{Dissipation error of regularized models}

The incorrect dissipation rate of shear and acoustics encountered with every regularized collision model can \textit{a priori} be explained by two potential sources of error:
\begin{enumerate}
    \item the hydrodynamic behavior of the discrete velocity Boltzmann equations (DVBE), from which regularized models are derived,
    \item a completely numerical effect, induced by the time and space discretization of the DVBE.
\end{enumerate} 
In order to answer this question, it is therefore essential to find out the DVBE from which regularized models are derived. Such a work has been properly done for the BGK collision model~\cite{He1998}, but, to the author's knowledge, no \textit{a priori} derivation of regularized schemes from continuous equations has been achieved yet. In the following, the focus will be put on the PR scheme, which can be written as a multiple relaxation time (MRT) model~\cite{Latt2007}. Note that all the equations below are considered \textit{dimensional} involving the characteristic time $\Delta t$ and length $\Delta x$.

\subsubsection{\textit{A priori} derivation of the PR scheme}
\label{sec:apriori_derivation_PR}

Let us start with the following system of equations, continuous in space and time, and where the velocity space has been discretized:
\begin{align}
\label{eq:regularized_DVBE}
    \frac{\partial f_i}{\partial t} + \boldsymbol{e_i} \cdot \frac{\partial f_i}{\partial \boldsymbol{x}} = \left[ \mathbf{H}^{-1} \mathbf{R} \mathbf{H} \right]_{ij} \left( f_j - f_j^{eq, N} \right),
\end{align}
where $\mathbf{H}$ is the matrix of Hermite polynomials eventually completed by a Gram-Schmidt orthogonalization procedure, and 
\begin{align}
    \mathbf{R} = \left( \frac{2}{\Delta t} - \frac{1}{\tau}\right) \mathbf{P}^{(2)} - \frac{2}{\Delta t} \mathbf{I},
\end{align}
where $\mathbf{I}$ is the identity matrix and $\mathbf{P}^{(2)}$ is the projection matrix onto the second-order terms.
%\begin{align}
%    \mathbf{P}^{(2)} = \mathrm{diag}(0,0,0,1,1,1,0,...).
%\end{align}
For instance, with the D2Q9 lattice, one has
\begin{align}
    & \mathbf{H}_{ij} = (\mathcal{H}^{(0)}_j, \mathcal{H}^{(1)}_{x,j}, \mathcal{H}^{(1)}_{y,j}, \mathcal{H}^{(2)}_{xx,j}, \mathcal{H}^{(2)}_{xy,j}, \mathcal{H}^{(2)}_{yy,j}, \nonumber \\
     & \qquad \qquad \qquad \mathcal{H}^{(3)}_{xxy,j}, \mathcal{H}^{(3)}_{xyy,j}, \mathcal{H}^{(4)}_{xxyy,j})^T, \\
     & \mathbf{P}^{(2)} = \mathrm{diag}(0,0,0,1,1,1,0,0,0), \\
     & \mathbf{R} = -\mathrm{diag}\left(\frac{2}{\Delta t}, \frac{2}{\Delta t}, \frac{2}{\Delta t}, \frac{1}{\tau}, \frac{1}{\tau}, \frac{1}{\tau}, \frac{2}{\Delta t}, \frac{2}{\Delta t}, \frac{2}{\Delta t}\right).
     \label{eq:matrix_R_D2Q9}
\end{align}
Note that, in absence of body-force term,  the first three coefficients of $\mathbf{R}$ have no influence on the model since they are related to collision invariants. It can be shown (\textit{cf.} \ref{app:apriori_derivation_PR}) that the PR model is nothing more than a particular time and space discretization of Eq.~(\ref{eq:regularized_DVBE}), using a trapezium rule and an appropriate change of variables. Hence, Eq.~(\ref{eq:regularized_DVBE}) is the DVBE from which the PR scheme can be derived, it will be referred to as the PR-DVBE in the following.

\subsubsection{Linear analysis of the PR-DVBE}

Thanks to the above \textit{a priori} derivation of the PR scheme, it %can be shown
is confirmed that this collision model is actually part of the multiple-relaxation time (MRT) family. A rather surprising observation can be drawn: the higher-order relaxation times of the continuous equations that the PR scheme intends to solve (namely the PR-DVBE) depend on a numerical parameter: the time step $\Delta t$. However, it is known that, when the ratio between the physical relaxation time ($\tau$) and that of the high-order moment ($2/\Delta t$ here) becomes too large, hyperviscosity may occur~\cite{Geier2015, PAM2019}, \textit{i.e.} high-order dissipation which is not expected by the Navier-Stokes equations, but consequent of an incorrect hydrodynamic limit of the DVBE. This phenomenon could explain the incorrect dissipation rate observed with the PR models, and \textit{a fortiori} with every regularized model, as well as any MRT model. To highlight the presence, or not, of hyperviscosity in the continuous model, it is proposed in this section to perform a linear stability analysis of Eq.~(\ref{eq:regularized_DVBE}). A linearization of the partial differential equations, followed by an injection of the plane monochromatic waves of Eq.~(\ref{eq:fluctuations_monochromatic_waves}), lead to the following eigenvalue problem
\begin{align}
    \omega \mathbf{\widehat{F}} = \mathbf{M}^{\mathrm{PR-DVBE}}\, \mathbf{\widehat{F}},
\end{align}
with
\begin{align}
    M^{\mathrm{PR-DVBE}}_{ij} = \left[ \boldsymbol{k} \cdot \boldsymbol{e_i}\, \delta_{ij} + \mathrm{i} \left[ \mathbf{H}^{-1}\mathbf{R} \mathbf{H} \right]_{il} \left( \delta_{lj} - J_{lj}^{eq, N} \right) \right],
\end{align}
where an implicit summation is done on the index $l$. Exactly as for the analysis of LB scheme, solving this eigenvalue problem gives access to the propagation ($\omega_r$) and dissipation properties ($\omega_i$) of the $V$ modes of the linearized system.

\begin{figure}[ht]
\begin{minipage}{0.4\textwidth}
\hspace{7mm}\includegraphics[scale=0.8]{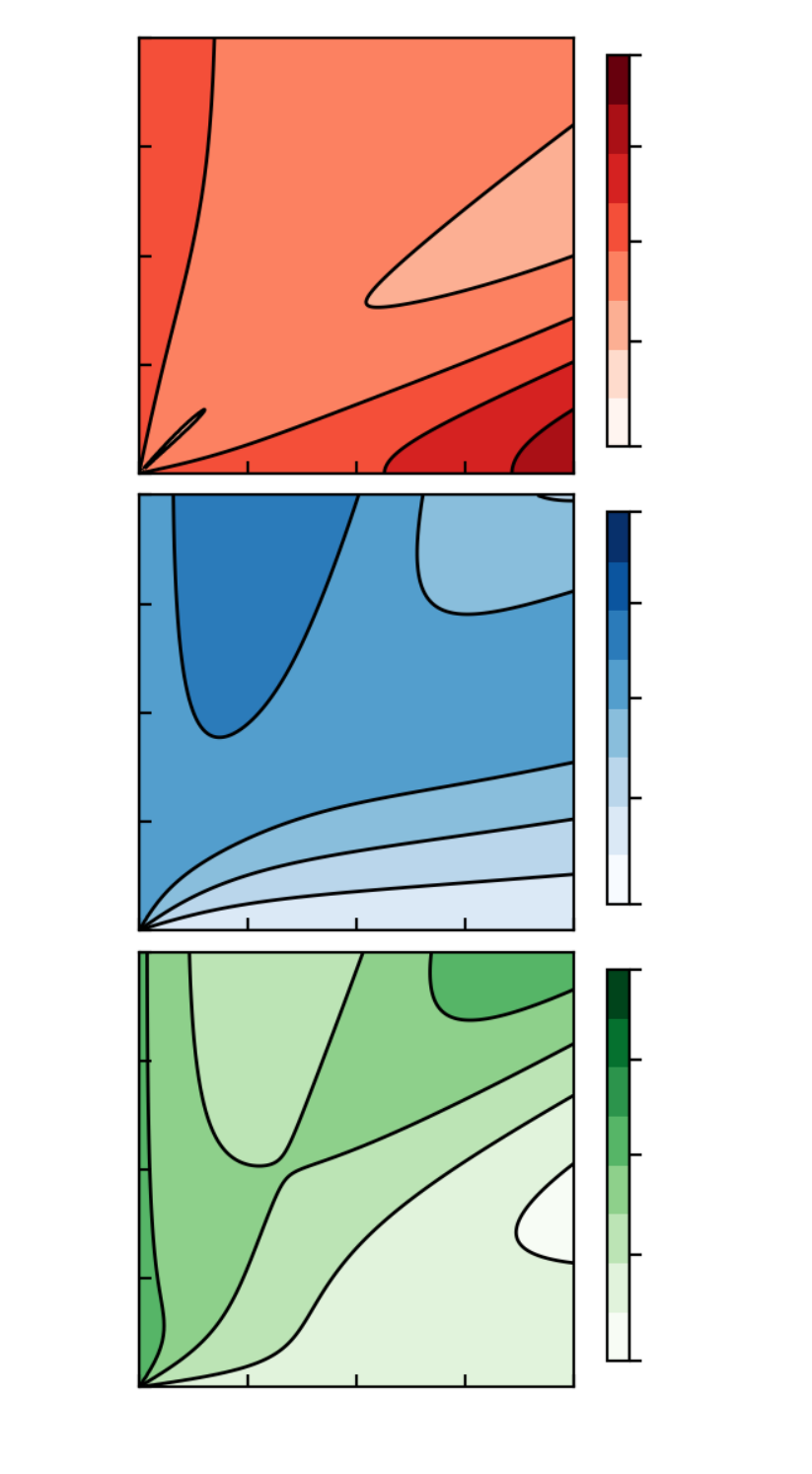}
\put(-113,0){$k_x \Delta x$}
\put(-161,11){$0$}
\put(-141,11){$\pi/4$}
\put(-115,11){$\pi/2$}
\put(-92,11){$3\pi/4$}
\put(-59,11){$\pi$}
\put(-208,67){$k_y \Delta x$}
\put(-166,18){$0$}
\put(-176,43){$\pi/4$}
\put(-176,68){$\pi/2$}
\put(-181,93){$3\pi/4$}
\put(-166,119){$\pi$}
\put(-208,175){$k_y \Delta x$}
\put(-166,125){$0$}
\put(-176,149){$\pi/4$}
\put(-176,175){$\pi/2$}
\put(-181,201){$3\pi/4$}
\put(-166,226){$\pi$}
\put(-208,283){$k_y \Delta x$}
\put(-166,233){$0$}
\put(-176,257){$\pi/4$}
\put(-176,283){$\pi/2$}
\put(-181,308){$3\pi/4$}
\put(-166,334){$\pi$}
\put(-37,25){\scriptsize{$0.9$}}
\put(-37,49){\scriptsize{$0.95$}}
\put(-37,73){\scriptsize{$1$}}
\put(-37,95){\scriptsize{$1.05$}}
\put(-37,116){\scriptsize{$1.1$}}
%\put(-37,101){\scriptsize{$10^{3}$}}
%\put(-37,116){\scriptsize{$10^{4}$}}
\put(-37,132){\scriptsize{$0.9$}}
\put(-37,156){\scriptsize{$0.95$}}
\put(-37,180){\scriptsize{$1$}}
\put(-37,202){\scriptsize{$1.05$}}
\put(-37,223){\scriptsize{$1.1$}}
\put(-37,239){\scriptsize{$0.9$}}
\put(-37,263){\scriptsize{$0.95$}}
\put(-37,286){\scriptsize{$1$}}
\put(-37,309){\scriptsize{$1.05$}}
\put(-37,331){\scriptsize{$1.1$}}
\put(-18,70){$\nu_e/\nu$}
\put(-18,178){$\nu_e/\nu$}
\put(-18,286){$\nu_e/\nu$}
\caption{Spectral maps of the dissipation properties of the PR-DVBE (\textit{cf.} Eq.~(\ref{eq:regularized_DVBE})) on the D2Q9 lattice with $N=4^*$, $\overline{\mathrm{Ma}}=0.2$, $\tau=10^{-5}$. Top: shear, middle: downstream acoustics, bottom: upstream acoustics.\label{fig:LSA_PR_DVBE}}
\end{minipage}
\end{figure}
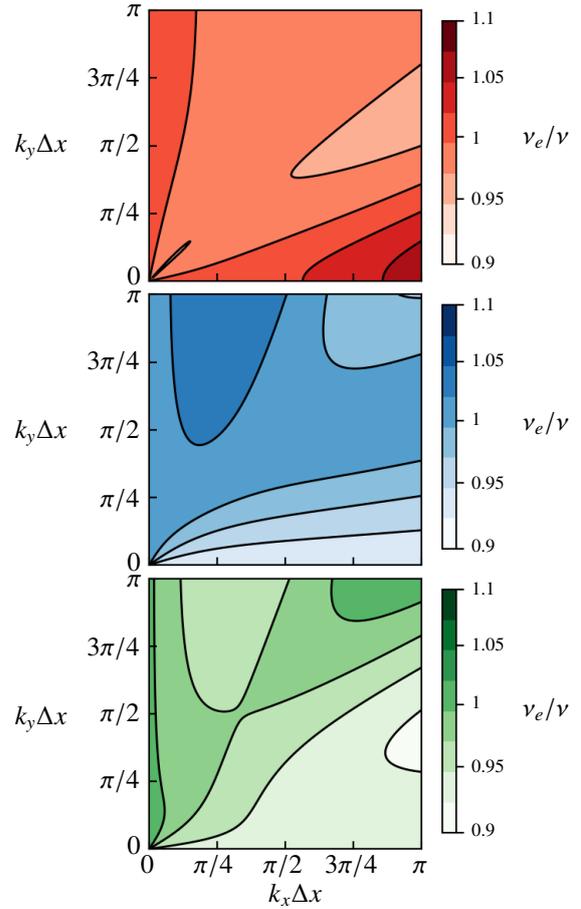

Maps of dissipation rates are displayed on Fig.~\ref{fig:LSA_PR_DVBE} for each of the physical waves that can be identified  thanks to an eigenvector analysis: the shear and the acoustic waves. The parameters set for this analysis are: a D2Q9 lattice with $N=4^*$, a horizontal mean flow at $\overline{\mathrm{Ma}}=0.2$ and a dimensionless relaxation time $\tau/\Delta t=10^{-5}$. Even if a slight deviation of the effective viscosity $(\nu_e = -\omega_i / ||\boldsymbol{k}||^2$) compared to the expected one ($\nu$) can be observed, its order of magnitude remains low, so that hyperviscous effects can reasonably be neglected. This conclusion could have been guessed from a previous work~\cite{PAM2019}, showing that high-order Knudsen effects (referred to as \textit{Prandtl degeneracy}) only have an influence on the NS physics when
\begin{align}
	\frac{\tau}{\tau_N} \gg ||\boldsymbol{k}|| \tau c_s \frac{\Delta x}{\Delta t},
\end{align}
where $\tau_N$ is the relaxation time applied to non collision invariants. With the PR model, $\tau_N = \Delta t/2$, so that this condition becomes
\begin{align}
	||\boldsymbol{k}|| \Delta x \gg 2/c_s \approx 3.5,
\end{align}
which is not the case on Fig.~\ref{fig:LSA_PR_DVBE}, where both $k_x$ and $k_y$ are restricted to $\pi/\Delta x$.

Hence, the over-dissipation previously observed with any regularized model, as well as the instability occurrence, cannot reasonably be attributed, at least for the PR model, to any hyperviscous effects occuring on the DVBE from which they \textit{a priori} derive. It is therefore necessary to focus on the numerical error induced by time and space discretizations of the regularized scheme.

\subsubsection{Numerical error of the PR scheme}

Since the hydrodynamic behavior of the PR-DVBE seems to be in good agreement with the NS equations, the dissipation issue encountered with the discrete scheme can only originate from a numerical error induced by the time and space discretization of the DVBE (Eq.~(\ref{eq:regularized_DVBE})). To this extent, let us focus on the \textit{a priori} derivation of the PR scheme of \ref{app:apriori_derivation_PR}. Exactly as in the case of the BGK collision model, a $O(\Delta t^3)$ error appears due to the second-order precision of the trapezium rule. However, in the case of the PR scheme, this term is multiplied by the matrix $\mathbf{R}$ containing terms of order $O(1/\Delta t)$, as shown in Eq.~(\ref{eq:matrix_R_D2Q9}). Even though the three first components of this matrix, related to collision invariants, have rigorously no influence on the numerical scheme (in absence of body-force term), the three last ones directly impact the time evolution of third- and fourth-order moments. For this reason, the numerical error induced by the time and space discretization of the PR scheme appears to be one order lower, in $\Delta t$, than that of the BGK scheme. This is a direct consequence of equilibrating high-order Hermite moments with this model.

However, numerous past investigations of regularized collision models have evidenced an order of convergence close to that of the BGK one, \textit{i.e.} second-order accuracy in space and time~\cite{Latt2007, Montessori2014a, Montessori2014b, Malaspinas2015, Mattila2015, Li2019}. This may appear in disagreement with the \textit{a priori} derivation of \ref{app:apriori_derivation_PR}, indicating a first-order accuracy for the PR scheme. It therefore seems that the corresponding error has no pragmatic effect on the order of convergence on the macroscopic equations. A possible explanation for this observation is that the numerical error does not directly affect mass and momentum equations, but the evolution of higher-order moments only. More in-depth investigations of the convergence error of the numerical scheme and that of the corresponding macroscopic equations will be the purpose of future work.

In any case, the spatio-temporal discretization of the regularized schemes turns out to be essentially different from that of the BGK one. This observation may account for numerical instabilities of different nature, namely
\begin{itemize}
    \item modal interactions (eigenvalue collisions) for the BGK model,
    \item an unexpected amplification of isolated modes for regularized models.
\end{itemize}

If such a link can be drawn between the time and space discretization and numerical instabilities, this phenomenon has to be expected on any model setting high-order relaxation rates to a pre-defined value. In particular, MRT models are likely to share similar numerical properties. It is all the more conceivable since the PR scheme is nothing more than a particular MRT model based on Hermite moments, while the RR$4^*$ model with $N=4^*$ on the D2Q9 lattice has recently been shown equivalent to a centered Hermite moment MRT model~\cite{Coreixas2019}. In order to shed some light on these points, the methodology employed in the present work is being considered for the analyses of various collision models. In addition, the possible link between time and space errors of the numerical schemes and potential dissipation issues will be the purpose of future work.

\section{Conclusion}

Linear stability analyses of several regularized collision models have been rigorously performed for the athermal lattice Boltzmann method. 
Two kinds of regularizations have been particularly investigated: the so-called \textit{projected regularization} (PR), which %is the former model by Ladd \& Verberg~\cite{Ladd2001} and Latt \& Chopard~\cite{Latt2006}, and
was successively used by Skordos~\cite{SKORDOS_PRE_49_1993}, Ladd \& Verberg~\cite{Ladd2001} and Latt \& Chopard~\cite{Latt2006}, and
the \textit{recursive regularization} (RR), where off-equilibrium distribution functions are computed by recurrence~\cite{Malaspinas2015,BROGI_JASA_142_2017, Coreixas2017}. Following a previous work performed on the BGK collision model, a particular attention has been paid to the eigenvectors of the linear analyses, which bear important information on the macroscopic content of a given LB mode. In particular, such a study has made it possible to systematically identify the modes carrying a physical information (acoustics or shear) and investigate their isotropy properties in two dimensions. Moreover, numerical cases of plane monochromatic waves have confirmed the results obtained by linear stability analyses. 

All in all, a rigorous methodology, that can be applied to any LB model, has been introduced. It provides objective comparisons of collision models and allows drawing fair conclusions on their stability in the linear regime. Since it seems that most phenomena have a numerical origin, their stability property are hardly predictable in an \textit{a priori} manner. In this context, the main interest of such a work is to provide hints in the hope of clearly identifying the origins of the numerical properties of each model, so as to help build more robust LB schemes.

Regarding regularized models, the results strongly warn on the importance of investigating numerical properties in every direction, since a large anisotropy has been highlighted. This is especially true for the dissipation of acoustic waves, which have a very good behavior in the mesh-aligned directions, but are severely attenuated, or even amplified, in diagonal ones. Two major conclusions shared by all the regularized models investigated in the present article have been highlighted: (1) a mode filtering property and (2) an incorrect dissipation rate (over-dissipation or amplification) of the physical waves travelling in some directions. The first property has been explained by a reduction of the rank of the system, as a result of the pre-collision regularization procedure. 
%This mode filtering is even more efficient with the FR regularized models where only modes carrying a physical wave remain (three modes in two-dimensional athermal LBM). 
This mode filtering helps reducing the occurrence of \textit{eigenvalue collisions}, that are responsible for the strong instability issues of the BGK collision model, and therefore contributes in an enhanced stability. For instance, with both the D2Q9 and D2V17 lattices, the most stable models (in term of maximal achievable Mach number) turn out to be a form of the recursive regularization. Unfortunately, the second property led to unphysical dissipation rates, in some cases of several orders of magnitude larger than the expected kinematic viscosity, in other cases responsible for linear instabilities. By discussing on the way the PR scheme can be \textit{a priori} obtained from its DVBE counterpart, it seems that a numerical error, one order lower (in time) than with the BGK model, may be at the origin of a particular source of instability. Indeed, while the numerical instabilities of the BGK model are caused by modal interactions only, the amplification of isolated modes has been observed with regularized models. It is likely that the fact of setting high-order relaxation rates to pre-defined values is at the origin of such a phenomenon. 

  %It could be attributed, at least for the PR model, to a first-order accuracy of the numerical scheme, which could be shown in two different ways. %Regarding the RR and FR models, such an error could not be theoretically exhibited. A better understanding of their dissipative properties, which are likely to originate from a time and space discretization error, could be the purpose of future work.  
%: by performing an \textit{a priori} derivation of the PR algorithm from a continuous system of equations, thanks to a Taylor expansion of the numerical scheme and by computing errors in real simulations of shear waves. 

Discussing on the numerical error of regularized schemes has in fact raised as many questions as it answered, especially regarding their order of convergence. In order to shed the light on this particular point, a study on the link between the numerical errors of LB schemes and potential dissipation issues is ongoing. Furthermore, as a perspective for future work, similar analyses can be performed on three-dimensional lattices for practical purposes. One would then expect a lower robustness due to the larger number of remaining modes, even after the regularization filter. It would also be interesting, if not essential, to reproduce such analyses to other LB models in order to find out how they achieve a potential stability gain. In this context, an investigation of the hybrid recursive regularized (HRR) collision model~\cite{Jacob2018} is being under study, both for athermal and compressible flows. More generally, other time and space discretizations of the discrete velocity Boltzmann equation may also be investigated, \textit{e.g.} fractional propagation schemes~\cite{Qian1997b, Fan2006a}. %, as well as, obviously, other collision models: MRT~\cite{DHumieres1994, Lallemand2000, DHumieres2002}, TRT~\cite{Ginzburg2008, Ginzburg2010}, cascaded models~\cite{Geier2006}, cumulant ones~\cite{Geier2015}, entropic LBM~\cite{Karlin1998, Boghosian2001, Ansumali2003, Karlin2014}, \textit{etc.} Especially regarding the MRT family, 
Obviously, the behavior of other collision models (either based on a particular moment space~\cite{DHumieres1994, Lallemand2000, DHumieres2002,Ginzburg2008, Ginzburg2010,Geier2006,Geier2015,LI_PRE_100_2019} or relying on an entropy maximization principle~\cite{Karlin1998, Boghosian2001, Ansumali2003, Karlin2014,ATIF_PRL_119_2017}) should also be investigated. For all the above collision models, when the relaxation rate of a given moment is such that it is imposed at its equilibrium value during the collision process, it is expected that similar properties as that of the regularized collision models would be recovered.% This idea seems to be supported by the recent findings, e.g., the anisotropic numerical diffusion of the cumulant approach~\cite{GEHRKE_CF_156_2017}. \newline

\section*{Acknowledgments}
The authors want to thank T. Astoul, F. Renard and P.-A. Masset for fruitful conversations regarding linear stability of regularized scheme and their accuracy order. This work has been partially funded by the DGAC project OMEGA3/ALBATROS. 

\appendix
\onecolumn

\section{Lattices}
\label{app:lattices}

D2Q9 and D2V17 lattices, used in the present article, are described below. For a sake of clarity, all velocities obtained by cyclic permutations with respect to each Cartesian axis are omitted, the number of velocities belonging to a same group (=same velocity norm) being denoted by $p$. The parameters of the D2V17 lattice are taken from~\cite{Siebert2008a}. 

\begin{center}
\begin{tabular}{ccccccc}
\hline
\hline
Lattice & Quadrature order & $\boldsymbol{e_i}$ & Group & $p$ & $w_i$ & $c_s$ \\ 
\hline
     &             & $(0, 0)$ & 1 & 1 & $4/9$ & \\
D2Q9 & $Q=5$ & $(1, 0)$ & 2 & 4 & $1/9$ & $1/\sqrt{3}$ \\
     &             & $(1, 1)$ & 3 & 4 & $1/36$ & \\
\hline
      &                & $(0, 0)$ & 1 & 1 & $(575+193\sqrt{193})/8100$   & \\
      &                & $(1, 0)$ & 2 & 4 & $(3355-91\sqrt{193})18000$  &  \\
D2V17 &   $Q=7$        & $(1, 1)$ & 3 & 4 & $(655+17\sqrt{193})/27000$ & $\sqrt{5(25+\sqrt{193})/72}$\\
      &                & $(2, 2)$ & 6 & 4 & $(685-49\sqrt{193})/54000$ & \\
      &                & $(3, 0)$ & 7 & 4 & $(1445-101\sqrt{193})/162000$ & \\
\hline
\hline
\end{tabular}
\end{center}

\section{Matrices for the linear systems}
\label{app:linear_matrices}

This appendix aims at detailing the derivation of matrices presented in the linear system of Sec.~\ref{sec:LSA},
\begin{align}
\label{eq:app_eigenvalue_problem}
    e^{-\mathrm{i} \omega} \mathbf{F} = \mathbf{M}\mathbf{F},
\end{align}
for the BGK and the regularized models introduced in Sec.~\ref{sec:LBM}. Whatever the collision model, it starts by linearizing the collide \& stream scheme as
\begin{align}
    f'_i(\boldsymbol{x}+\boldsymbol{e_i}, t+1) = \left. \frac{\partial f^*_i}{\partial f_j}\right|_{f_j=\overline{f_j}} f'_j,
\end{align}
where $f^*_i$ are post collision populations, whose expression depends on the adopted collision model. After injecting the monochromatic plane wave of Eq.~(\ref{eq:fluctuations_monochromatic_waves}), one has
\begin{align}
\label{eq:app_linearized_eq}
    e^{\mathrm{i} \boldsymbol{k} \cdot \boldsymbol{e_i}} e^{-\mathrm{i} \omega} \widehat{f_i} = \left. \frac{\partial f^*_i}{\partial f_j}\right|_{f_j=\overline{f_j}} \widehat{f_j}.
\end{align}

\subsection{BGK collision model}

With the BGK collision model, post-collision populations are computed as
\begin{align}
    f_i^*(\boldsymbol{x}, t) = f_i(\boldsymbol{x}, t) - \frac{1}{\overline{\tau}} \left( f_i(\boldsymbol{x}, t) - f_i^{eq, N}(\boldsymbol{x}, t) \right).
\end{align}
Hence,
\begin{align}
\label{eq:app_Jcoll_BGK}
    \left. \frac{\partial f^*_i}{\partial f_j}\right|_{f_j=\overline{f_j}} = \delta_{ij} - \frac{1}{\overline{\tau}} \left( \delta_{ij} - J^{eq, N}_{ij} \right),
\end{align} 
where $\mathbf{J}^{eq,N}$, the Jacobian matrix of the equilibrium distribution functions, can be computed as
\begin{align}
    J^{eq, N}_{ij} = \left. \frac{\partial f^{eq,N}_i}{\partial f_j}\right|_{f_j=\overline{f_j}}.
\end{align}
The equilibrium distribution functions of interest in the present article can be written in the general form
\begin{align}
    f_i^{eq,N} = w_i \sum_{n=0}^N \frac{1}{n! c_s^{2n}} \boldsymbol{a}_{eq}^{(n)}:\boldsymbol{\mathcal{H}}_i^{(n)}.
\end{align}
In the above expression, coefficients $\boldsymbol{a}_{eq}^{(n)}$ only are implicit function of all populations $(f_j)$ through macroscopic variables $(\rho,\boldsymbol{u})$. For this reason, one has
\begin{align}
    J_{i,j}^{eq,N} = w_i \sum_{n=0}^N \frac{1}{n! c_s^{2n}} \boldsymbol{\Lambda}_{eq,j}^{(n)}:\boldsymbol{\mathcal{H}}_i^{(n)},
\end{align}
with
\begin{align}
    \left(\mathbf{\Lambda}_{eq, j}^{(n)}\right)_{\alpha_1..\alpha_n} = \left. \frac{\partial a_{eq, \alpha_1..\alpha_n}^{(n)}}{\partial f_j} \right|_{f_j = \overline{f_j}} = \left. \frac{\partial a_{eq, \alpha_1..\alpha_n}^{(n)}}{\partial \rho}\right|_{(\overline{\rho}, \overline{\rho} \overline{\boldsymbol{u}})} \frac{\partial \rho}{\partial f_j} + \left. \frac{\partial a_{eq, \alpha_1..\alpha_n}^{(n)}}{\partial (\rho u_\beta)}\right|_{(\overline{\rho}, \overline{\rho} \overline{\boldsymbol{u}})} \frac{\partial (\rho u_\beta)}{\partial f_j},
\end{align}
where an implicit summation is done over the index $\beta$ . It should be noted that partial derivatives of $\boldsymbol{a}_{eq}^{(n)}$ over $\rho$ are done at $(\rho \boldsymbol{u})$ constant rather than at $\boldsymbol{u}$ constant. Knowing the definitions of $\rho = \sum{f_i}$ and $\rho \boldsymbol{u}=\sum{\boldsymbol{e_i}f_i}$, one has 
\begin{align}
\label{eq:app_linearized_macros}
    \frac{\partial \rho}{\partial f_j} = 1, \qquad \frac{\partial (\rho \boldsymbol{u})}{\partial f_j} = \boldsymbol{e_j}.
\end{align}
Moreover, equilibrium coefficients $\boldsymbol{a}_{eq}^{(n)}$ can be expressed as
\begin{align}
    {a}_{eq, \alpha_1...\alpha_n}^{(n)} = \rho u_{\alpha_1}..u_{\alpha_n} = \frac{j_{\alpha_1}..j_{\alpha_n}}{\rho^{n-1}},
\end{align}
after denoting $\boldsymbol{j}= \rho \boldsymbol{u}$, so that
\begin{align}
    \left. \frac{\partial a_{eq, \alpha_1..\alpha_n}^{(n)}}{\partial \rho}\right|_{(\overline{\rho}, \overline{\rho} \overline{\boldsymbol{u}})} = -\frac{(n-1) \overline{j_{\alpha_1}}..\overline{j_{\alpha_n}}}{\overline{\rho}^n}=-(n-1) \overline{u_{\alpha_1}}.. \overline{u_{\alpha_n}}, \\
    \left. \frac{\partial a_{eq, \alpha_1..\alpha_n}^{(n)}}{\partial j_\beta}\right|_{(\overline{\rho}, \overline{j})} = \frac{1}{\overline{\rho}^{n-1}} \sum_{i=1}^n \overline{j_{\alpha_1}}..\overline{j_{\alpha_{i-1}}}\overline{j_{\alpha_{i+1}}}..\overline{j_{\alpha_n}} \delta_{\alpha_i \beta} = \sum_{i=1}^n \overline{u_{\alpha_1}}..\overline{u_{\alpha_{i-1}}}\overline{u_{\alpha_{i+1}}}..\overline{u_{\alpha_n}} \delta_{\alpha_i \beta}.
\end{align}
Note that these expressions do not involve the mean density $\overline{\rho}$ any more. Hence, the equilibrium Jacobian matrix $\mathbf{J}^{eq,N}$ can be computed with
\begin{align}
    \left(\mathbf{\Lambda}_{eq, j}^{(n)}\right)_{\alpha_1..\alpha_n} = -(n-1)\overline{u_{\alpha_1}}.. \overline{u_{\alpha_n}} + \sum_{i=1}^n \overline{u_{\alpha_1}}..\overline{u_{\alpha_{i-1}}}\overline{u_{\alpha_{i+1}}}..\overline{u_{\alpha_n}} e_{j, \alpha_i}.
\end{align}
For instance, some of these coefficients are provided below:
\begin{align}
    &\Lambda_{eq,j}^{(0)} = 1, \qquad \left( \boldsymbol{\Lambda}_{eq,j}^{(1)} \right)_\alpha = e_{j,\alpha}, \\
    &\left( \boldsymbol{\Lambda}_{eq,j}^{(2)} \right)_{xx} = -\overline{u_x}^2 + 2 \overline{u_x} e_{j,x}, \qquad \left( \boldsymbol{\Lambda}_{eq,j}^{(2)} \right)_{xy} = -\overline{u_x} \overline{u_y} + \overline{u_x}e_{j,y} + \overline{u_y}e_{j,x}, \qquad \left( \boldsymbol{\Lambda}_{eq,j}^{(2)} \right)_{yy} = -\overline{u_y}^2 + 2 \overline{u_y} e_{j,y}, \\
    & \left( \boldsymbol{\Lambda}_{eq,j}^{(3)} \right)_{xxx} = -2\overline{u_x}^3+3\overline{u_x}^2e_{j,x}, \qquad \left( \boldsymbol{\Lambda}_{eq,j}^{(3)} \right)_{xxy} = -2\overline{u_x}^2\overline{u_y} + \overline{u_x}^2 e_{j, y} + 2\overline{u_x} \overline{u_y} e_{j,x}, \\
    & \left( \boldsymbol{\Lambda}_{eq,j}^{(3)} \right)_{xyy} = -2\overline{u_x}\overline{u_y}^2 + \overline{u_y}^2 e_{j, x} + 2\overline{u_x} \overline{u_y} e_{j,y}, \qquad \left( \boldsymbol{\Lambda}_{eq,j}^{(3)} \right)_{yyy} = -2\overline{u_y}^3+3\overline{u_y}^2e_{j,y}, \\
    & \left( \boldsymbol{\Lambda}_{eq,j}^{(4)} \right)_{xxyy} = -3\overline{u_x}^2 \overline{u_y}^2 + 2\overline{u_x}^2 \overline{u_y} e_{j,y} + 2 \overline{u_x} \overline{u_y}^2 e_{j,x}.
\end{align}
Finally, injecting Eq.~(\ref{eq:app_Jcoll_BGK}) in Eq.~(\ref{eq:app_linearized_eq}) leads to the eigenvalue problem of Eq.~(\ref{eq:app_eigenvalue_problem}) with
\begin{align}
    M_{ij} = e^{-\mathrm{i} \boldsymbol{k} \cdot \boldsymbol{e_i}} \left[ \delta_{ij} - \frac{1}{\overline{\tau}} \left(\delta_{ij} - J^{eq,N}_{ij} \right) \right].
\end{align}

\subsection{Regularization by projection (PR)}

In the PR approach, post-collision populations can be re-written as
\begin{align}
    f_i^{*, \mathrm{PR}} = f_i^{eq,N} + \left( 1-\frac{1}{\overline{\tau}} \right) \left( f_k - f^{eq,N}_k \right) h_{ik},
\end{align}
where an implicit summation is done over the index $k$ and with
\begin{align}
    h_{ik} = \frac{w_i}{2c_s^4} \, \boldsymbol{\mathcal{H}}_i^{(2)}:\boldsymbol{\mathcal{H}}_k^{(2)}.
\end{align}
Thus, the Jacobian matrix of post-collision populations can be computed as
\begin{align}
    \left. \frac{\partial f^{*,\mathrm{PR}}_i}{\partial f_j}\right|_{f_j=\overline{f_j}} = J^{eq,N}_{ij} + \left( 1-\frac{1}{\overline{\tau}} \right)  \left( \delta_{kj} - J^{eq,N}_{kj} \right) h_{ik}.
\end{align}
The eigenvalue problem of Eq.~(\ref{eq:app_eigenvalue_problem}) can therefore be obtained with
\begin{align}
    M_{ij}^{\mathrm{PR}} = e^{-\mathrm{i} \boldsymbol{k} \cdot \boldsymbol{e_i}} \bigg[ J^{eq,N}_{ij} + \left( 1-\frac{1}{\overline{\tau}} \right)  \left( \delta_{kj} - J^{eq,N}_{kj} \right) h_{ik} \bigg].
\end{align}

\subsection{Recursive regularization (RR$N_r$)}

With the recursive regularization, the post-collision populations can be written as
\begin{align}
    f_i^{*, \mathrm{RR}N_r} = f_i^{*, \mathrm{PR}} + \left( 1-\frac{1}{\overline{\tau}} \right) \, \sum_{n=3}^{N_r} \frac{w_i}{n! c_s^{2n}}\, \boldsymbol{a}_1^{(n)}:\boldsymbol{\mathcal{H}}_i^{(n)}.
\end{align}
In the second right-hand-side term of the above relation, only coefficients $\boldsymbol{a}_1^{(n)}$ are (implicit) functions of the discrete populations $(f_j)$, through the recursive formula of Eq.~(\ref{eq:regul_recursive_formula}). A linearization of these terms therefore involves the derivative $n^\mathrm{th}$-order tensors $\mathbf{\Lambda}_{1, j}^{(n)}$ defined as 
\begin{align}
    \left(\mathbf{\Lambda}_{1, j}^{(n)}\right)_{\alpha_1..\alpha_n} = \left. \frac{\partial a_{1, \alpha_1..\alpha_n}^{(n)}}{\partial f_j} \right|_{f_j = \overline{f_j}}.
\end{align}
Off-equilibrium coefficients of interest in this article can be obtained as follows with the recursive relation of Eq.~(\ref{eq:regul_recursive_formula}):
\begin{align} 
    & a_{1,xxx}^{(3)} = 3 u_x a_{1,xx}^{(2)}, \qquad a_{1,xxy}^{(2)} = 2u_x a_{1,xy}^{(2)} + u_y a_{1,xx}^{(2)}, \qquad a_{1,xyy}^{(3)} = 2u_y a_{1,xy}^{(2)} + u_x a_{1,yy}^{(2)}, \qquad a_{1, yyy}^{(3)} = 3 u_y a_{1,yy}^{(2)}, \\
    & \qquad \qquad \qquad \qquad \qquad \qquad \qquad a_{1,xxyy}^{(4)} = u_y^2 a_{1,xx}^{(2)} + 4 u_x u_y a_{1,xy}^{(2)} + u_x^2 a_{1,yy}^{(2)},
\end{align}
where
\begin{align}
    a_{1,xx}^{(2)} = \sum_k (f_k - f_k^{eq, N}) \mathcal{H}_{k,xx}^{(2)}, \qquad a_{1,xy}^{(2)} = \sum_k (f_k - f_k^{eq, N}) \mathcal{H}_{k,xy}^{(2)},\qquad a_{1,yy}^{(2)} = \sum_k (f_k - f_k^{eq, N}) \mathcal{H}_{k,yy}^{(2)}.
\end{align}
Eventually linearizing these coefficients yields
\begin{align}
    \left(\mathbf{\Lambda}_{1, j}^{(3)}\right)_{xxx} &= 3 \overline{u_x} \left( \mathcal{H}_{j,xx}^{(2)} - \sum_k J_{kj}^{eq,N} \mathcal{H}_{k,xx}^{(2)} \right),\\% \sum_k (\delta_{kj}-J_{kj}^{eq,N}) \mathcal{H}_{k,xx}^{(2)}, \\
    \left(\mathbf{\Lambda}_{1, j}^{(3)}\right)_{xxy} &= 2\overline{u_x} \left( \mathcal{H}_{j,xy}^{(2)} - \sum_k J_{kj}^{eq,N} \mathcal{H}_{k,xy}^{(2)} \right) + \overline{u_y} \left( \mathcal{H}_{j,xx}^{(2)} - \sum_k J_{kj}^{eq,N} \mathcal{H}_{k,xx}^{(2)} \right), \\ %2 \overline{u_x} \sum_k (\delta_{kj}-J_{kj}^{eq,N}) \mathcal{H}_{k,xy}^{(2)} + \overline{u_y} \sum_k (\delta_{kj}-J_{kj}^{eq,N}) \mathcal{H}_{k,xx}^{(2)},\\
    \left(\mathbf{\Lambda}_{1, j}^{(3)}\right)_{xyy} &= 2\overline{u_y} \left( \mathcal{H}_{j,xy}^{(2)} - \sum_k J_{kj}^{eq,N} \mathcal{H}_{k,xy}^{(2)} \right) + \overline{u_x} \left( \mathcal{H}_{j,yy}^{(2)} - \sum_k J_{kj}^{eq,N} \mathcal{H}_{k,yy}^{(2)} \right), \\ %2 \overline{u_y} \sum_k (\delta_{kj}-J_{kj}^{eq,N}) \mathcal{H}_{k,xy}^{(2)} + \overline{u_x} \sum_k (\delta_{kj}-J_{kj}^{eq,N}) \mathcal{H}_{k,yy}^{(2)}, \\
    \left(\mathbf{\Lambda}_{1, j}^{(3)}\right)_{yyy} &= 3 \overline{u_y} \left( \mathcal{H}_{j,yy}^{(2)} - \sum_k J_{kj}^{eq,N} \mathcal{H}_{k,yy}^{(2)} \right),\\% 3 \overline{u_y} \sum_k (\delta_{kj}-J_{kj}^{eq,N}) \mathcal{H}_{k,yy}^{(2)}, \\
    \left(\mathbf{\Lambda}_{1, j}^{(4)}\right)_{xxyy} &= \overline{u_y}^2 \left( \mathcal{H}_{j,xx}^{(2)} - \sum_k J_{kj}^{eq,N} \mathcal{H}_{k,xx}^{(2)} \right) + 4 \overline{u_x}\overline{u_y} \left( \mathcal{H}_{j,xy}^{(2)} - \sum_k J_{kj}^{eq,N} \mathcal{H}_{k,xy}^{(2)} \right) + \overline{u_x}^2 \left( \mathcal{H}_{j,yy}^{(2)} - \sum_k J_{kj}^{eq,N} \mathcal{H}_{k,yy}^{(2)} \right).
\end{align}
Finally, the eigenvalue problem of Eq.~(\ref{eq:app_eigenvalue_problem}) can be recovered with
\begin{align}
    M_{ij}^{\mathrm{RR}N_r} = M_{ij}^{\mathrm{PR}} + e^{-\mathrm{i} \boldsymbol{k} \cdot \boldsymbol{e_i}} \left( 1-\frac{1}{\overline{\tau}} \right) \sum_{n=3}^{N_r} \frac{w_i}{n! c_s^{2n}}\, \mathbf{\Lambda}_{1, j}^{(n)}:\boldsymbol{\mathcal{H}}_i^{(n)}.
\end{align}

\subsection{Analytical regularization (AR)}

In a so-called ``analytical regularization'', as described in this article, post-collision populations are reconstructed as
\begin{align}
    f_i^{*, \mathrm{AR}} = f_i^{eq, N} + \left( 1 - \frac{1}{\overline{\tau}} \right) \frac{w_i}{2c_s^{4}}\, \boldsymbol{a}_1^{(2), \mathrm{AR}}:\boldsymbol{\mathcal{H}}_i^{(2)}, 
\end{align}
where $\boldsymbol{a}_1^{(2), \mathrm{AR}}$ is the analytically computed coefficient obtained with a Chapman-Enskog expansion, whose components are given by
\begin{align}
    a_{1, \alpha \beta}^{(2), \mathrm{AR}} = -\overline{\tau} \rho c_s^2 \left( \frac{\partial u_\alpha}{\partial x_\beta} + \frac{\partial u_\beta}{\partial x_\alpha} \right), %\bigg( & \frac{u_\alpha(\boldsymbol{x} + \boldsymbol{e_\beta} )-u_\alpha(\boldsymbol{x}-\boldsymbol{e_\beta})}{2} + \frac{u_\beta(\boldsymbol{x} + \boldsymbol{e_\alpha} )-u_\beta(\boldsymbol{x}-\boldsymbol{e_\alpha})}{2}  \bigg),
\end{align}
where $\alpha, \beta \in \{x, y\}$. As previously, this function is an implicit function of all populations $(f_j)$ through macroscopic quantities $\rho$ and $\boldsymbol{u}$. It can be linearized as
\begin{align}
    a_{1, \alpha \beta}^{(2), \mathrm{AR}} (f_j) = a_{1, \alpha \beta}^{(2), \mathrm{AR}}(\overline{f_j}) + \underbrace{\left. \frac{\partial a_{1, \alpha \beta}^{(2), \mathrm{AR}}}{\partial f_j} \right|_{f_j = \overline{f_j}}}_{\left(\mathbf{\Lambda}_{1, j}^{(2), \mathrm{AR}}\right)_{\alpha \beta}} f'_j + O({f'_j}^2).
\end{align}
Note that the first right-hand-side term vanishes since the gradient of the mean flow is null by definition. Computing the first-order fluctuations leads to
\begin{align}
\label{eq:app_FR_a12_detailed}
    a_{1, \alpha \beta}^{(2), \mathrm{AR}} (f_j) = - \overline{\tau} c_s^2 \overline{\rho} \left( \left.\frac{\partial u_\alpha}{\partial f_j}\right|_{f_j=\overline{f_j}} \cdot \frac{\partial f'_j}{\partial x_\beta} + \left. \frac{\partial u_\beta}{\partial f_j}\right|_{f_j=\overline{f_j}} \cdot \frac{\partial f'_j}{\partial x_\alpha} \right) + O({f'_j}^2).
\end{align}
On the one hand, by denoting $\boldsymbol{j} = \rho \boldsymbol{u}$, one has
\begin{align}
   \left. \frac{\partial u_\alpha}{\partial f_j}\right|_{f_j=\overline{f_j}} = \left.\frac{\partial (j_\alpha/\rho)}{\partial f_j}\right|_{f_j=\overline{f_j}} = \frac{1}{\overline{\rho}} \left. \frac{\partial j_\alpha}{\partial f_j}\right|_{f_j=\overline{f_j}} - \frac{\overline{j_\alpha}}{\overline{\rho}^2} \left. \frac{\partial \rho}{\partial f_j}\right|_{f_j=\overline{f_j}} = \frac{e_{j,\alpha} - \overline{u_\alpha}}{\overline{\rho}},
\end{align}
where Eq.~(\ref{eq:app_linearized_macros}) has been used to establish the last equality. On the other hand, injecting the fluctuation forms of Eq.~(\ref{eq:fluctuations_monochromatic_waves}) into Eq.~(\ref{eq:app_FR_a12_detailed}) yields
\begin{align}
	\frac{\partial f'_j}{\partial x_\alpha} = \mathrm{i} k_\alpha f'_j, \qquad \frac{\partial f'_j}{\partial x_\beta} = \mathrm{i} k_\beta f'_j. \\
    %f'_j(\boldsymbol{x} + \boldsymbol{e_\alpha}) - f'_j(\boldsymbol{x}-\boldsymbol{e_\alpha}) = \widehat{f_j} \exp(i(\boldsymbol{k} \cdot \boldsymbol{x}-\omega t)) \left( e^{ik_\alpha} - e^{-ik_\alpha} \right) = -2i \sin(k_\alpha)f'_j.
\end{align}
This leads to
\begin{align}
    \left(\mathbf{\Lambda}_{1, j}^{(2), \mathrm{AR}}\right)_{\alpha \beta} = -\mathrm{i} \overline{\tau} c_s^2 \left( (e_{j_\alpha} - \overline{u_\alpha}) k_\beta + (e_{j_\beta} - \overline{u_\beta}) k_\alpha \right).
\end{align}
%Regarding higher-order coefficients $\left( \boldsymbol{a}_1^{(n), \mathrm{FR}} \right)_{n \geq 3}$, their linearization involves the derivative tensors $\boldsymbol{\Lambda}_{1,j}^{(n), \mathrm{FR}}$ defined as
%\begin{align}
%        \left(\mathbf{\Lambda}_{1, j}^{(n), \mathrm{FR}}\right)_{\alpha_1..\alpha_n} = \left. \frac{\partial a_{1, \alpha_1..\alpha_n}^{(n), \mathrm{FR}}}{\partial f_j} \right|_{f_j = \overline{f_j}}.
%\end{align}
%Using the recursive relations of Eq.~(\ref{eq:regul_recursive_formula}), and knowing that $\boldsymbol{a}_{1}^{(2), \mathrm{FR}} (\overline{f_j}) = 0$, one has
%\begin{align}
%    \left( \Lambda_{1,j}^{(3), \mathrm{FR}} \right)_{xxx} = 3 \overline{u_x} \left( \Lambda_{1,j}^{(2), \mathrm{FR}} \right)_{xx}, \qquad \left( \Lambda_{1,j}^{(3), \mathrm{FR}} \right)_{xxy} = 2 \overline{u_x} \left( \Lambda_{1,j}^{(2), \mathrm{FR}} \right)_{xy} + \overline{u_y} \left( \Lambda_{1,j}^{(2), \mathrm{FR}} \right)_{xx}, \\
%    \left( \Lambda_{1,j}^{(3), \mathrm{FR}} \right)_{xyy} = 2 \overline{u_y} \left( \Lambda_{1,j}^{(2), \mathrm{FR}} \right)_{xy} + \overline{u_x} \left( \Lambda_{1,j}^{(2), \mathrm{FR}} \right)_{yy}, \qquad \left( \Lambda_{1,j}^{(3), \mathrm{FR}} \right)_{yyy} = 3 \overline{u_y} \left( \Lambda_{1,j}^{(2), \mathrm{FR}} \right)_{yy}, \\
%    \left( \Lambda_{1,j}^{(4), \mathrm{FR}} \right)_{xxyy} = \overline{u_y}^2 \left( \Lambda_{1,j}^{(2), \mathrm{FR}} \right)_{xx} + 4 \overline{u_x} \overline{u_y} \left( \Lambda_{1,j}^{(2), \mathrm{FR}} \right)_{xy} + \overline{u_x}^2 \left( \Lambda_{1,j}^{(2), \mathrm{FR}} \right)_{yy}.
%\end{align}
Finally, one has 
\begin{align}
    \left. \frac{\partial f_i^{*, \mathrm{AR}}}{\partial f_j} \right|_{f_j = \overline{f_j}} = J_{ij}^{eq,N} + \left( 1-\frac{1}{\overline{\tau}} \right) \frac{w_i}{2c_s^{4}}\, \mathbf{\Lambda}_{1, j}^{(2), \mathrm{AR}}:\boldsymbol{\mathcal{H}}_i^{(2)},
\end{align}
so that the eigenvalue problem of Eq.~(\ref{eq:app_eigenvalue_problem}) is recovered with
\begin{align}
    M_{ij}^{\mathrm{AR}} = e^{\mathrm{i} \boldsymbol{k} \cdot \boldsymbol{e_i}} \left[ J_{ij}^{eq,N} + \left( 1-\frac{1}{\overline{\tau}} \right) \frac{w_i}{2c_s^{4}}\, \mathbf{\Lambda}_{1, j}^{(2), \mathrm{AR}}:\boldsymbol{\mathcal{H}}_i^{(2)} \right].
\end{align}

\section{Convergence study in LSA of the D2Q9-BGK scheme}
\label{app:convergence_D2Q9_BGK}

\begin{figure*}[ht]
\begin{minipage}{1.\textwidth}
\centering
\hspace{-8mm}
\includegraphics[scale=0.95]{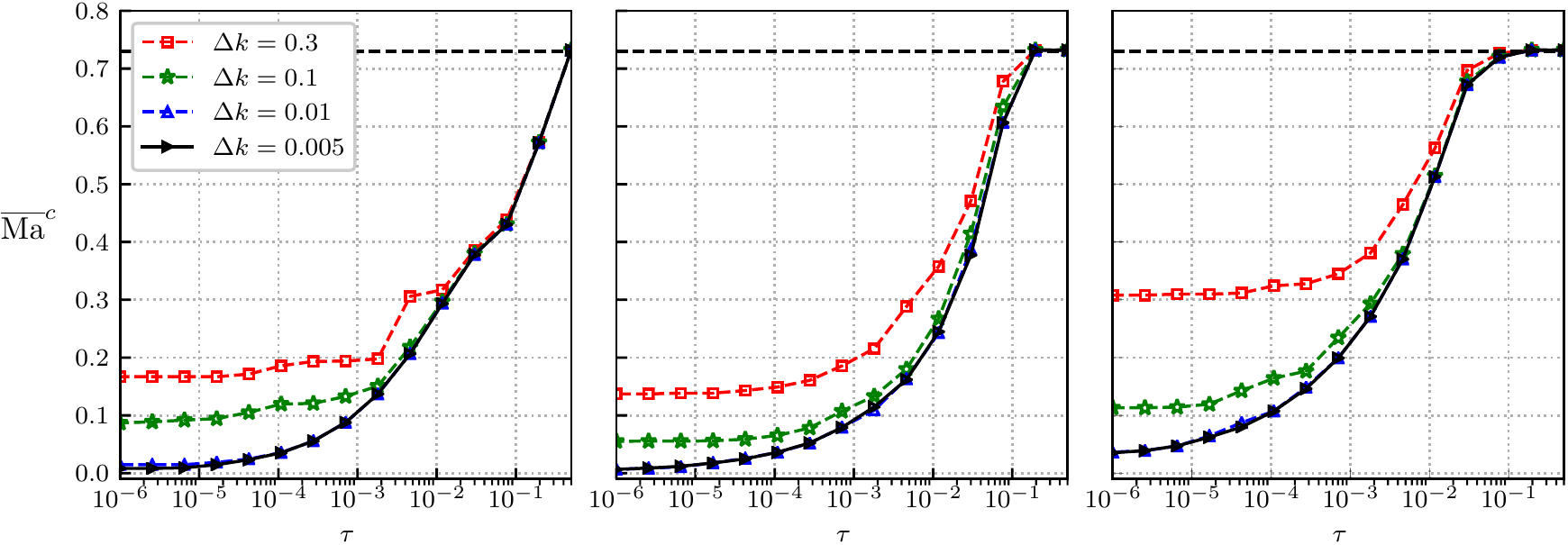}
\begin{minipage}{1.\textwidth}
\hspace{1.5cm}
\begin{minipage}{0.2\textwidth}
\subcaption{$N=2$}
\end{minipage}
\begin{minipage}{0.43\textwidth}
\subcaption{$N=3^*$}
\end{minipage}
\begin{minipage}{0.2\textwidth}
\subcaption{$N=4^*$}
\end{minipage}
\end{minipage}
\caption{Convergence of the critical mean flow Mach number $\overline{\mathrm{Ma}}^c$ of the BGK with the D2Q9 lattice and several equilibrium distribution orders $N$, considering any mean flow orientation. Several spectral discretizations with a step $\Delta k$ are considered. The dashed line represents the theoretical limit of lattice Boltzmann models with a second-order equilibrium: $\overline{\mathrm{Ma}}^c = \sqrt{3}-1 \approx 0.73$~\cite{PAM2019}. \label{fig:convergence_Mac_BGK_D2Q9_allflows}}
\end{minipage}
\end{figure*}

This apppendix aims at providing information regarding the convergence of the linear stability analyses performed in this work, especially regarding the critical Mach number obtained. With the BGK collision model, Sec.~\ref{sec:LSA_results} highlights the presence of thin instability peaks in the spectral space. Therefore, a sufficiently resolved spectral resolution is required to capture them, so as to obtain a correct estimation of the maximum reachable Mach number. It is recalled here that each study is performed for any wavenumber so as $k_x \in [-\pi, \pi]$ and $k_y \in [0, \pi]$ with a step $\Delta k$. Fig.~\ref{fig:convergence_Mac_BGK_D2Q9_allflows} displays the maximal Mach number obtained with the BGK-D2Q9 model for several resolutions $\Delta k$, considering mean flow orientations in $[0^\circ, 45^\circ]$ with a step of $1^\circ$. A convergence in the results can be noticed for $\Delta k < 0.01$.

\begin{figure*}[ht]
\begin{minipage}{1.\textwidth}
\centering
\hspace{-8mm}
\includegraphics[scale=0.95]{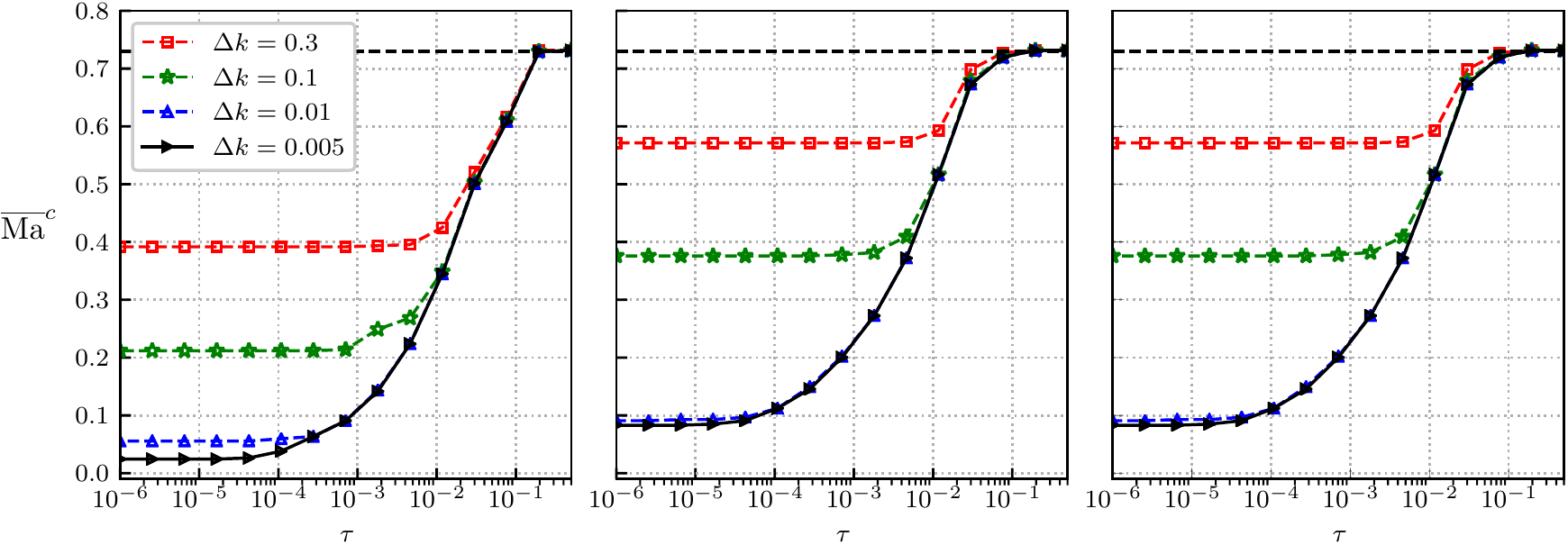}
\begin{minipage}{1.\textwidth}
\hspace{1.5cm}
\begin{minipage}{0.2\textwidth}
\subcaption{$N=2$}
\end{minipage}
\begin{minipage}{0.43\textwidth}
\subcaption{$N=3^*$}
\end{minipage}
\begin{minipage}{0.2\textwidth}
\subcaption{$N=4^*$}
\end{minipage}
\end{minipage}
\caption{Convergence of the critical mean flow Mach number $\overline{\mathrm{Ma}}^c$ of the BGK with the D2Q9 lattice and several equilibrium distribution orders $N$, considering $x$-aligned mean flows only. Several spectral discretizations with a step $\Delta k$ are considered. The dashed line represents the theoretical limit of lattice Boltzmann models with a second-order equilibrium: $\overline{\mathrm{Ma}}^c = \sqrt{3}-1 \approx 0.73$~\cite{PAM2019}. \label{fig:convergence_Mac_BGK_D2Q9_Horizontalflow}}
\end{minipage}
\end{figure*}

Fig.~\ref{fig:convergence_Mac_BGK_D2Q9_Horizontalflow} provides similar results considering $x$-aligned mean base flows only. It highlights the importance of investigating any flow orientation in order to obtain a correct estimation of the stability property of a given scheme, as a consequence of the large anisotropy of numerical errors.

\section{\textit{A priori} derivation of the PR scheme}
\label{app:apriori_derivation_PR}

Note that every quantity presented in this appendix is \textit{dimensional}, as adopted in Sec.~\ref{sec:discussion}. \newline

Let us start with the following system of equations, continuous in time and space:
\begin{align}
\label{eq:app_regularized_DVBE}
    \frac{\partial f_i}{\partial t} + \boldsymbol{e_i} \cdot \frac{\partial f_i}{\partial \boldsymbol{x}} = \Omega_i^{\mathrm{PR}} = \left[ \mathbf{H}^{-1} \mathbf{R} \mathbf{H} \right]_{ij} \left( f_j - f_j^{eq, N} \right),
\end{align}
where $\mathbf{H}$ and $\mathbf{R}$ are $(V \times V)$ matrices given in Sec.~\ref{sec:apriori_derivation_PR}. Integrating Eq.~(\ref{eq:app_regularized_DVBE}) along a characteristic line and using a trapezium rule yields
\begin{align}
    & f_i(\boldsymbol{x}+ \boldsymbol{e_i}\Delta t, t+\Delta t) - f_i(\boldsymbol{x}, t) \nonumber \\
    & \hspace{1cm}= \left[ \mathbf{H}^{-1} \mathbf{R}\mathbf{H} \right]_{ij} \left( \frac{\Delta t}{2} \left( f_j^{neq, N}(\boldsymbol{x}+\boldsymbol{e_i}\Delta t, t+\Delta t) + f_j^{neq, N}(\boldsymbol{x}, t) \right) + O(\Delta t^3) \right),
\end{align}
where $f_i^{neq,N}=f_i - f_i^{eq,N}$. As with the BGK collision model~\cite{He1998}, a new variable can be introduced:
\begin{align}
\label{eq:app_variable_change_PR}
    g_i = f_i - \frac{\Delta t}{2} \Omega_i^{\mathrm{PR}},
\end{align}
leading to the following equation
\begin{align}
\label{eq:app_DVBE_PR2_integrated}
    g_i (\boldsymbol{x}+\boldsymbol{e_i}\Delta t, t+\Delta t) - g_i(\boldsymbol{x}, t) = \left[ \mathbf{H}^{-1} \mathbf{R}\mathbf{H} \right]_{ij} \left( \Delta t f^{neq, N}_j(\boldsymbol{x}, t) + O(\Delta t^3) \right).
\end{align}
In order to obtain an explicit collide and stream scheme, the right-hand side term of (\ref{eq:app_DVBE_PR2_integrated}) has to be expressed as a function of $g_i$:
\begin{align}
     & \Omega^{\mathrm{PR}}_i = \left[ \mathbf{H}^{-1} \mathbf{R} \mathbf{H} \right]_{ij} \left( f_j - f_j^{eq, N} \right) = \left[ \mathbf{H}^{-1} \mathbf{R} \mathbf{H} \right]_{ij} \left( g_j + \frac{\Delta t}{2} \Omega^{\mathrm{PR}}_j - f_j^{eq, N} \right), \\
     & \hspace{1cm} \Rightarrow \left[ \mathbf{H}^{-1} \left( \mathbf{I} - \frac{\Delta t}{2} \mathbf{R} \right) \mathbf{H} \right]_{ij} \Omega^{\mathrm{PR}}_j = \left[ \mathbf{H}^{-1} \mathbf{R} \mathbf{H} \right]_{ij} \left( g_j - f_j^{eq, N} \right).
\end{align}
After some math, this leads to
\begin{align}
    \Delta t \, \Omega_i^{\mathrm{PR}} = \left[ \mathbf{H}^{-1} \mathbf{R^D} \mathbf{H} \right]_{ij} \left( g_j - f_j^{eq, N} \right),
\end{align}
with 
\begin{align}
    \mathbf{R^D} = \left(1 - \frac{\Delta t}{\tau+ \Delta t/2} \right) \mathbf{P}^{(2)} - \mathbf{I} = \left( 1-\frac{\Delta t}{\overline{\tau}} \right)  \mathbf{P}^{(2)} - \mathbf{I},
\end{align}
where $\overline{\tau}=\tau + \Delta t/2$ and $\mathbf{P}^{(2)}$ is the projection matrix onto second-order moments. For instance, with the D2Q9 lattice, $\mathbf{P}^{(2)} = \mathrm{diag}(0, 0, 0, 1, 1, 1, 0, 0, 0)$. The following numerical scheme, explicit for $g_i$, is obtained:
\begin{align}
\label{eq:app_PR_dimensional}
    g_i (\boldsymbol{x}+\boldsymbol{e_i}\Delta t, t+\Delta t) &= g_i(\boldsymbol{x}, t) + \left[ \mathbf{H}^{-1} \mathbf{R^D}\mathbf{H} \right]_{ij} \left( g_j - f_j^{eq, N} \right) + O(\mathbf{R}\, \Delta t^3) \\
    & = g_i(\boldsymbol{x}, t) + \left[ \left( 1 - \frac{\Delta t}{\overline{\tau}} \right) \mathbf{H}^{-1} \mathbf{P}^{(2)}\mathbf{H} - \mathbf{I} \right]_{ij} \left( g_j - f_j^{eq, N} \right)+ O(\mathbf{R}\, \Delta t^3) \\
    & = f_i^{eq,N} + \left( 1 - \frac{1}{\overline{\tau}} \right) \left[ \mathbf{H}^{-1} \mathbf{P}^{(2)}\mathbf{H} \right]_{ij}\left( g_j - f_j^{eq, N} \right)+ O(\mathbf{R}\, \Delta t^3).
\end{align}
Dropping the $O(\Delta t^3)$ error, the projected regularized scheme of Eq.~(\ref{eq:general_form_regul}) is recovered on the discrete distributions $g_i$. Note that this matrix form of the PR scheme is well-known in the literature~\cite{Latt2007}.

\bibliographystyle{acm}

\end{document}